\documentclass[usenatbib, MNRAS]{emulateapj}
\usepackage{xspace}

\usepackage{hyperref}

\usepackage{epstopdf}
\usepackage{inputenc}
\usepackage[T1]{fontenc}
\usepackage{lmodern}
\usepackage{inputenc}
\usepackage{epsfig}
\usepackage{epstopdf}
\usepackage{amsmath}
\usepackage{lmodern}
\usepackage[T1]{fontenc}
\usepackage{natbib}
\usepackage{relsize}

\newcommand{\kms}{km~s$^{-1}$ }
\newcommand{\kmsp}{km~s$^{ -1}$}
\newcommand{\oi}{O~{\sc I}~}
\newcommand{\ci}{C~{\sc I}~}
\newcommand{\cip}{C~{\sc I}}
\newcommand{\nv}{N~{\sc V}~}
\newcommand{\cii}{C~{\sc II}~}

\newcommand{\siii}{Si~{\sc II}~}
\newcommand{\siiii}{Si~{\sc III}~}

\newcommand{\siiv}{Si~{\sc IV}~}
\newcommand{\vcen}{v$_{\text{cen}}$~}
\newcommand{\vn}{v$_{90}$~}
\newcommand{\oip}{O~{\sc I}}

\newcommand{\nvp}{N~{\sc V}}
\newcommand{\ciip}{C~{\sc II}}
\newcommand{\siiip}{Si~{\sc II}}

\newcommand{\siiiip}{Si~{\sc III}}

\newcommand{\civp}{C~{\sc IV}}
\newcommand{\civ}{C~{\sc IV}~}
\newcommand{\siivp}{Si~{\sc IV}}
\newcommand{\vcenp}{$v_{\text{cen}}$}
\newcommand{\vnp}{$v_{90}$}

\newcommand{\mstar}{$M_\ast$ }
\newcommand{\mstarp}{$M_\ast$}

\newcommand{\ciis}{C{\sc II$^\ast$} }
\newcommand{\ciisp}{C{\sc II$^\ast$}}

\newcommand{\sfr}{M$_\odot$~yr$^{-1}$ }
\newcommand{\sfrp}{M$_\odot$~yr$^{-1}$}

\shorttitle{Photo-ionized Outflows}
\shortauthors{Chisholm et al.}

\begin{document}

\title{Shining A Light On Galactic Outflows: Photo-Ionized Outflows}
\author{John Chisholm}
\affil{Astronomy Department, University of Wisconsin, Madison, 475
  N. Charter St., WI 53711, USA}
\email{chisholm@astro.wisc.edu}

\author{Christy A. Tremonti}
\affil{Astronomy Department, University of Wisconsin, Madison, 475
  N. Charter St., WI 53711, USA}

\author{Claus Leitherer}
\affil{Space Telescope Science Institute, 3700 San Martin Drive, Baltimore, MD 21218, USA}

\author{Yanmei Chen}
\affil{Department of Astronomy, Nanjing University, Nanjing 210093, China}

\author{Aida Wofford}
\affil{UPMC-CNRS, UMR7095, Institut d\rq{}Astrophysique de Paris, F-75014 Paris, France}

\begin{abstract}

We study the ionization structure of galactic outflows in 37 nearby, star forming galaxies with the Cosmic Origins Spectrograph on the {\it Hubble Space Telescope}. We use the O~{\sc I}, Si~{\sc II}, Si~{\sc III}, and Si~{\sc IV} ultraviolet absorption lines to characterize the different ionization states of  outflowing gas. We measure the equivalent widths, line widths, and outflow velocities of the four transitions, and find shallow scaling relations between them and galactic stellar mass and star formation rate. Regardless of the ionization potential, lines of similar strength have similar velocities and line widths, indicating that the four transitions can be modeled as a co-moving phase. The Si equivalent width ratios (e.g. Si~{\sc IV}/Si~{\sc II}) have low dispersion, and little variation with stellar mass; while ratios with O~{\sc I} and Si vary by a factor of 2 for a given stellar mass. Photo-ionization models reproduce these equivalent width ratios, while shock models under predict the relative amount of high ionization gas. The photo-ionization models constrain the ionization parameter (U) between $-2.25$~<~log(U)~<~-1.5, and require that the outflow metallicities are greater than 0.5~Z$_\odot$. We derive ionization fractions for the transitions, and show that the range of ionization parameters and stellar metallicities leads to a factor of 1.15-10 variation in the ionization fractions. Historically, mass outflow rates are calculated by converting a column density measurement from a single metal ion into a total Hydrogen column density using an ionization fraction, thus mass outflow rates are sensitive to the assumed ionization structure of the outflow.


\end{abstract}
\section{INTRODUCTION}

Star formation is an inefficient process. Gravity collapses cold gas on relatively short timescales, and left in isolation gas should be converted into stars on about a free-fall time. This naive picture predicts that all of the gas in the universe should be rapidly converted into stars. However, observations show that only 1-10\% of the gas is converted into stars \citep{larson74, kennicutt, moster10}, and these stars are formed in multiple free-fall times \citep{kennicutt}. Therefore, star formation is both a slow, and an inefficient process. 

The standard solution to inefficient star formation has been stellar feedback. Massive stars emit high energy photons throughout their lives, which heat and accelerate the surrounding gas. Meanwhile, at the end of their lives, massive stars explode as supernovae, depositing large amounts of energy into the surrounding gas. This energy and momentum heats the gas, drives turbulence, and slows down the star formation. Accounting for stellar feedback leads to more realistic cosmological star formation histories, with star formation peaking at early times, but continuing at a moderate pace to late times \citep{oppenheimer06, hopkins12a, hopkins14}. 

When star formation is highly concentrated, the energy and momentum from the high mass stars ejects  gas out of the star forming region, in a large scale galactic outflow \citep{heckman90, veilleux, erb15}. Theoretically, these galactic outflows move gas from the star forming regions into the galactic halo, where it can then fall back down onto the galaxy as a galactic fountain \citep{shapiro}, or completely escape the galactic potential \citep{heckman2000, martin2005, rupkee2005, weiner, chisholm}. Whether the gas is recycled back into the galaxy, or lost through an outflow depends on the velocity of the outflow, and the mass of the galaxy, where lower mass galaxies are more susceptible to losing gas. This may create the observed Mass-Metallicity relation \citep{tremonti04, andrews13}, by preferentially removing the metals from low-mass galaxies \citep{heckman90, finlator08, dave12}. 

The inefficiency of star formation is related to how much mass is transported out of the galaxy through a galactic outflow. Various theories predict different \lq\lq{}scaling relations\rq\rq{}, or how the outflow properties (like mass outflow rate and outflow velocity) scale with host galaxy properties \citep{castor75, mckee77, weaver77, ferrara2000, springel03, murray05, hopkins12a}. In practice, mass outflow rates have proven difficult to observe.  Firstly, the least contaminated outflow tracer in the optical is the Na~{\sc I} doublet. With an ionization potential of 5.1~eV, Na~{\sc I} requires large dust opacities to shield it from photo-ionization \citep{chen10}. This has forced previous studies to focus on dusty, massive starbursts, which decreases the dynamical range of the sample \citep{heckman2000, martin2005, rupkee2005, chen10}. Secondly, the outflow geometry is highly unknown. Typically outflows are assumed to be biconical, largely due to the spectacular nearby outflow in M82, but the opening angle, inner and outer outflow radius, and clumpiness of the outflows are all uncertain.

 Finally, to calculate a mass outflow rate the fraction of the total mass in each transition (the ionization fraction) and the metallicity of the outflow are required. Both of these numbers are crucial because they convert the measured column density of a given transition (say Na~{\sc I}) into a total hydrogen column density. Typically the ionization fractions are either ignored \citep{weiner} or set as constant values \citep{heckman2000, martin2005, rupkee2005}, but this adds a factor of 40 scatter to the mass outflow rates calculated using Na~{\sc I} \citep{murray07}. A detailed observational analysis of the ionization structure of galactic outflows is needed before accurate mass outflow rates can be calculated.

In \citet{chisholm} (hereafter, Paper {\sc I}) we studied how the outflow velocity of \siii gas scales with the stellar mass (\mstarp) and star formation rate (SFR) of the host galaxies. We found that the \siii velocity scales shallowly, but significantly, with both \mstar and SFR. Here, we expand the analysis by studying four UV transitions: \oip, \siiip, \siiiip, \siivp. Using these transitions we explore the ionization structure of the galactic outflows, and use models to determine the ionization mechanism of the outflows. Establishing the ionization mechanism enables robust calculations of the mass outflow rates for future work. We first review the data reduction process, the measurement of the absorption lines, and the measurement of host galaxy properties (\autoref{data}). We then summarize the different physical conditions probed by each transition (\autoref{phases}). In \autoref{results} we use the equivalent widths, velocities, and velocity distributions of the absorption lines to study how the outflow properties scale with host galaxy properties, and how these scaling relations differ between the transitions. We discuss the kinematic implications for the outflows (\autoref{outflows}). Finally we use photo-ionization and shock ionization models to describe the ionization structure of the outflows, and conclude that the photo-ionization models best reproduce the data (\autoref{ionize}). In \autoref{conclusion} we summarize our results. 

In this paper we use $\Omega_M = .28$, $\Omega_\Lambda = .72$ and H$_0$ = 70~\kms~Mpc$^{-1}$ \citep{wmap}.

\section{DATA}
\label{data}
\subsection{Sample}
\begin{figure}[t]
\includegraphics[width = 0.5\textwidth]{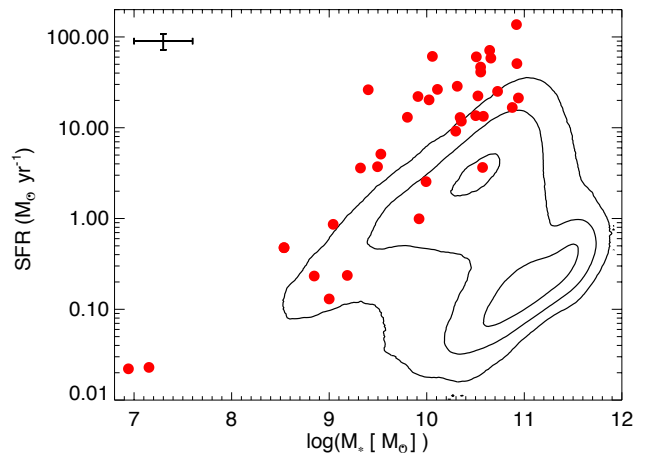}
\caption{Plot of the calculated SFR and stellar mass of the COS sample. Filled red circles show the 37 galaxies that make up the sample. A representative error bar for the stellar mass and star formation rate is given on the left. The contours shown are from the full Sloan Digital Sky Survey DR7, and enclose 25\%, 68\% and 85\% of the SDSS sample using the JHU-MPA calculation of stellar masses and SFRs \citep{kauffmann2003, brinchmann2004}. The sample covers both normal star-forming galaxies on the SDSS main-sequence, and starburst galaxies that lie roughly 1~dex off the main-sequence.}
\label{fig:masssfr}
\end{figure}

We form a sample of 37 galaxies by combining nine previous COS-GO/GTO proposals that target star forming or starbursting galaxies with the G130-M, or G160-M, grating on the Cosmic Origins Spectrograph (COS) \citep{cos} on the Hubble Space Telescope (HST). The sample size is reduced from Paper {\sc I} because we impose a stricter detection threshold (see below), which eliminates many of the lowest signal-to-noise galaxies.  \autoref{tab:sample} gives the proposal IDs, PIs, and references for previous papers using the data. We also give the number of galaxies from each proposal used in the sample, and the typical rest-frame wavelength coverage of each proposal. The typical rest-frame wavelength varies between proposals because different proposals target galaxies with a variety of redshifts, and use different COS setups. \autoref{fig:masssfr} compares the stellar masses (\mstarp) and star formation rates (SFR) of the sample (in red points) with galaxies drawn from the Sloan Digital Sky Survey (SDSS; black contours). The sample covers a wide range of \mstar and SFR, including spirals with typical SFRs, massive interacting starbursts, and dwarf irregulars. The stellar masses, SFRs and morphologies of the sample are given in \autoref{tab:samplegal}.

\begin{deluxetable*}{ccccc}
\tablewidth{0pt}
\tablecaption{Previous Proposals Used}
\tablehead{
\colhead{(1)} &
\colhead{(2)} &
\colhead{(3)}  &
\colhead{(4)} &
\colhead{(5)} \\ 
\colhead{Proposal ID} &
\colhead{PI} &
\colhead{References}  &
\colhead{Number}  &
\colhead{Restframe Wavelength Coverage}  \\
\colhead{} &
\colhead{} &
\colhead{}  &
\colhead{}  &
\colhead{(\AA)} 
}
\startdata
11522 & J. Green   & \citet{france2010, wofford2013} & 3 & 1100-1400\\
& & \citet{ostlin, Rivera} & \\ 
11579 &  A. Aloisi & \citet{james} & 7 & 1130-1430 \\
11727  & T. Heckman & \citet{heckman2011} & 5 & 1000-1550 \\
12027 &  J. Green &  \citet{wofford2013} & 2 & 1100-1400 \\
12173 &   C. Leitherer &  \citet{claus2012} & 4 & 1130-1420 \\
12583  &  M. Hayes & \citet{hayes, ostlin}  & 5& 1120-1420 \\
& & \citet{Rivera} & & \\ 
12604  & J. Fox & \citet{fox2013, richter2013, fox14} &  1 & 1120-1600 \\
12928  &  A. Henry & \citet{jaskot}  & 2 & 1000-1500\\
13017 &  T. Heckman & \citet{alexandroff, heckman15} & 8 & 1000-1600\\

\enddata
\tablecomments{Table of the 9 COS-GO/GTO proposals used to form the sample. Original HST proposal ID numbers and PIs are given in the first two columns, with associated references in the third column. The number of galaxies per proposal in the final sample of 37 is shown in Column 4. The approximate restframe wavelength regime of each sample is given in the last column. This depends on both the instrument set-up and the redshift of the project.}
\label{tab:sample}
\end{deluxetable*}

\begin{deluxetable*}{llcccc}
\tablewidth{0pt}
\tablecaption{Lines Used}
\tablehead{
\colhead{(1)} &
\colhead{(2)} &
\colhead{(3)}  &
\colhead{(4)} &
\colhead{(5)} & 
\colhead{(6)} \\
\colhead{Element} &
\colhead{$f$-value} &
\colhead{Formation Potential} &
\colhead{Ionization Potential} & 
\colhead{Abundance} &
\colhead{Reference}  \\
\colhead{} &
\colhead{} &
\colhead{(eV)} &
\colhead{(eV)} &
\colhead{} &
\colhead{} 
}
\startdata
{C~{\sc I} 1193.031} & 0.0409& 0 & 11.3 & 1.1 $\times$ 10$^{-4}$ & 1 \\
{C~{\sc I} 1260.735} & 0.0394&  0 & 11.3 & 1.1 $\times$ 10$^{-4}$ & 1  \\
{C~{\sc I} 1277.245} & 0.0923& 0 & 11.3 & 1.1 $\times$ 10$^{-4}$ & 1 \\
{C~{\sc I} 1328.834} & 0.0580&  0 & 11.3 & 1.1 $\times$ 10$^{-4}$ & 1 \\
\hline
H~{\sc I} 1215.67 & 0.416 & 0 & 13.6 & 1 &2 \\
{O~{\sc I} 1302.168} & 0.052 & 0 & 13.6 & 5.9 $\times$ 10$^{-4}$ & 3 \\
\hline
Fe~{\sc II} 1143.220 & 0.019 & 7.9 & 16.2 &5.2 $\times$ 10$^{-6}$ & 4\\
Fe~{\sc II} 1144.926 & 0.083 &  7.9 & 16.2  &5.2 $\times$ 10$^{-6}$ & 4\\
Fe~{\sc II} 1260.525 & 0.020 & 7.9 & 16.2  &5.2 $\times$ 10$^{-6}$ & 4\\
{Si~{\sc II} 1190.42} & 0.277 & 8.2 & 16.3 & 3.2 $\times$ 10$^{-5}$ & 5\\
{Si~{\sc II} 1193.28} & 0.575 &  8.2 & 16.3 & 3.2 $\times$ 10$^{-5}$ & 5 \\
{Si~{\sc II} 1260.42} & 1.22&  8.2 & 16.3 &3.2 $\times$ 10$^{-5}$ & 5\\
{Si~{\sc II} 1304.37} & 0.0928&  8.2 & 16.3 &3.2 $\times$ 10$^{-5}$ & 5\\
S~{\sc II} 1250.578 & 0.00602 &  10.4 & 23.3 &1.5 $\times$ 10$^{-5}$ & 6 \\
S~{\sc II} 1253.805 & 0.0121 &  10.4 & 23.3 &1.5 $\times$ 10$^{-5}$ & 6 \\
S~{\sc II} 1259.518 & 0.0182 &  10.4 & 23.3 &1.5 $\times$ 10$^{-5}$ & 6 \\
C~{\sc II} 1334.532 & 0.129 &  11.3 & 24.4& 1.1 $\times$ 10$^{-4}$ & 3\\
C~{\sc II*} 1335.708 & 0.115 &  11.3 & 24.4& 1.1 $\times$ 10$^{-4}$ & 3\\
\hline
{Si~{\sc III} 1206.51} & 1.67 &  16.3 & 33.5 &3.2 $\times$ 10$^{-5}$ & 7 \\
S~{\sc III} 1190.206 & 0.0258 &  23.3 &34.9  &1.5 $\times$ 10$^{-5}$ & 6 \\

{Si~{\sc IV} 1393.755} & 0.513& 33.5 & 45.1 &3.2 $\times$ 10$^{-5}$ & 8\\
{Si~{\sc IV} 1402.770} & 0.255&  33.5 & 45.1 &3.2 $\times$ 10$^{-5}$ & 8\\
C~{\sc IV} 1548.202 & 0.19 & 47.9 & 64.5 &1.1 $\times$ 10$^{-4}$ & 1\\
C~{\sc IV} 1550.774 & 0.0952 &  47.9 & 64.5 &1.1 $\times$ 10$^{-4}$ & 1 \\
\hline
{N~{\sc V} 1238.821}  & 0.156 &  77.5 & 97.9 & 6.2 $\times$ 10$^{-5}$ & 9 \\
{N~{\sc V} 1242.804} & 0.078 &  77.5 & 97.9 &6.2 $\times$ 10$^{-5}$ & 9\\
\enddata
\tablecomments{Table of the lines studied in the COS wavelength regime. The horizontal lines distinguish groupings of ionization:  above the first line the gas is ionized by photons with energies less than 1~Rydberg, while  between the first and second line the gas is ionized by photon energy of exactly 1~Rydberg. Between the second and third line the gas can be either photo-ionized or neutral, while the transitions below the third line are completely photo-ionized. Transitions below the last line probe gas above the second He ionization potential, and are likely not photo-ionized  (see \autoref{phases}). The first column gives the names of the transitions and their wavelengths, with their $f$-value in the second column.  Columns 3 and 5 give the formation, and ionization potentials of each transition. Column 5 gives the observed solar gas-phase, dust-depleted abundances by number for each element, relative to H \citep{jenkins, draine}. Note these abundances are the total for each element, and not for the particular ionization stage. We use the latest atomic data from NIST \citep{nist} with the references as: 1) \citet{carbon}, 2) \citet{hydrogen} 3) \citet{oxygen} 4) \citet{iron}  5) \citet{si2} 6)\citet{s} 7)  \citet{si3} 8) \citet{si4} 9) \citet{nitrogen}.}
\label{tab:lines}
\end{deluxetable*}

\begin{figure}
\includegraphics[width = 0.5\textwidth]{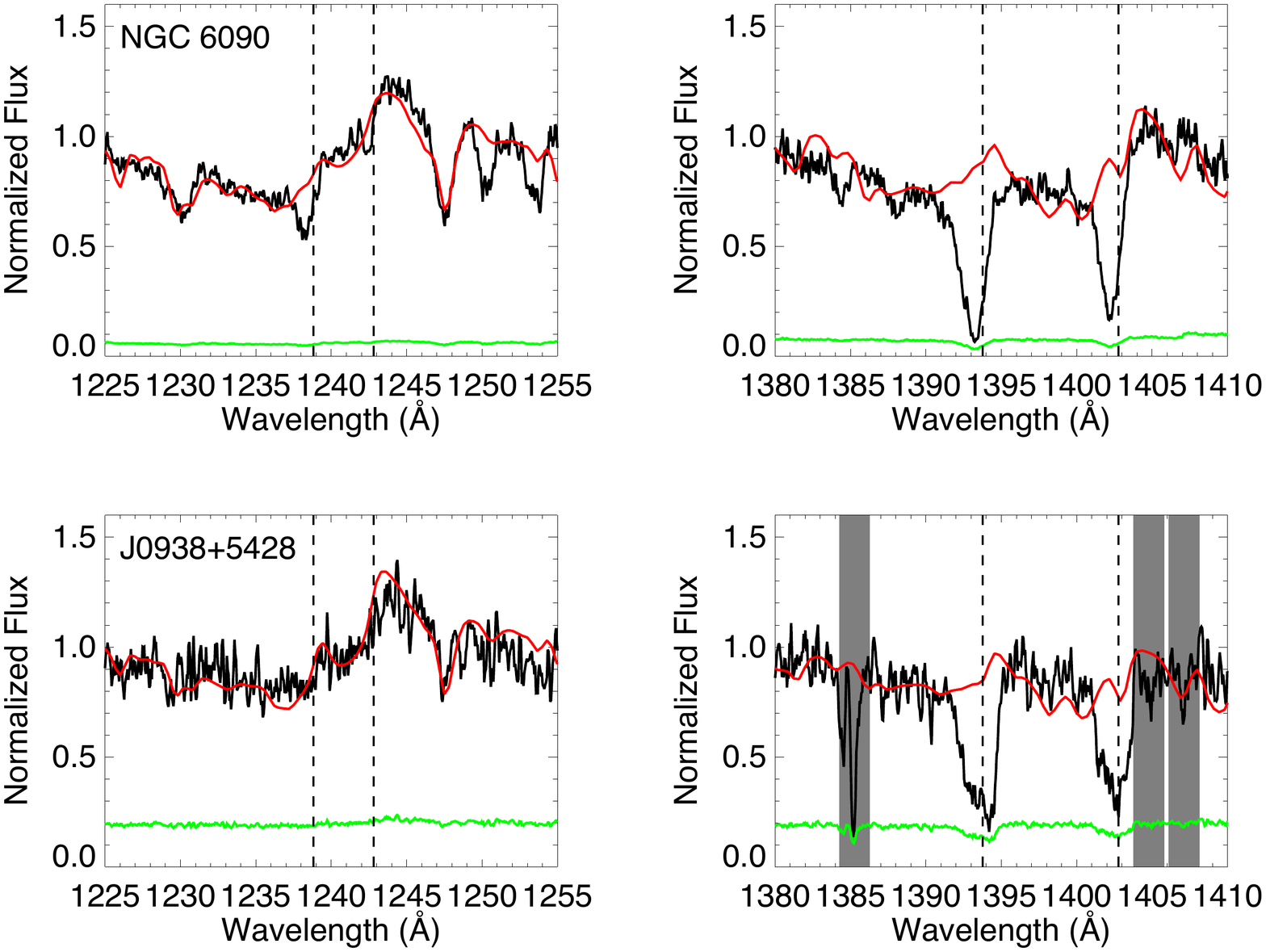}
\caption{ Example of Starburst99 stellar wind fits for two galaxies: NGC~6090 (top panels) and J0938+5428 (lower panels). The left panels are the fits to the \nvp~1240~\AA\ doublets, and the right panels are the fits to the \siivp~1400~\AA\ doublets. The Starburst99 stellar continuum fit is over-plotted on the galaxy spectrum, the lower line is the error on the measured flux, and shaded regions show possible Milky Way absorption features. Dashed lines show the zero-velocity of the \nv and \siiv line centers. \siiv shows deep, broad absorption, while the \nv absorption is weak and blueshifted for NGC~6090, but not observed for J0938+5428.}
\label{fig:windfits}
\end{figure}

\subsection{COS data}
\label{cos}
A full description of the COS data reduction and analysis is available in Paper {\sc I}, and here we provide an overview of the steps taken to derive the outflow velocities, equivalent widths, and line widths of the sample. We start by downloading the spectra from the MAST server and processing the data through the CalCOS pipeline, version 2.20.1. The individual spectra are then aligned and combined using the methods outlined in \citet{wakker}. The spectra are then deredshifted using the redshifts from the SDSS, or from NASA/IPAC Extragalactic Database (NED)\footnote{The NASA/IPAC Extragalactic Database (NED) is operated by the Jet Propulsion Laboratory, California Institute of Technology, under contract with the National Aeronautics and Space Administration.}. We normalize the flux to the median flux between 1310-1320\AA, bin the spectra by 5 pixels (12~\kmsp), and smooth by 3 pixels. 

We then fit the stellar continuum with a linear combination of single age Starburst99 stellar continuum models \citep{claus99, claus2010}. We use the fully theoretical spectral libraries from the Geneva group with high-mass loss \citep{geneva94}, which are computed using the WM-basic code  \citep{claus99, claus2010}. To make the individual Starburst99 models, we assume a stellar metallicity from the literature (see Paper~{\sc I} for the metallicities used), a single burst star formation law, and an array of ages between 1~Myr and 20~Myr with time steps between 1 and 5~Myr. After 20~Myr the UV stellar continuum is increasingly dominated by B-type stars, and evolves slowly thereafter \citep{demello}. During the fit, we account for continuum dust attenuation by reddening the Starburst99 models with a Calzetti extinction law \citep{calzetti}, and fitting for the E(B-V). We use MPFIT\footnote{The entire package can be found at http://purl.com/net/mpfit}, a non-linear least squares fitting routine to find the best-fit Starburst99 model and E(B-V) \citep{mpfit}. We then normalize the observed spectrum with the stellar continuum model. 

Fitting the stellar continuum with a linear combination of multiple single aged stellar spectra is important to remove the strong stellar wind features from the \nv 1240~\AA, \siiv 1400~\AA, and \civ 1550~\AA\ lines. \autoref{fig:windfits} shows two examples of the wind regions around \nv and \siivp. The fits of the lines show how the stellar continuum is reproduced, and how the \siiv lines are distinguishable from the stellar winds. Meanwhile, NGC~6090 shows weak \nv absorption (upper left panel). The stellar continuum residuals have a median of 3\%, but the wind lines could be more heavily effected. We conservatively estimate the errors on the stellar continuum fits to be 10\%.

Milky Way lines can severely effect the measured absorption. At low redshifts the Milky Way lines blend with similar transitions from the target galaxy, while at higher redshifts redder Milky Way lines are coincident with bluer lines from the target galaxy. As described in Paper~{\sc I}, we fit and remove a large suite of Milky Way absorption lines. While fitting the Milky Way lines we convolve the fits with a Gaussian, and fit for a full width at half maximum (FWHM) of the Milky Way lines. We use the Milky Way FWHM as the effective spectral resolution. We do not use the COS line-profile because the objects are extended, and the spectral resolution is degraded by the extended distribution of the light by as much as 200~\kms \citep{france09}. The median FWHM of the sample is 62~\kmsp, with a range between 36 and 136~\kmsp, where compact sources have lower FHWMs. This fitted FWHM is used in subsequent fits of the target galaxy\rq{}s absorption lines to account for the instrumental broadening.

As energy is injected into outflows, turbulence and instabilities disperse the entrained gas into a wide range of velocities. Since the gas will not be normally distributed, we use MPFIT to fit the stellar and Milky Way subtracted spectra with a total of up to 10 Voigt components, with initial spacing of 175~\kmsp, and freedom to move up to 88~\kmsp. The number of components used depends on the observed velocity range. 

The Voigt fits are parameterized by a line-center velocity ($v_0$), a covering fraction (C$_f$), a fixed b-parameter (equal to one-half the fitted instrumental FWHM), and a column density. Since many of these parameters can be degenerate with each other at our resolution, we tie the $C_f$ and $v_0$ for ions without multiple transitions (\siiiip, \ciip, and \ciisp) to the \siii transitions. Additionally, there are many weak lines that are blended with the \siii transitions, and we simultaneously fit these features to account for their contributions to the \siii lines. These blended lines include \oip, Fe~{\sc II}, C~{\sc I}, S~{\sc II} and S~{\sc III}. Therefore, each galaxy has three groupings where we tie the parameters together within the groups: (1) \siiip, \siiiip, \oip, C~{\sc I}, Fe~{\sc II}, S~{\sc II}, C~{\sc II}, C~{\sc II$^\ast$}, and S~{\sc III}; (2) \siiv and \civp; and (3)  \nvp. 

\subsection{Measurement of Equivalent Width, Velocity, $\Delta$, and Covering Fraction}

\begin{figure}
\includegraphics[width = 0.5\textwidth]{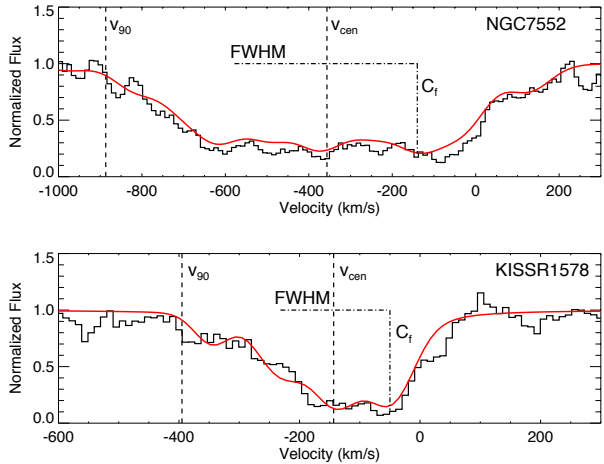}
\caption{Two examples of \siiv line profiles for NGC~7552 (top) and KISSR~1578 (bottom), with the red line giving the multiple component fit to the data. The central velocity (\vcenp) and the velocity at 90\% of the continuum (\vnp) are marked by dashed vertical lines. The horizontal and vertical dotted-dashed lines mark the FWHM (FWHW~=~2.355$\Delta$) and the covering fraction (C$_f$) of the line, respectively. NGC~7552 has a large velocity width, with a nearly constant absorption depth between -600 and -50~\kmsp. C$_f$ is measured at the line center, but is shown offset to illustrate how the product of C$_f$ and FWHM (or $\Delta$) is the equivalent width of the profile. }
\label{fig:profiles}
\end{figure}

Once the line profiles are calculated, we measure the equivalent width, the velocities of the profiles, the velocity widths, the covering fraction at line center, and the optical depth of the fitted profile. We use the fitted line profiles because the fitted profiles account for blended nearby lines, and use all available transitions to constrain the line profile. We calculate the equivalent width (W; in \AA) from the fitted line profile as
\begin{equation}
W = \int^{\lambda_\text{max}}_{\lambda_\text{min}}\left(1-{F_C\left(\lambda\right)}\right) d\lambda
\end{equation}
Where F$_C$ is the multiple-component fit to the continuum normalized data, and the range between $\lambda_\text{max}$ and $\lambda_\text{min}$ is a by-eye region specified for each transition. We calculate the error on W by bootstrapping the F$_C$ with the error on the flux, and creating 1000 simulations of W with a random kernel drawn from the flux error array. These 1000 simulations produce a distribution of W with 1000 values, and we use this distribution to quantify the error on W. We then add this error in quadrature with the 10\% continuum normalization error to calculate the total error on W \citep{Sembach1992}.

\autoref{fig:profiles} shows two examples of line profiles from our sample. These lines show a characteristic \lq\lq{}saw-tooth\rq\rq{} profile, where the blue absorption edge is much more gradual than the red edge \citep{weiner, martin12}. \citet{martin12} postulate that this gradual edge is because  outflows are continually accelerated, and larger velocities corresponds to larger radii. In a mass-conserving flow, either the column density or the covering fraction must drop with radius as a fixed amount of gas is spread over a larger area (a process sometimes termed \lq\lq{}geometric dilution\rq\rq{}). Thus, while the highest velocity gas makes a small contribution to the observed line profile, it may contribute substantially to the mass and energy budget of the outflow. Additionally, the highest velocity gas is most likely escape the potential, and deplete the gaseous reservoir \citep{chisholm}.

We estimate the velocities in two ways: the equivalent width weighted velocity (the central velocity, or \vcenp), and the velocity at 90\% of the continuum (\vnp; the maximum velocity). \vn is the velocity at which the absorption line reaches 90\% of the continuum level, and is measured from the fitted line profile. We measure \vcen from the fitted line profiles as:
\begin{equation}
v_\text{cen} =  \frac{\int^{v_1}_{v_2} \left[1-F_c(v)\right]v~dv}{\int^{v_1}_{v_2} \left[1-F_c(v)\right]dv}
\end{equation}
where the integration regime is the same by-eye regime as for W. The two velocity measurements have different advantages and disadvantages. Inflowing and zero velocity absorption can decrease \vcenp, while resonance emission can fill the profile in at low velocities, increasing \vcenp. We do not observe strong Si~{\sc II}$^\ast$ emission in many galaxies \citep{prochaska2011, martin12, rubin13}, indicating that the resonance emission does not heavily impact the measured velocities. This is likely because the low redshift sight-lines only probe a small volume of the outflow. Moreover, in Paper {\sc I} we discuss how inflowing and low velocity absorption influences \vcenp, and find that only a small fraction of galaxies are affected. Meanwhile, \vn is sensitive to the continuum level, but in Paper {\sc I} we find the \vn relations to have lower scatter than the \vcen relations. Therefore, we focus on the \vn relations, but report both sets of relations.

The velocity width of the line ($\Delta$) is the second moment of the equivalent width, with respect to velocity \citep{Sembach1992}, or
\begin{equation}
\Delta  =\sqrt{\frac{\int^{v_1}_{v_2} \left[1-F_c(v)\right](v-v_\text{cen})^2dv}{\int^{v_1}_{v_2} \left[1-F_c(v)\right]dv}}
\label{eq:b}
\end{equation}
The errors for the two velocities and $\Delta$ are bootstrapped, similarly to W.

Typically, the continuum source is assumed to be completely covered by the absorbing gas, but this is not the case for a clumpy medium, or for certain geometries (galactic outflows are typically assumed to be biconical, with an opening angle that only covers a fraction of the starburst region).  We measure C$_f$ at the line center of the doublets (\siiip, \siivp, \civp, and \nvp), following the method from \citet{hamann} as:
\begin{equation}
\text{C}_f =  \frac{I_B^2 -2I_B + 1}{I_R -2I_B + 1}
\label{eq:cf}
\end{equation} 
Where I$_R$ and I$_B$ are the intensity at line center for the red line (stronger line) and the blue line (weaker line). It is important to measure the C$_f$ from a doublet because the degeneracy between optical depth and C$_f$ can impact the residual intensity level, especially for unsaturated lines. 

Finally, we calculate the optical depth at line center of the doublets ($\tau_0$). Following \citep{hamann} we calculate the optical depth by solving the radiative transfer equation, where the intensity is given by:
\begin{equation}
I = 1- C_f  + C_f e^{-\tau_0}
\end{equation}
and the optical depth of the weaker doublet transition (the redder transition) is given by
\begin{equation}
\tau_0 = \text{ln}\left(\frac{\text{C}_f}{I_R + \text{C}_F - 1}\right)
\label{eq:taucf}
\end{equation}
where C$_f$ is found in \autoref{eq:cf}. This measures the degree of saturation for the transitions. 

In \autoref{fig:profiles} we mark the derived quantities from two \siiv line profiles. The vertical dot-dashed line in \autoref{fig:profiles} marks the C$_f$ of each of the lines, but it is offset from where it is measured, for clarity.  The equivalent width is the area of the absorption feature, and for a moderately saturated line the equivalent width will be the height times the width of the line, or the product of $\Delta$ and C$_f$. It is important to emphasize that C$_f$ and $\tau_0$ are measured at the center of the lines, and do not describe the gas in the wings of the profile. 

An absorption feature is detected if W is significant at the 3$\sigma$ significance level, while we classify a detection as an outflow if \vcen is less than zero, at the one sigma significance level. We note that we did not have an equivalent width cut in Paper 1, therefore this sample has 10 fewer low-signal galaxies. 

We tabulate the measured \vcen (in \autoref{tab:samplevcen}), \vn (in \autoref{tab:samplev90}), and W (in \autoref{tab:sampleew}) for the six transitions that are commonly detected. The values of $\tau_0$, $\Delta$, and C$_f$ are tabulated in \autoref{tab:samplesi4} for the \siiv transition. Finally, the W ratios are given in \autoref{tab:sampleewrat}.

\subsection{Host galaxy properties}
\label{masssfr}

We use a collection of ancillary multi-wavelength data to calculate the host galaxy properties for the sample. Here we give a brief summary of the methods used to calculate the properties, but a detailed overview is given in Paper {\sc I}. 

We calculate the star formation rate (SFR) using a combination of {\it GALEX} \citep{galex} and {\it WISE}  \citep{wise} luminosities. The luminosities are aperture corrected \citep{jarrett2011, wisesupp}, k-corrected \citep{kcorrect, chisholm}, and foreground extinction corrected \citep{galexatlas}. We then use the luminosities to calculate the ultraviolet (UV) and infrared (IR) SFR \citep{jarrett2013}, and convert these into a total SFR using the relation from \citet{buat11}, which uses a Chabrier initial mass function. This method accounts for the dust-obscured SFR, the dominant SFR component for dusty, high SFR, galaxies. \citet{jarrett2013} conservatively estimates the SFR errors using this method to be 20\%.

The stellar mass (\mstarp) is calculated using the {\it WISE} 3.4~$\mu$m luminosities,  and a 3.4 and 4.5~$\mu$m color dependent mass-to-light ratio \citep{querejeta}. The color dependent mass-to-light ratio accounts for the contribution of dust in the 3.4~$\mu$m luminosity, which can be significant for dusty galaxies. To note: this stellar mass calibration is different than the calibration of Paper 1, which used a constant mass-to-light ratio. The different mass-to-light ratio typically decreases the measured \mstar by 0.1-0.5~dex, and impacts the dusty starbursts more than the quiescent galaxies.  The log(\mstarp/M$_\odot$) measurements have an error of 0.3~dex, which incorporates the uncertainty of the dust emission. The Green Pea galaxies have extreme nebular properties, with large emission line equivalent widths, and hot dust temperatures \citep{cardamone, izotov, henry}. For this reason, we use the nebular emission corrected \mstar and H$\alpha$ SFR measurements from \citet{izotov} and \citet{henry} for the two Green Pea galaxies in the sample. All of the \mstar measurements use a Chabrier initial mass function. 

Morphologies are documented from SDSS imaging, and HST imaging, when available \citep{overzier09}. We have four morphological categories: Irregular, Spiral, Merger/Interacting, and Compact. We classify galaxies as mergers or interacting if there are tidal tails, double nuclei, interacting companions, or other major asymmetries. The compact galaxies are usually unresolved even by HST, and in Paper~{\sc I} we postulate that the extreme compact galaxies have likely undergone a major merger. The sample is nearly split evenly between the four categories, leading to small sample sizes within the individual categories. 

All calculated galactic parameters are tabulated in \autoref{tab:samplegal}, below. 

\section{AVAILABLE TRANSITIONS AND PHASES}
\label{phases}

\begin{deluxetable*}{lcccccc}
\tablewidth{0pt}
\tablecaption{Line Detection}
\tablehead{
\colhead{(1)} &
\colhead{(2)} &
\colhead{(3)}  &
\colhead{(4)} &
\colhead{(5)}  &
\colhead{(6)} & 
\colhead{(7)} \\
\colhead{Line} &
\colhead{Coverage} &
\colhead{Detection}  &
\colhead{Outflow} &
\colhead{Detection}  &
\colhead{Outflow} & 
\colhead{W of Constant H} \\
\colhead{} &
\colhead{(Number)} &
\colhead{(Number)}  &
\colhead{(Number)} &
\colhead{(\%)}  &
\colhead{(\%)} &
\colhead{(\AA )}
}
\startdata
C~{\sc I}~1328 & 43 & 1  & 0  & 2 $\pm$ 3 & 0 $\pm$ 0 &  0.10 \\
O~{\sc I}~1302 & 33 & 29 & 19 & 88 $\pm$ 6 & 66 $\pm$ 9 &  0.46\\
Si~{\sc II}~1260 & 45 & 44 & 37  & 98 $\pm$ 5 & 84 $\pm$ 6 & 0.55  \\
Si~{\sc III}~1206& 35 & 30 & 26 & 86 $\pm$ 6 & 87 $\pm$ 6 &  0.69 \\
Si~{\sc IV}~1394 & 38 & 34 & 31&89 $\pm$ 6 & 91  $\pm$ 5 &   0.28 \\
C~{\sc IV}~1548 & 12 & 11 & 11 & 92$\pm$ 12 &100 $\pm$ 30 & 0.44 \\
N~{\sc V}~1239 & 36 & 5 &  5  &14 $\pm$ 4 & 100 $\pm$ 44 &  0.13 \\
\enddata
\tablecomments{Table of the detected transitions in the sample. The first column gives the ionic species, the second column gives the total number of galaxies with this transition without contaminating features (geocoronal emission or Milky Way absorption). The third and fifth columns give the number of galaxies where the transition is detected, at the 3$\sigma$ significance, and the detection faction. The fourth and sixth columns gives the number, and fraction, of those detections that have \vcen less than 0, at the 1$\sigma$ significance. The seventh column gives the expected equivalent width (W) if each transition probed a constant Hydrogen column density, with a value of 10$^{18}$~cm$^{-2}$. The H column densities are converted into the ionic column density using the relative abundances of each element in \autoref{tab:lines}, and into an equivalent width assuming that the line is optically thin and all of the gas is in the given transition. These values are illustrative and are not meant to predict equivalent widths. A more rigorous analysis is made in \autoref{ionize}.}
\label{tab:detection}
\end{deluxetable*}
\begin{figure}
\includegraphics[width = 0.5\textwidth]{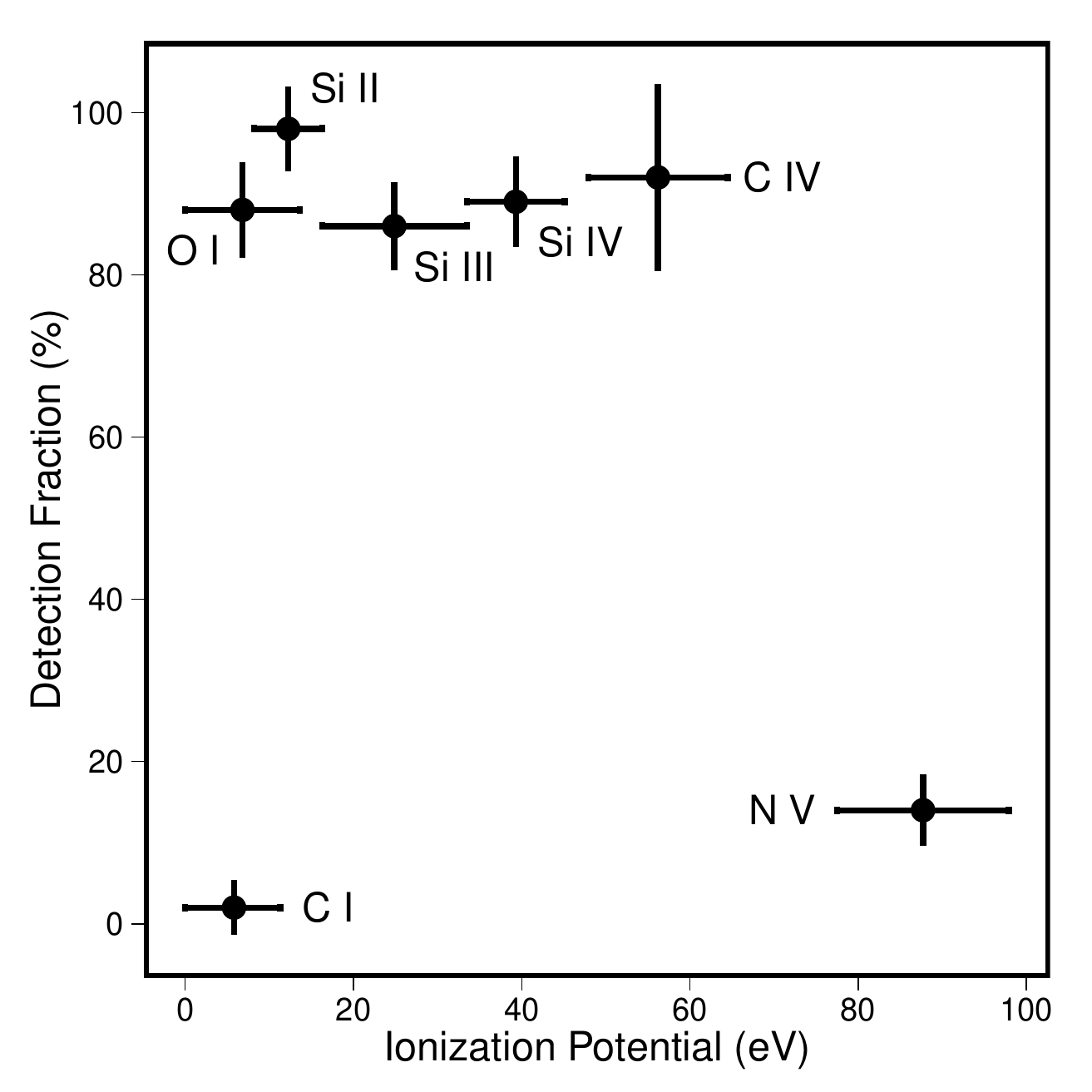}
\caption{Detection fraction of transitions plotted versus the ionization potentials of the transition. Each point corresponds to a specific transition, as shown by the labels. The ionization potentials are given as the full range of the possible ionization potentials; with the minimum corresponding to the formation potential, and the maximum corresponding to the ionization potential (see \autoref{tab:lines}).}
\label{fig:detect}
\end{figure}

We use the full COS G130-M and G160-M wavelength bandpass between 1190 and 1600 to probe a diverse set of atomic transitions. These transitions probe gas in different physical conditions, and in different ionization stages. Most photons with energy above 13.6~eV will ionize Hydrogen gas, meaning transitions with formation potentials below 13.6~eV trace neutral gas (see \autoref{tab:lines}). Helium is the second most abundant element, and has a first ionization potential of 24.6~eV, and a second IP of 54.4~eV. Nearly all of the photons with energies greater than 54.4~eV will ionize H and He, and transitions probing these energies are not likely formed through photo-ionization \citep{spitzer}.

We divide the transitions into categories based on the ionization stages they trace: neutral, partially photo-ionized, and coronal. The partially photo-ionized phase may have a mix of neutral or coronal gas, and their exact formation mechanism depends on the local conditions of the gas and radiation field (see \autoref{ionize} below for modeling of their formation). Additionally, we split the neutral gas by whether it is ionized by energies of less than 13.6~eV (1~Rydberg; \cip), and whether it is ionized by exactly 1~Rydberg (\oip).  The strongest and least contaminated tracers of each phase are:  \ci 1328~\AA\ for the low ionization neutral gas, \oi 1302~\AA\ for higher ionization neutral gas, \siii 1260~\AA,  \siiii 1206~\AA,  \siiv 1394~\AA\,  \civ 1548~\AA  for partially photo-ionized gas, and \nv 1239~\AA\ for the coronal gas. Slightly higher redshifts, and non-standard instrument setups cover the molecular phase with the H$_2$ 1049~\AA\ Lyman band transitions. However, we do not detect H$_2$ absorption in any of the sample (or in a stacked spectrum), which is likely due to the large dust attenuation required to shield molecular gas from UV photons. \cii and \ciis are strong transitions, but we do not use them because we cannot constrain the line profiles of singlets as well as doublets, and both lines trace similar ionization potentials as \siii and \siiiip.

We focus on seven transitions typically covered by the COS bandpass: \cip, \oip, \siiip, \siiiip, \siivp, \civp, and \nvp. Each transition is affected by Milky Way absorption, geocoronal emission, or chip gaps differently. The \oi and \siii 1304 transitions are typically contaminated by the \oi 1302+1306~\AA\ geocoronal lines, while \ci 1277~\AA\ is occasionally located within a chip gap. To avoid the chip gaps, we use the slightly weaker \ci 1328~\AA\ line to quantify \cip. Even though most transitions are within the COS bandpass, the coverage number is different for each transition based on redshifts and instrument set-ups (see \autoref{tab:detection}).

The strengths of the transitions, and relative abundances of the elements, play a role in the interpretation of the trends we measure, as discussed fully in \autoref{ionize}. Before we model the full ionization structure of each transition, we crudely estimate a W detection limit to approximate the strengths of each transition. To do this, we use a constant H column density of 10$^{18}$~cm$^{-2}$, and convert this into the column density of the various transitions using the relative abundances (\autoref{tab:lines}). We then assume the line is optically thin and calculate the expected equivalent width of the line, as shown in the last column of \autoref{tab:detection}. This method is crude (and in \autoref{ionize} we do a more rigorous modeling), but these number give a sense of the relative strength of each transition, due to differences in $f$-values and relative abundances.

The continuum subtraction plays a role in the detection limits, especially for the higher ions that have substantial stellar wind components. We find that the equivalent width of weaker features can be effected by up to 25\% by continuum errors. This largely effects the weaker \siiv lines in dwarf starbursts. These lines are detected at the 1-2$\sigma$ level, but not the 3$\sigma$ level, and are not included in the \siiv sample. This reduces the dynamic range for the \siiv sample. 

Using the estimated Ws in \autoref{tab:detection}, we expect the \oip, \siiip, \siiiip, \siivp, and \civ lines to be the strongest, while the \ci and \nv transitions will be more difficult to detect. However, this is merely to illustrate an idealized scenario of how the Ws compare between the transitions.

\section{RESULTS}
\label{results}
Here we study how the equivalent widths (W) and outflow velocities depend on host galaxy properties. To derive the significances of each trend we use the Kendall\rq{}s $\tau_K$ (denoted as $\tau_K$ to avoid confusion with the optical depth) parameter which is a nonparametric measure of the correlation between two variables, with +1 indicating perfect correlation, -1 indicating perfect anti-correlation, and 0 indicating no correlation. We use a hypothesis test, with the null hypothesis that the relations are uncorrelated, to give the significance of the Kendall\rq{}s $\tau_K$ value in terms of the standard deviation of a normal distribution ($\sigma$). Higher significances are indicated by higher $\sigma$ levels. Only trends that are greater than 3$\sigma$ are considered highly significant. We also use the coefficient of determination (often called R$^2$), which is the ratio of the error sum of squares to the total sum of squares of the relation. R$^{2}$ measures the amount of variation attributable to a given model, and can take values from zero (none of the variation) to one (all of the variation). We use the M-estimator robust regression technique to derive the listed trends, which minimizes the likelihood function, but deemphasizes outliers \citep{feigelson}.

First we explore which gas is present in the outflows, as described by the detection and outflow fraction (\autoref{detect}). The detection fraction crudely probes the phase structure of the gas, and sets the basis for a further study of the ionization structure in \autoref{ionize}. We then explore how the W scales with host galaxy properties, what drives increases in W (\autoref{ewscaling}), and the variance of W from transition to transition (\autoref{ewratios}). We finish by discussing the outflow velocities of the different transitions, how they scale with host galactic properties (\autoref{vscale}), and how they vary between the transitions (\autoref{vphases}).
\subsection{Outflow Detection Rate}
\label{detect}

We define the detection of a line if W is significant at the 3$\sigma$ level. The detection rate is given in  \autoref{fig:detect} and Column 5 of \autoref{tab:detection}. There are four transitions that are frequently detected: \oip, \siiip, \siiiip, and \siivp. While \civp, and \nv are detected, the sample sizes are not enough to draw statistically significant relations (with only 11 and 5 detections, respectively). This means that we are unable to draw meaningful conclusions about the coronal phase of the outflow. In most of the paper we will only focus on the four frequently detected transitions. The average detection rate of these four transitions is constant with ionization potential near 90\%.

While the absorption lines are not always detected, when the lines are detected the velocity is typically less than zero. We find an outflow fraction that is greater than 66\% for all of the transitions, and is fairly constant over ionization potential, with a slight rise from 66\% for \oi to 91\% for the \siiv lines. The detection fraction and outflow fraction are consistent with previous studies using low redshift observations of Na~{\sc I} \citep{heckman2000, rupkee2005, martin2005}, and higher redshift observations of Mg~{\sc II} \citep{weiner} and Fe~{\sc II} \citep{erb12, kornei12, rubin13}.

\subsection{Equivalent Width Scaling Relations}
\label{ewscaling}
\begin{figure*}
\includegraphics[width = \textwidth]{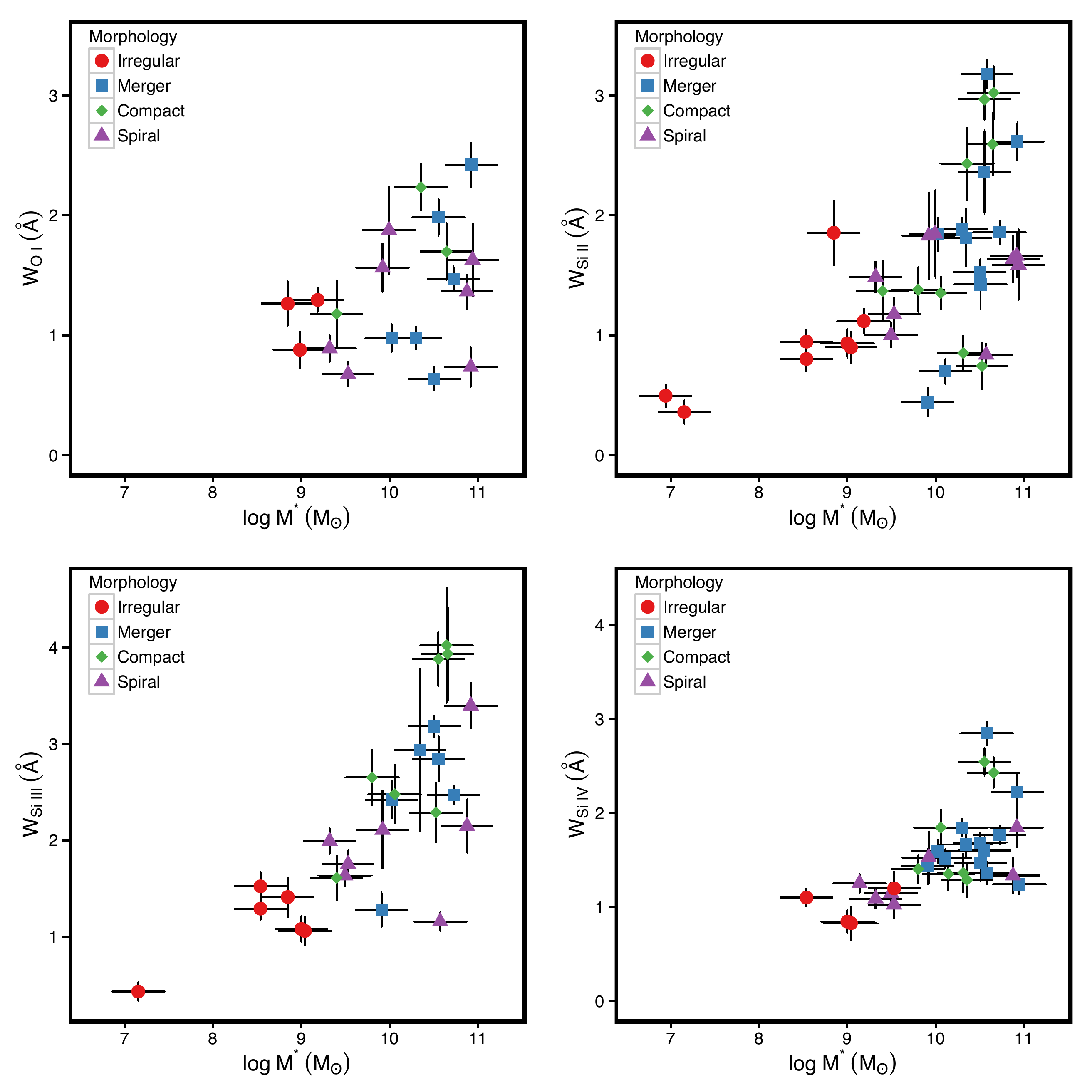}
\caption{The equivalent width (W; units of \AA) plotted against the stellar mass (log(M/M$_\odot$)). The four panels are four different transitions, from left to right, top to bottom: \oi~1302~\AA, \siiip~1260~\AA, \siiii~1206~\AA, and \siiv~1393~\AA. The transitions are only plotted if W is detected at the 3$\sigma$ significance, and \vcen is less than zero at the 1$\sigma$ significance level. The points are classified by the galaxy\rq{}s morphology, as given in the legends. Note that the fits are measured in log(W). }
\label{fig:ewm}
\end{figure*}

\begin{figure*}
\includegraphics[width = \textwidth]{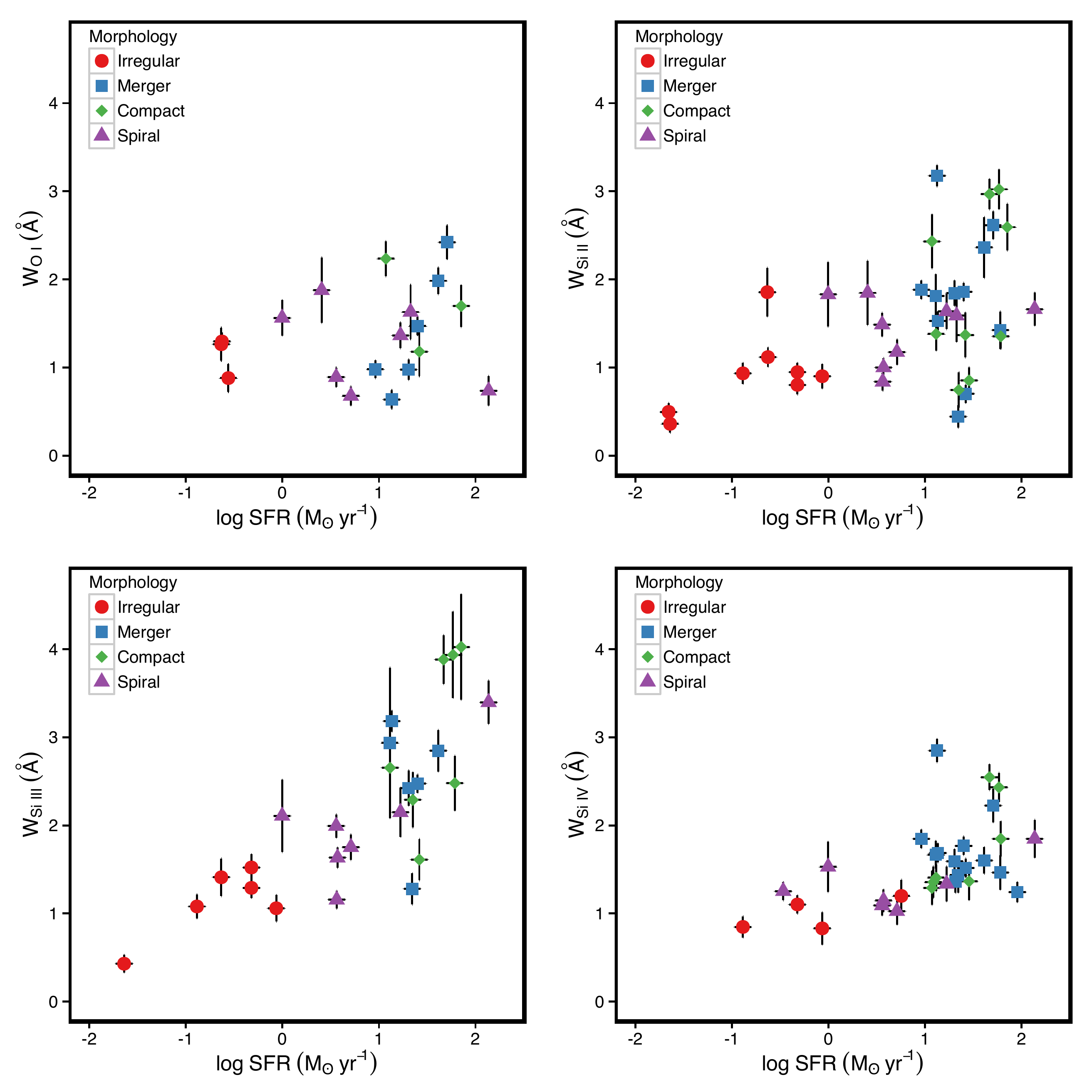}
\caption{Similar to \autoref{fig:ewm}, but for the relationship between equivalent width (W; units of \AA) and the star formation rate (log(SFR[M$_\odot$~yr$^{-1}$])). The four panels are four different transitions, from left to right, top to bottom: \oi~1302~\AA, \siiip~1260~\AA, \siiii~1206~\AA, and \siiv~1393~\AA. The transitions are only plotted if W is detected at the 3$\sigma$ significance, and \vcen is less than zero at the 1$\sigma$ significance level. The points are classified by the galaxy\rq{}s morphology, as given in the legends. Note that the fits are measured in log(W). }
\label{fig:ewsfr}
\end{figure*}

\begin{figure*}
\includegraphics[width = \textwidth]{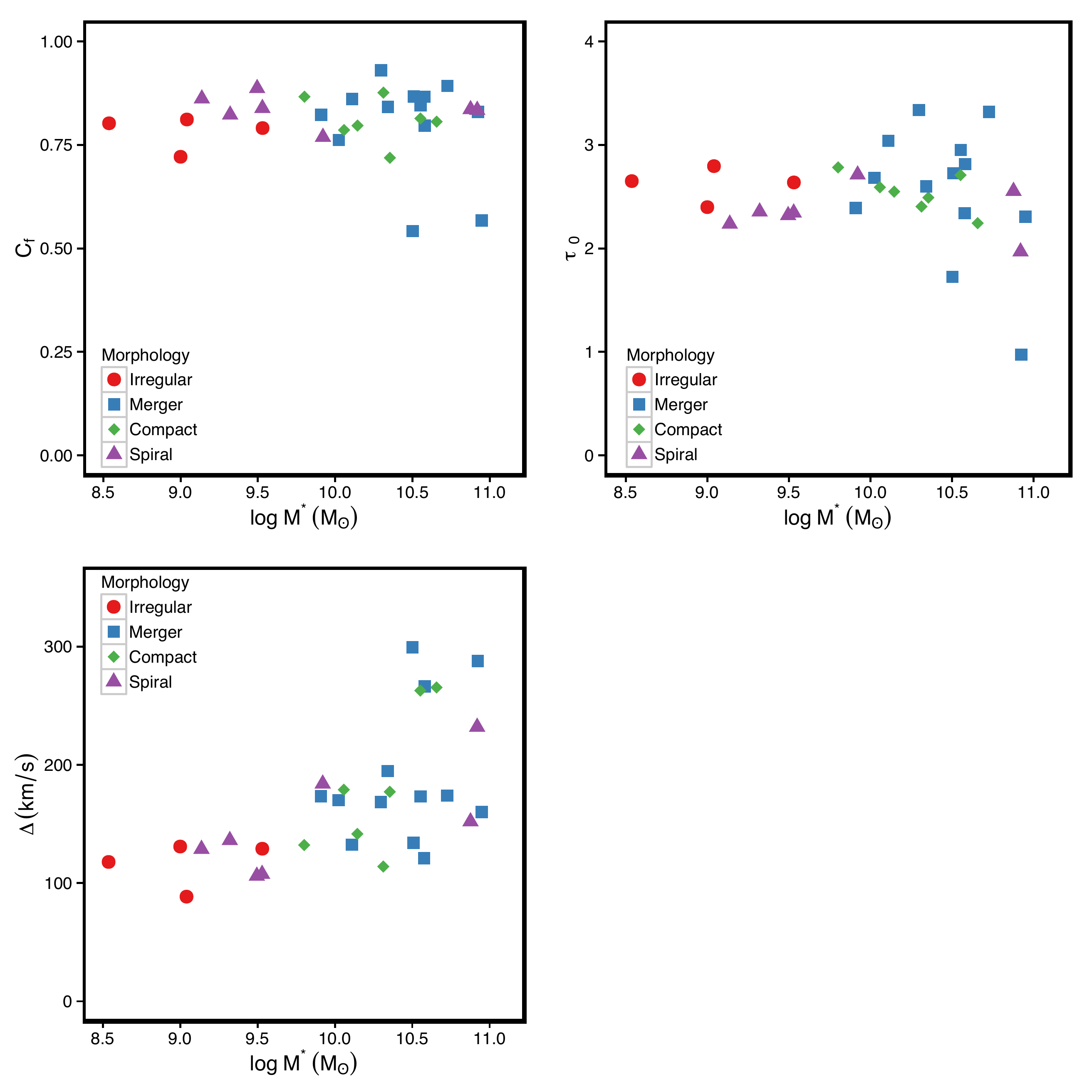}
\caption{The 3 components that determine the equivalent width of the \siivp~1402~\AA\ line profile: the covering fraction (C$_f$), the optical depth ($\tau_0$), and the line width ($\Delta$). The top left panel shows that the covering fraction has a flat distribution with a median value of 0.82, and a scatter of 0.05. The optical depth distribution (top right) also is flat with \mstarp.  Finally, the bottom left panel shows a 3.4$\sigma$ relation between $\Delta$ and \mstar that accounts for nearly the entire variation between \mstar and W (see \autoref{tab:si4relations}). }
\label{fig:doublet}
\end{figure*}

\begin{deluxetable*}{lllllll}
\tablewidth{0pt}
\tablecaption{Equivalent Width-\mstar Relations}
\tablehead{
\colhead{(1)} &
\colhead{(2)} &
\colhead{(3)}  &
\colhead{(4)} &
\colhead{(5)}  &
\colhead{(6)} & \\
\colhead{Line} &
\colhead{Slope} &
\colhead{Intercept}  &
\colhead{Kendall\rq{}s $\tau$} &
\colhead{Significance}  &
\colhead{Number}  
}
\startdata
O~{\sc I}~1302\AA & 0.10 $\pm$ 0.06 & 0.10 $\pm$ 0.04 & 0.24 & 1.4$\sigma$ & 19 \\
Si~{\sc II}~1260\AA & 0.17 $\pm$ 0.03 & 0.17 $\pm$ 0.03 & 0.42  & 3.6$\sigma$ & 37 \\
Si~{\sc III}~1206\AA & 0.21 $\pm$ 0.03 & 0.34   $\pm$ 0.02 & 0.57 & 4.1$\sigma$ & 26\\
Si~{\sc IV}~1393\AA & 0.14 $\pm$ 0.03 & 0.15 $\pm$ 0.02 &0.48 & 3.8$\sigma$ & 31
\enddata
\tablecomments{Measured relations between the logarithm of the equivalent width and log(\mstarp/10$^{10}$~M$_\odot$). The transitions are given in the first column, the slope and the intercept are given in the second and third columns. The Kendall\rq{}s $\tau_K$ value and significance level are given in the forth and fifth columns. The final column gives the number of galaxies for each transition. Only the \siiip, \siiiip, and \siiv relations are highly significant. The \siiii relation has a stronger correlation than \siiv because the \siiii sample has a larger dynamic range. If 2 marginally detected lines (less than 3$\sigma$ but greater than 1$\sigma$) are included in the \siiv sample the $\tau_K$ increases to 0.52, at a 4.3$\sigma$ significance level.}
\label{tab:ewm}
\end{deluxetable*}

\begin{deluxetable*}{lllllll}
\tablewidth{0pt}
\tablecaption{Equivalent Width-SFR Relations}
\tablehead{
\colhead{(1)} &
\colhead{(2)} &
\colhead{(3)}  &
\colhead{(4)} &
\colhead{(5)}  &
\colhead{(6)} \\
\colhead{Line} &
\colhead{Slope} &
\colhead{Intercept}  &
\colhead{Kendall\rq{}s $\tau$} &
\colhead{Significance}  &
\colhead{Number}  
}
\startdata
O~{\sc I}~1302\AA & 0.03 $\pm$ 0.06 & 0.11 $\pm$ 0.05 & 0.18 & 1.1$\sigma$ & 19 \\
Si~{\sc II}~1260\AA  & 0.14 $\pm$ 0.03 & 0.17 $\pm$ 0.03 & 0.28  & 2.4$\sigma$ & 37 \\
Si~{\sc III}~1206\AA & 0.19 $\pm$ 0.03 & 0.34   $\pm$ 0.03 & 0.61 & 4.4$\sigma$ & 26\\
Si~{\sc IV}~1393\AA & 0.11 $\pm$ 0.03 & 0.16 $\pm$ 0.02 &0.48 & 3.8$\sigma$ & 31
\enddata
\tablecomments{Similar to \autoref{tab:ewm} but for log(SFR/10~M$_\odot$~yr$^{-1}$). The relations are comparable to the \mstar relations found in \autoref{tab:ewm}, but the significances for \siii and \oi are substantially lower.}
\label{tab:ewsfr}
\end{deluxetable*}

\begin{deluxetable*}{lllllll}
\tablewidth{0pt}
\tablecaption{\siiv Relations}
\tablehead{
\colhead{(1)} &
\colhead{(2)} &
\colhead{(3)} &
\colhead{(4)}  &
\colhead{(5)} &
\colhead{(6)}  \\
\colhead{Y} &
\colhead{X} &
\colhead{Slope} &
\colhead{Intercept}  &
\colhead{Kendall\rq{}s $\tau$} &
\colhead{Significance} 
}
\startdata
log($\tau_0$) & log(\mstarp/10$^{10}$~M$_\odot$)  & -0.00 $\pm$ 0.02 & 0.40 $\pm$ 0.01 & -0.05 &-0.42$\sigma$ \\
log(C$_f$) & log(\mstarp/10$^{10}$~M$_\odot$)  & 0.00 $\pm$ 0.01 & -0.09 $\pm$ 0.01 & 0.02 & 0.19$\sigma$ \\
log($\Delta$) & log(\mstarp/10$^{10}$~M$_\odot$)  & 0.14 $\pm$ 0.03 &  2.19 $\pm$ 0.02 & 0.46 & 3.6$\sigma$ 
\enddata
\tablecomments{Measured relations between the three components that determine the \siiv~1402~\AA\ line, and their scaling with log(\mstarp/10$^{10}$~M$_\odot$). The variables studied are given in the first two columns, the slopes and the intercepts are given in the third and forth columns. The Kendall\rq{}s $\tau_K$ value and significance level are given in the final two columns. The only parameter with a strong correlation with \mstar is $\Delta$. }
\label{tab:si4relations}
\end{deluxetable*}

The equivalent width measures the amount of energy removed from the continuum.  W is an incredibly complex measurement: it encodes information on the column density, the velocity distribution, and the covering fraction of the gas. This makes the interpretation of W difficult. First we explore how W scales with host galaxy properties, and then we study what drives this evolution.

The W scaling relations with \mstar are presented in \autoref{fig:ewm}. For \siii~1260~\AA, \siiii~1206~\AA, and \siiv~1393~\AA\ the relations are strong with 3.6, 4.1 and 3.8$\sigma$ significance levels. The higher significance level for the \siiii transition versus the \siiv transition is because the \siiii transition has a larger dynamic range that extends down to log(\mstarp/M$_\odot$) of 7.5, while the lowest mass \siiv galaxy is 8.96. \siiv is a weaker transition, and only \siii and \siiii are detected in the lower mass galaxies. There are two marginally detected low-mass galaxies (greater than 1$\sigma$ but less than 3$\sigma$) that increase the significance level of the \siiv W-\mstar trend to 4.3$\sigma$. The three significant Si relations are given as:
\begin{equation}
\begin{aligned}
W_\text{Si~{\sc II}}  &= \left(1.48 \pm 0.09 ~\text{\AA\ } \right)  \left(\frac{M_\ast}{10^{10} ~\text{M}_\odot}\right)^{0.17 \pm 0.03} \\
{W}_\text{Si~{\sc III}}  &= \left(2.18 \pm 0.13 ~\text{\AA\ } \right)  \left(\frac{M_\ast}{10^{10} ~\text{M}_\odot}\right)^{0.21 \pm 0.03}\\
W_\text{Si~{\sc IV}}  &= \left(1.42 \pm 0.08 ~\text{\AA\ } \right)  \left(\frac{M_\ast}{10^{10} ~\text{M}_\odot}\right)^{0.14 \pm 0.03}
\label{eq:wrelations}
\end{aligned}
\end{equation}
The normalizations in \autoref{eq:wrelations} go from highest to lowest for \siiiip, \siiip, and \siivp, qualitatively consistent with the trend of W estimations given in \autoref{tab:detection}. However, quantitatively they are quite different than calculated in \autoref{tab:detection}. The differences show that both the strength of the line and the fraction of gas in each transition (the ionization fraction) determines the equivalent width ratios. In \autoref{ionize} we explore the mechanisms that determine the ionization fractions.

We also study how W varies with SFR, and gas phase metallicity. The SFR trends are similar to the \mstar trends (see \autoref{tab:ewsfr}), but the significances are lower for the \oi and \siii transitions. Additionally, the gas phase metallicity is known to vary with \mstar (the \lq\lq{}mass-metallicity relation\rq\rq{}; \citealp{tremonti04, andrews13}), and this relationship might be a natural driver of the W-\mstar (and SFR) relationships. We use the gas-phase metallicities from the literature (see Paper {\sc I}) to find that the correlation between the \siiv W and gas phase metallicity is weaker than for \mstar alone (2.8$\sigma$ for the metallicity and 3.7$\sigma$ for \mstarp).

The interpretation of W is difficult because it has complicated degeneracies. A large W can indicate a high column density gas distributed over a small velocity range, or a low column density of gas widely distributed in velocity space.  Additionally, W will depend on the fraction of the source covered by the outflow (where the uncovered part of the source decreases W). This leads to a complicated triple degeneracy with column density, velocity distribution, and covering fraction. Below, we use the \siiv doublet to explore what drives the observed relation between W and \mstar. We use \siiv because the equivalent width is strongly correlated with \mstar and SFR, while the doublet allows for calculation of the covering fraction (C$_f$; \autoref{eq:cf}), the line center optical depth ($\tau_0$; \autoref{eq:taucf}), and the velocity width ($\Delta$; \autoref{eq:b}), which we cannot measure with the \siiii singlet. We can then determine the primary driver of the shallow  W trends with \mstarp.

The upper right panel of \autoref{fig:doublet} shows a flat distribution of $\tau_0$ with \mstar (\autoref{tab:si4relations}). There are two mergers and one spiral with optical depths substantially lower than the rest of the galaxies (NGC~3256 has an optical depth less than one), but the rest of the transitions are optically thick. The median $\tau_0$ is 2.55, with a range from 0.97 to 3.3. The optical depths indicate that the lines are moderately saturated, and the optical depth will not drive strong evolution in W. 

\begin{figure}
\includegraphics[width = 0.5\textwidth]{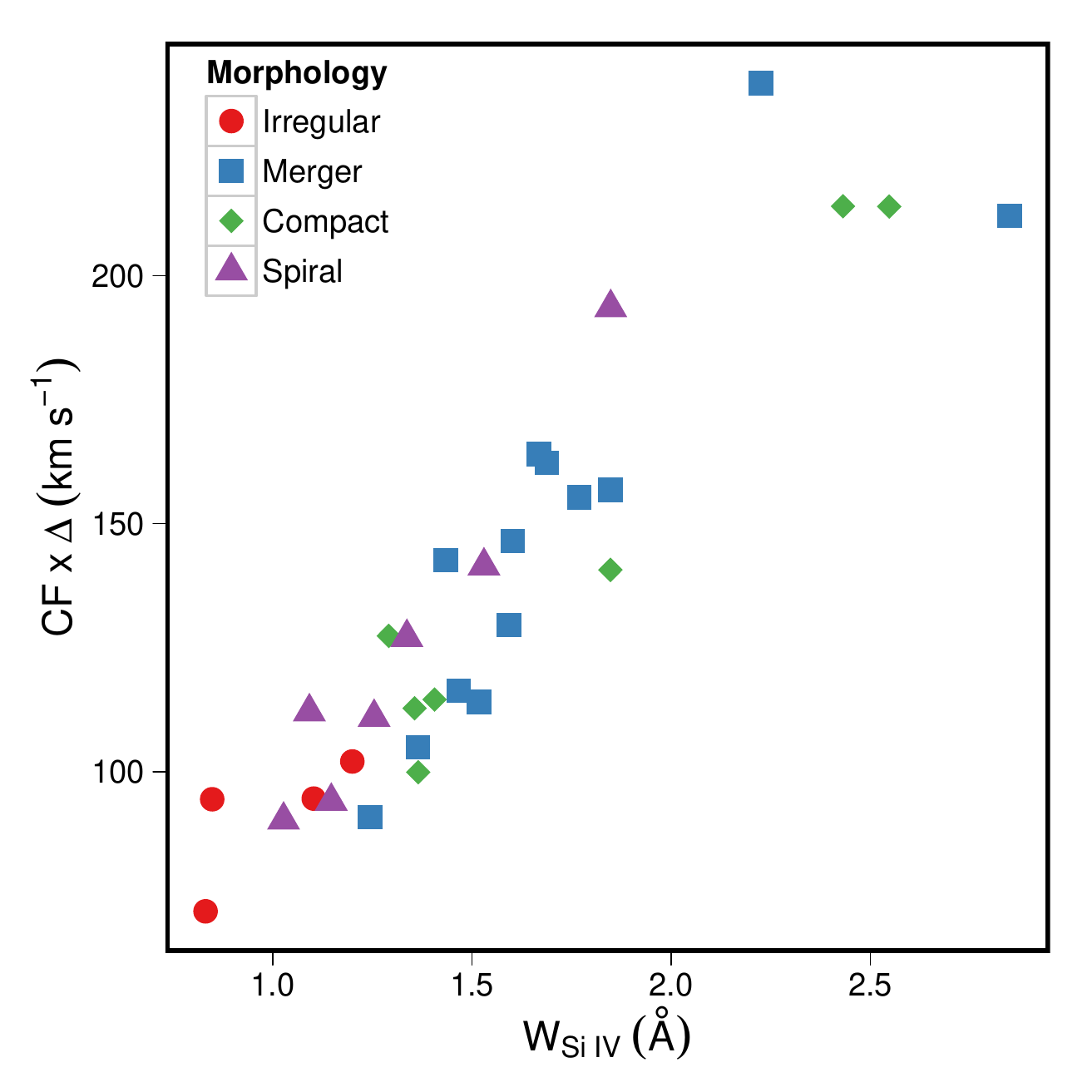}
\caption{Product of the line width ($\Delta$) and the covering fraction (C$_f$) versus the equivalent width (W) of the \siiv 1393~\AA\ line. For moderately saturated lines, the equivalent width is approximately equal to this product. The strong, nearly linear, correlation shows that $\Delta$ times C$_f$ accounts for 85\% of the variation of W. }
\label{fig:cfdel}
\end{figure}

For a moderately optically thick line, W is the product of $\Delta$ and C$_f$. This relationship is shown in \autoref{fig:cfdel}, where a tight (scatter of 4~\kmsp) and strong (6$\sigma$, $\tau_K$ = 0.75, and R$^2$ = 0.85) relationship describes 85\% of the variation of W. Therefore, either $\Delta$, C$_f$, or the product of the two drives the observed relationship between \mstar and W.

The lower left panel of \autoref{fig:doublet} shows the distribution of the line center C$_f$ with \mstarp. A nearly constant relationship is seen, with a median value of 0.82. The null correlation with \mstar (0.45$\sigma$) is also seen with the relationship between C$_f$ and W, which has a $\tau_K$ of 0.02 and a significance level of 0.18$\sigma$. Therefore,  C$_f$ does not drive the W values.  The non-unity C$_f$ may arise from emission line infilling of resonantly scattered emission lines \citep{prochaska2011, martin12, rubin13, guantun}. However, we only observe Si~{\sc II$^\ast$} emission lines in a small subset of our spectra  (all at higher redshifts), and we conclude that Si~{\sc II} resonant emission does not substantially effect the measured C$_f$, or line profiles. The degraded spectral resolution from the COS line-spread function and extended light distribution could contribute to the non-unity covering fraction \citep{prochaska2011}. 

Finally, the lower right panel of \autoref{fig:doublet} shows the scaling of $\Delta$ with \mstarp. A 3.6$\sigma$ relationship is seen between log($\Delta$) and log(\mstarp/M$_\odot$) with a scaling of (see \autoref{tab:si4relations}):
\begin{equation}
\Delta = \left(156 \pm 7\right) \left(\frac{\text{M}_\ast}{10^{10} ~\text{M}_\odot}\right)^{0.14 \pm 0.03}
\label{eq:delta}
\end{equation} 
The relation is fairly tight (15~\kmsp) until log(\mstarp/M$_\odot$) of 10.5, where three mergers and two compact galaxies increase the scatter. Furthermore, $\Delta$ and W are strongly correlated with a 5.0$\sigma$ significance, $\tau_K$ of 0.63, and an R$^2$ of 0.70. The \mstar exponent is consistent with the exponent of the W relations, and indicates that the line width (as measured by $\Delta$) drives the variation between W and \mstarp.

In summary, there are shallow, but highly significant, trends between W and \mstar for the \siii~1260~\AA, \siiii~1206~\AA\, and \siiv~1393~\AA\ transitions. The SFR has similar trends as \mstarp. A larger sample with a range of SFR at constant \mstar is needed to determine the primary driver of the trends. For the \siiv relations, $\Delta$ is primarily responsible for increasing W, such that larger W corresponds to larger $\Delta$. 

\subsection{Equivalent Widths Between the Phases}
\label{ewratios}

\begin{figure*}
\includegraphics[width = \textwidth]{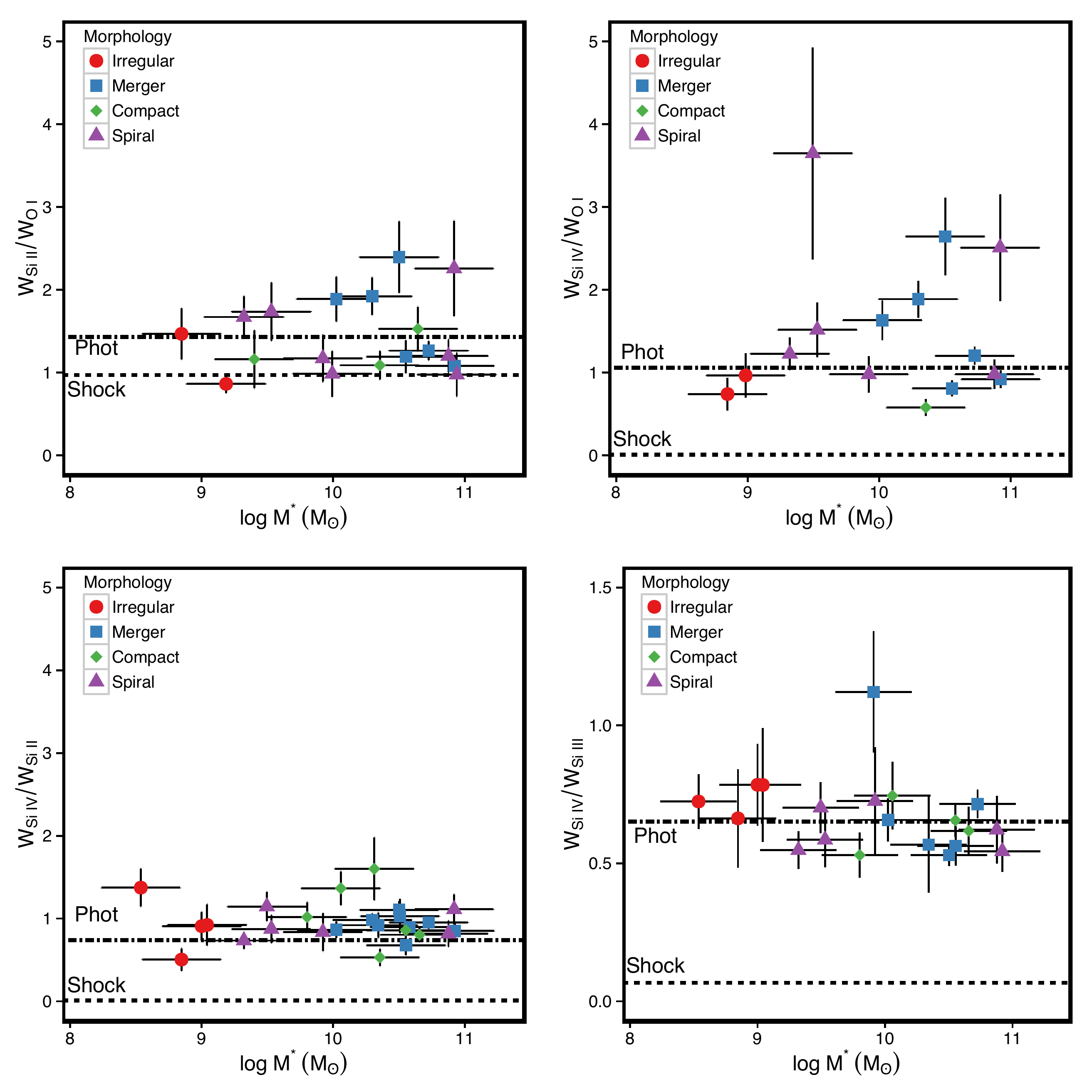}
\caption{Ratio of equivalent widths (W; in \AA) versus log(\mstarp/M$_\odot$) for four different transition ratios. The ratios are: \siiip/\oi (top left), \siivp/\oi (top right), \siivp/\siii (bottom left), and \siivp/\siiii (bottom right). The equivalent width ratios are flat, with large scatter for the ratios involving \oip. In \autoref{ionize} we use shock models (dashed lines) and photo-ionization models (dotted-dashed lines) to predict the W ratio values, with the photo-ionization models roughly matching the data. The dotted-dashed line is calculated using the column densities from a seven component Cloudy model with an ionization parameter of log(U)~=~-2.00, a hydrogen column density of log(n$_\text{o}$)~=~2.4, stellar metallicity of 0.2~Z$_\odot$, and an outflow metallicity of 0.5~Z$_\odot$ (see \autoref{ionize}). Other combinations of stellar metallicities, outflow metallicities, and ionization parameters are need to reproduce the full range of the observed ratios.}
\label{fig:ewratios}
\end{figure*}

\begin{figure*}
\includegraphics[width = \textwidth]{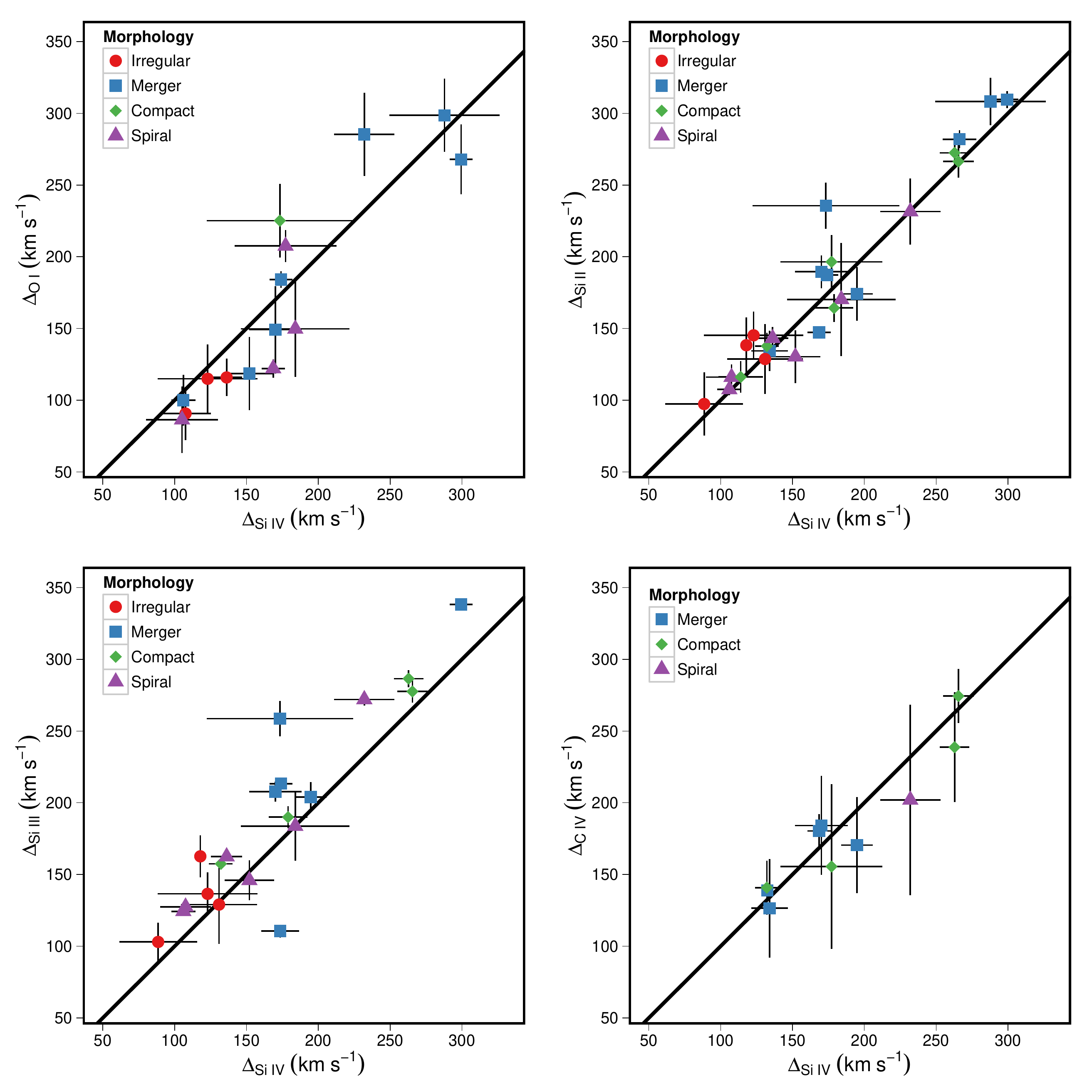}
\caption{Comparison of the velocity width ($\Delta$) between the \oi (upper left), \siii (upper right), and \siiii (lower left), \civ (lower right) and \siivp. The black line indicates a slope of 1 and an intercept of 0, showing that the $\Delta$s scatter around a linear relation. The \siiii lines are slightly offset from this line. They also have larger equivalent widths, illustrating that stronger lines have larger line widths.}
\label{fig:del}
\end{figure*}

The ratio of the equivalent widths are important diagnostics of the physical conditions of the gas. The ratios give the ionization structure, and determine how the gas is ionized. This sets the ionization corrections, and is vital for calculations of the mass outflow rates.

We show the W ratios in \autoref{fig:ewratios}. There are not significant trends in the ratios with \mstar for any of the transitions. The \siivp/\siiii and \siivp/\siii ratios have nearly constant values, with median values of 0.91 and 0.66, consistent with the W normalization ratios in \autoref{eq:wrelations}. Meanwhile, the \siivp/\oi and \siiip/\oi relations have considerable scatter (with a factor of 2 variation at constant \mstarp).  In \autoref{ionize} we explore the values of these ratios, and what mechanisms might set these values.

In \autoref{ewscaling} we find that the line width ($\Delta$) correlates strongest with W, and in \autoref{fig:del} we plot $\Delta$ for the \oip, \siiip, \siiiip, and \civ transitions versus the $\Delta$ for \siivp. The black lines in \autoref{fig:del} show a 1:1 line, with the $\Delta$s scattering about this line. The $\tau_K$ of these relations are very strong, with values of 0.95, 0.93, 0.83, and 0.85 for the \oip, \siiip, \siiiip, and \civ relations, respectively.  The \siiii $\Delta$ is larger than the \siiv $\Delta$, demonstrating that stronger transitions have wider velocity distributions.  Additionally, the C$_f$ is consistent across the transitions, with \siii having a median C$_f$ of 0.81 (compared to 0.82 for \siivp). 

\subsection{Velocity Scaling Relations}
\label{vscale}

In Paper {\sc I} we found the \siii outflow velocity to scale significantly, but shallowly, with \mstar and SFR. These \lq\lq{}scaling relations\rq\rq{} are important for galaxy evolution simulations and theory because they describe the amount of energy injected into the interstellar medium at scales that current simulations cannot resolve. Crucially, the different transitions probe different gas densities and temperatures, and the scaling relations for the different temperatures illustrate the energetics of the different phases.

\begin{figure*}
\includegraphics[width = \textwidth]{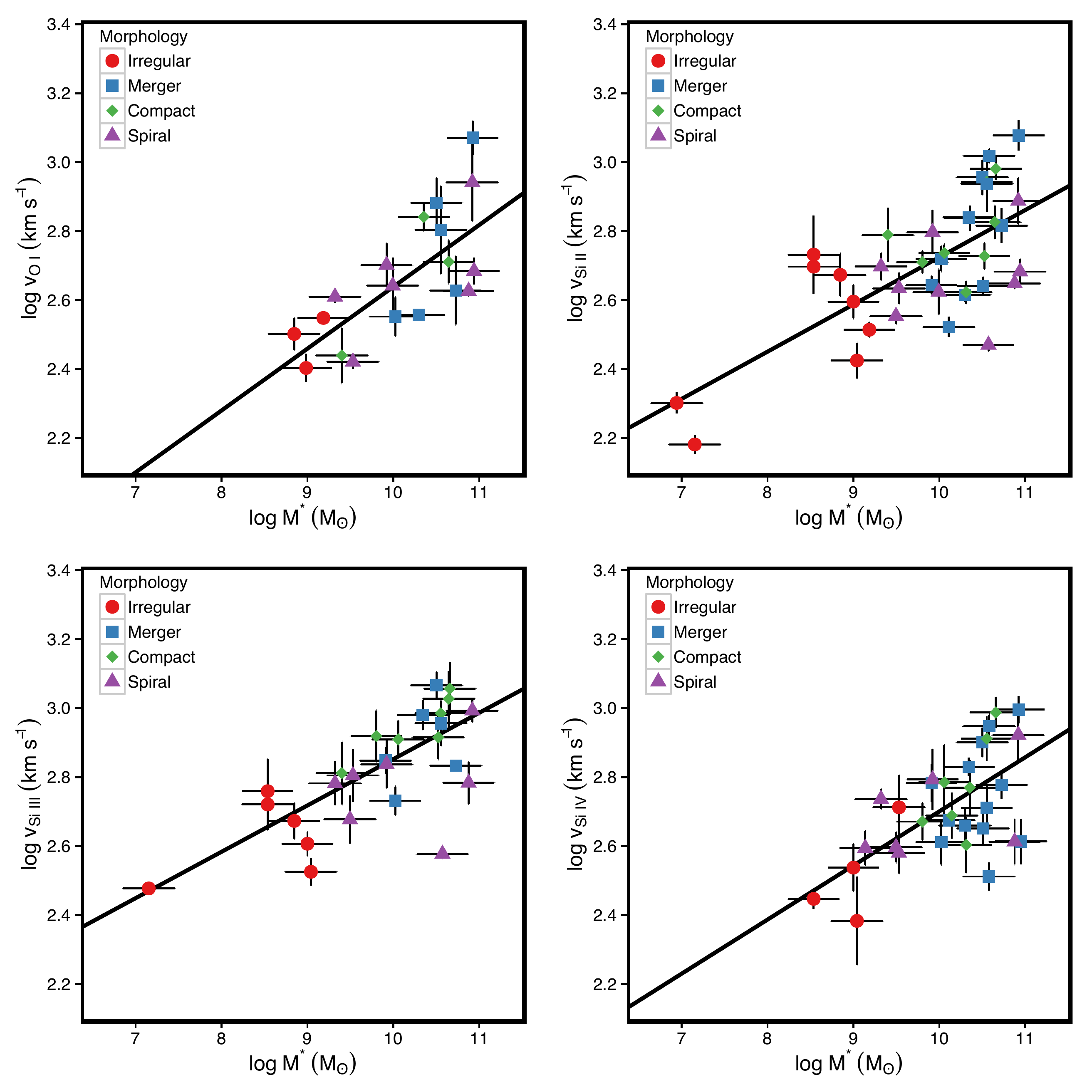}
\caption{Plot of the scaling of the velocity at 90\% of the continuum (\vn) with \mstarp. Black lines give the measured regression trends (see \autoref{tab:massvn}), where the slopes for all four relations are similar within the errors of the fits. }
\label{fig:massscaling}
\end{figure*}

\begin{deluxetable*}{lccccc}
\tablewidth{0pt}
\tablecaption{\mstarp-\vn Relations}
\tablehead{
\colhead{(1)} &
\colhead{(2)} &
\colhead{(3)}  &
\colhead{(4)} &
\colhead{(5)}  &
\colhead{(6)} \\ 
\colhead{Line} &
\colhead{Slope} &
\colhead{Intercept}  &
\colhead{Kendall\rq{}s $\tau$} &
\colhead{Significance}  &
\colhead{Number} 
}
\startdata
O~{\sc I} & 0.18 $\pm$ 0.04 & 2.64 $\pm$ 0.03 &0.53 & 3.2$\sigma$ & 19 \\
Si~{\sc II}  & 0.14 $\pm$ 0.03& 2.72 $\pm$ 0.03 & 0.42  & 3.6$\sigma$ & 37 \\
Si~{\sc III} & 0.13 $\pm$0.03 & 2.85 $\pm$ 0.02 & 0.51 & 3.6$\sigma$ & 26\\
Si~{\sc IV} & 0.16 $\pm$ 0.04 & 2.70 $\pm$ 0.02 & 0.41 & 3.2$\sigma$ & 31 \\
\enddata
\tablecomments{Measured relations between \mstar and \vn for various transitions. The slopes and the intercept are given in the second and third columns, and the Kendall\rq{}s $\tau_K$ value and significance level are given in the forth and fifth columns. The last column gives the number of galaxies for each fit.}
\label{tab:massvn}
\end{deluxetable*}
\begin{figure*}
\includegraphics[width = \textwidth]{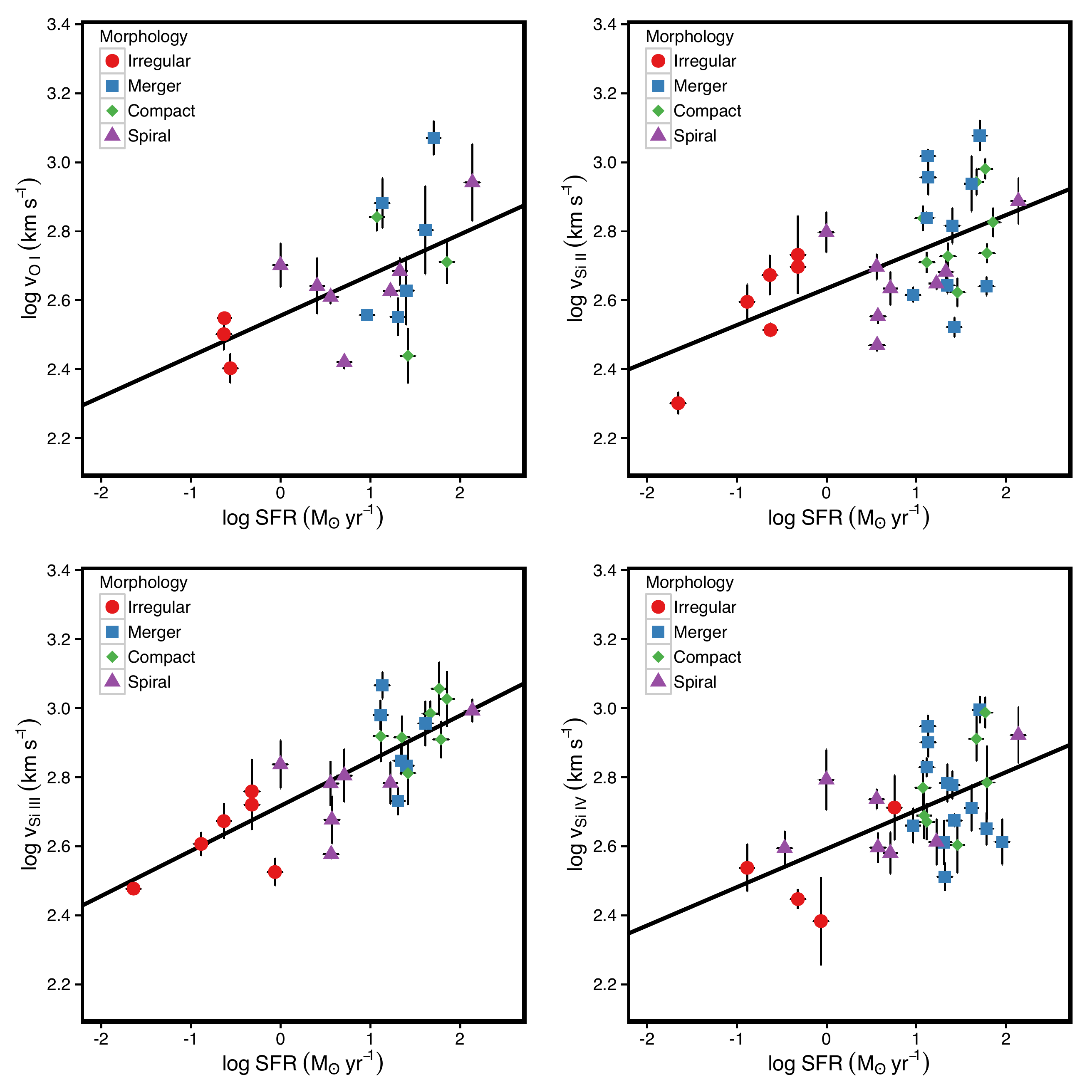}
\caption{Similar to \autoref{fig:massscaling}, but for the \vnp-SFR. Black lines give the measured  trends (see \autoref{tab:sfrvn}), where the slopes for all four relations are similar, within the errors. }
\label{fig:sfrscaling}
\end{figure*}

\begin{deluxetable*}{lccccc}
\tablewidth{0pt}
\tablecaption{SFR-\vn Relations}
\tablehead{
\colhead{(1)} &
\colhead{(2)} &
\colhead{(3)}  &
\colhead{(4)} &
\colhead{(5)}  &
\colhead{(6)} \\ 
\colhead{Line} &
\colhead{Slope} &
\colhead{Intercept}  &
\colhead{Kendall\rq{}s $\tau$} &
\colhead{Significance}  &
\colhead{Number} 
}
\startdata
O~{\sc I} & 0.12 $\pm$ 0.05 & 2.56$\pm$ 0.06 &0.40 & 2.4$\sigma$ & 19 \\
Si~{\sc II}  & 0.11 $\pm$ 0.03 & 2.74 $\pm$ 0.03 & 0.36  & 2.9$\sigma$ & 37 \\
Si~{\sc III} & 0.13 $\pm$0.02 & 2.85 $\pm$ 0.02 & 0.60 & 4.3$\sigma$ & 26\\
Si~{\sc IV} & 0.11 $\pm$ 0.04 & 2.70 $\pm$ 0.03 &0.32 & 2.6$\sigma$ & 31 \\
\enddata
\tablecomments{Similar to \autoref{tab:massvn}, but for the relations between \vn and SFR. }
\label{tab:sfrvn}
\end{deluxetable*}

\begin{deluxetable*}{lccccc}
\tablewidth{0pt}
\tablecaption{SFR-\vcen Relations}
\tablehead{
\colhead{(1)} &
\colhead{(2)} &
\colhead{(3)}  &
\colhead{(4)} &
\colhead{(5)}  &
\colhead{(6)} \\ 
\colhead{Line} &
\colhead{Slope} &
\colhead{Intercept}  &
\colhead{Kendall\rq{}s $\tau$} &
\colhead{Significance}  &
\colhead{Number} 
}
\startdata
O~{\sc I} & 0.14 $\pm$ 0.08 & 2.06$\pm$ 0.06 &0.27 & 1.6$\sigma$ & 19 \\
Si~{\sc II}  & 0.18 $\pm$ 0.05 & 2.18 $\pm$ 0.04 & 0.38  & 3.0$\sigma$ & 37 \\
Si~{\sc III} & 0.16 $\pm$0.05 & 2.22 $\pm$ 0.05 & 0.42 & 3.0$\sigma$ & 26\\
Si~{\sc IV} & 0.23 $\pm$ 0.08 & 2.09 $\pm$ 0.06 &0.32 & 2.6$\sigma$ & 31 \\
\enddata
\tablecomments{Similar to \autoref{tab:massvn}, but for the relations between \vcen and SFR.} 
\label{tab:sfrvcen}
\end{deluxetable*}

In \autoref{fig:massscaling} we show the scaling relations with the velocity at 90\% of the continuum (\vnp) and \mstarp, for \oip~1302~\AA\ (upper left), \siiip~1260~\AA\ (upper right), \siiiip~1206~\AA\ (lower left), and \siivp~1392~\AA\ (lower right). In each panel we over-plot the derived scaling relations from \autoref{tab:massvn}. Each of the four derived trends are highly significant, with significance levels ranging from 2.9-3.7$\sigma$. The \siiii relation has an R$^2$ of 0.51. Similar to the W relations, the significance of \siiv is lower than the \siiii because the dynamic range is lower. If marginal detections are included the relations are 3.4$\sigma$ significant with $\tau_K$ of 0.42. The scaling relations for the four transitions are tabulated in \autoref{tab:massvn}), and given as:
\begin{equation}
\begin{aligned}
\text{v}_\text{O~{\sc I}} &= \left(437 \pm 30~ \text{\kmsp} \right)~\left(\frac{M_\ast}{10 \times 10^{10}~M_\odot}\right)^{0.18 \pm 0.04} \\
\text{v}_\text{Si~{\sc II}} &= \left(524 \pm 36~ \text{\kmsp} \right) ~\left(\frac{M_\ast}{10 \times 10^{10}~M_\odot}\right)^{0.14 \pm 0.03} \\
\text{v}_\text{Si~{\sc III}} &= \left(707 \pm 32~ \text{\kmsp} \right) ~\left(\frac{M_\ast}{10 \times 10^{10}~M_\odot}\right)^{0.13 \pm 0.03} \\
\text{v}_\text{Si~{\sc IV}} &= \left(501 \pm 23~ \text{\kmsp}\right) ~\left(\frac{M_\ast}{10 \times 10^{10}~M_\odot}\right)^{0.16 \pm 0.04} \\
\label{eq:mstarvn}
\end{aligned}
\end{equation}
with the exponents consistent within the errors. Meanwhile, the normalizations of the four trends range from 437~\kms for the \oi relation to 707~\kms for the \siiii relation. The change in intercept matches the change in the normalization of W (\autoref{eq:wrelations}), such that stronger lines have larger outflow velocities. 

The scaling relations for \vn and SFR are shown in \autoref{fig:sfrscaling}. The scaling relations are given by:
\begin{equation}
\begin{aligned}
\text{v}_\text{O~{\sc I}} &= \left(360 \pm 47~ \text{\kmsp} \right)~\left(\frac{\text{SFR}}{10~\text{M$_\odot$ yr$^{-1}$}}\right)^{0.12 \pm 0.05} \\
\text{v}_\text{Si~{\sc II}} &= \left(551 \pm 35~ \text{\kmsp} \right) ~\left(\frac{\text{SFR}}{10~\text{M$_\odot$ yr$^{-1}$}}\right)^{0.11 \pm 0.03}  \\
\text{v}_\text{Si~{\sc III}} &= \left(705 \pm 32~ \text{\kmsp} \right) ~\left(\frac{\text{SFR}}{10~\text{M$_\odot$ yr$^{-1}$}}\right)^{0.13 \pm 0.02} \\
\text{v}_\text{Si~{\sc IV}} &= \left(506 \pm 34~ \text{\kmsp}\right) ~\left(\frac{\text{SFR}}{10~\text{M$_\odot$ yr$^{-1}$}}\right)^{0.11 \pm 0.04} \\
\label{eq:sfrvn}
\end{aligned}
\end{equation}
Where the slopes and intercepts have similar trends as the \mstar relations in \autoref{eq:mstarvn}. The powers of the SFR relations are remarkably consistent over all four of the transitions. The available data cannot determine whether the SFR, or the \mstarp, is the major driver of these trends, and a larger sample of galaxies at a constant \mstar but varying SFR are needed to disentangle these trends. For completeness, \autoref{tab:sfrvcen} gives the tabulated relations between SFR and \vcenp. As we discuss in Paper {\sc I}, the \vcenp-SFR relations are slightly steeper than the \vnp-SFR relations, but they have larger uncertainties on the slopes and greater scatter, perhaps due to the influence of inflowing gas on the line centroid.

\subsection{Velocity Differences Between the Transitions}
\label{vphases}

\begin{figure*}
\includegraphics[width = \textwidth]{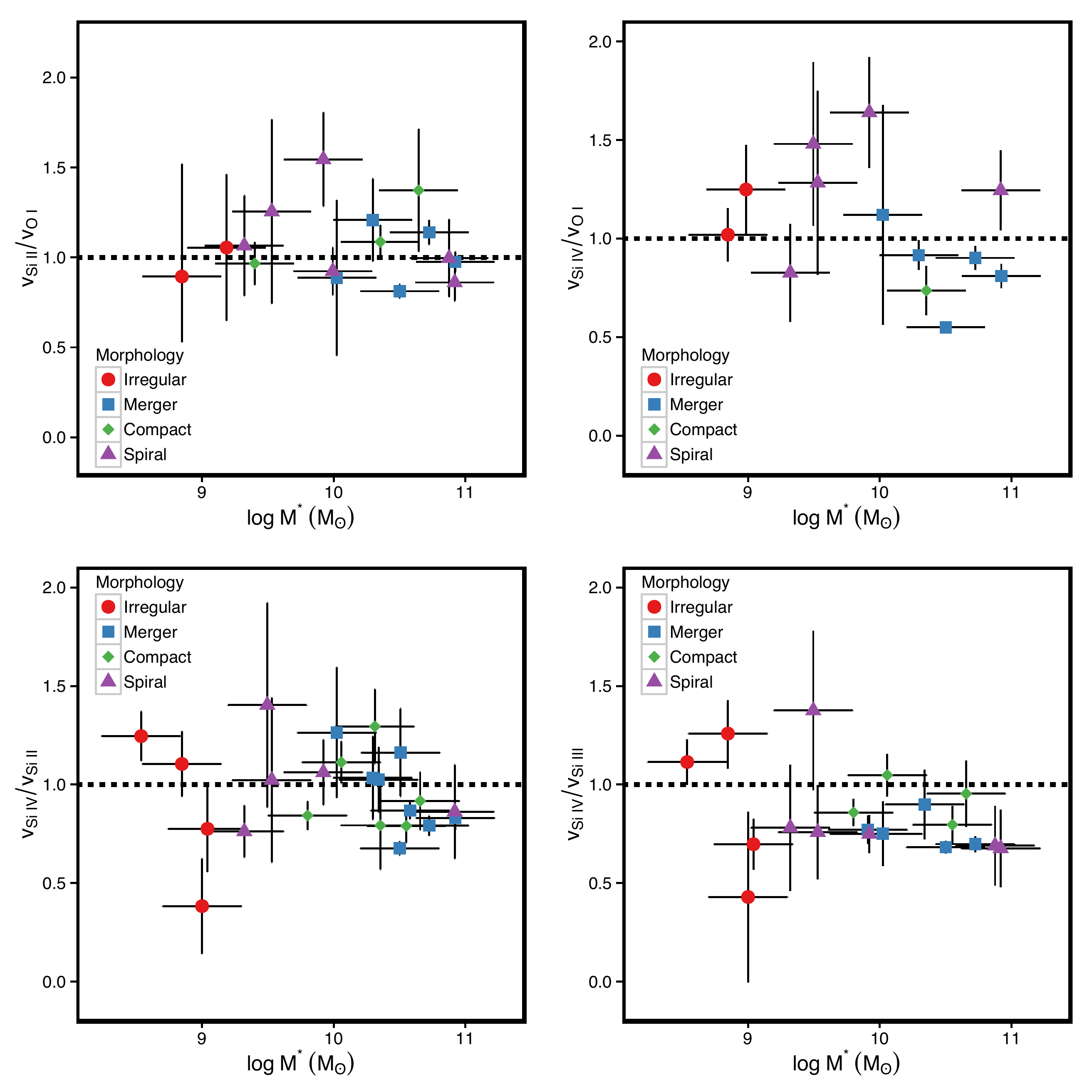}
\caption{Trends of the central velocity (\vcenp) ratio for four different transitions with log(\mstarp/M$_\odot$). The four transitions are \siiip/\oi (upper left), \siivp/\oip (upper right), \siivp/\siiii (lower left), and \siivp/\siiii (lower right). The trends show flat relations with \mstarp, with many scattering around the one-to-one line (dashed line). The \siivp/\siiii ratio has a median value near 0.75, demonstrating that \siiii has a larger \vcen than \siivp. This is consistent with the idea that stronger lines probe gas to higher velocities.}
\label{fig:vratios}
\end{figure*}

The difference in velocity between the various transitions explains how gas with different temperatures and densities moves relative to each other. By looking at the velocity differences between the phases we  explore whether the neutral and ionized gas are accelerated to similar velocities. 

We use the central velocity (\vcenp) to quantify the outflow velocity of the different transitions because the \vcen has a weaker dependence on the strength of the transition. \autoref{fig:vratios} shows the \siiip/\oip, \siivp/\oip, \siivp/\siiip, and  \siivp/\siiii \vcen ratios. All of the ratios are flat, with the \siiip/\oi and \siivp/\siii ratios having unity value (dashed lines in \autoref{fig:vratios}). The \siivp/\siiii ratio has a median value of 0.75~$\pm$~0.15. This effect is similar to the W ratios, and further illustrates that stronger transitions probe gas at higher velocities.

\section{DISCUSSION}

\subsection{Are the Transitions Co-Spatial?}
\label{outflows}
In \autoref{results} we explored how the outflow properties scale with host galaxy properties, and how the outflow properties change between the different transitions. Here we recap the results, and discuss whether all of the transitions probe the same gas. 

The various transitions have similar scaling relations for most outflow parameters. The equivalent widths (W), line widths ($\Delta$), and outflow velocities all scale similarly regardless of the physical conditions of the gas they probe (i.e. neutral, photo-ionized, etc.). Further, the $\Delta$ and outflow velocities of similar strength lines have similar values, regardless of the phase of gas (see \autoref{fig:del} and \autoref{fig:vratios}). This suggests that transitions of similar strength are probing gas that is likely co-spatial and co-moving.

The values of the parameters are driven largely by the strength of the transition: larger Ws lead to larger velocities and $\Delta$\rq{}s. This result is also found by \citet{grimes09}. This suggests that each transition probes up to a specific velocity, but there is gas at higher velocities with covering fractions (or column densities) below the detection limit that cannot be detected \citep{steidel10, martin09}. The true \lq\lq{}maximum\rq\rq{} velocity of the outflows is likely higher than the \vn values found here.

\subsection{How Are Outflows Ionized?}
\label{ionize}

\begin{deluxetable*}{lcccc}
\tablewidth{0pt}
\tablecaption{Model Equivalent Width Ratios}
\tablehead{
\colhead{(1)} &
\colhead{(2)} &
\colhead{(3)} &
\colhead{(4)}  &
\colhead{(5)} \\
\colhead{W Ratio} &
\colhead{Observed} &
\colhead{Shock Ionization} &
\colhead{Photo-Ionization Sub-Solar}  &
\colhead{Photo-Ionization Super-Solar}
}

\startdata
W$_\text{Si~IV}$/W$_\text{Si~II}$ & 0.91 $\pm$ 0.25 & 0.00  & 0.74 & 0.80 \\
W$_\text{Si~IV}$/W$_\text{O~I}$ & 1.2 $\pm$ 0.86 & 0.00  & 1.06 & 1.19 \\
W$_\text{Si~IV}$/W$_\text{Si~III}$ & 0.66 $\pm$ 0.14  & 0.00  & 0.66 & 0.69 \\
W$_\text{Si~II}$/W$_\text{O~I}$ & 1.23 $\pm$ 0.45 & 0.97  & 1.43 & 1.49  \\
W$_\text{C~I}$ & 0 & 0.48 & 0 & 0 \\
W$_\text{N~V}$ & 0.98 $\pm$ 0.53 & 0 & 0 & 0 \\
\enddata
\tablecomments{Results for the ionization modeling using the observed equivalent width (W) ratios for various transitions (as given in column 1). The second column gives the observed W ratios, with a one standard deviation spread. The third column gives the expected W ratios for the seven component shock models using column densities from \citep{allen}, a shock velocity of 500~\kmsp, a preshock density of 100~cm$^{-3}$, and a magnetic field of 1~$\mu$G. Other shock velocities similarly under-produce the high ionization lines.  Column four gives the Cloudy ratios using a sub-solar stellar metallicity Starburst99 model as the ionizing source, with seven absorbing components, an ionization parameter of log(U)~=~-2.25, a Hydrogen density of n$_H$ = 250~cm$^{-3}$, and an outflow metallicity of 0.5~Z$_\odot$. Column five gives the Cloudy equivalent width ratios for a model with seven absorption components, log(U)~=-1.75, n$_H$~=~250cm$^{-3}$, an outflow metallicity of 0.5~Z$_\odot$, and a stellar metallicity of 2~Z$_\odot$.}
\label{tab:ewmodels}
\end{deluxetable*}

\begin{deluxetable*}{cccc}
\tablewidth{0pt}
\tablecaption{Ionizing Photons From Starburst99 Models}
\tablehead{
\colhead{Stellar Metallicity} &
\colhead{H~{\sc II} Photons} &
\colhead{He~{\sc II} Photons} &
\colhead{He~{\sc III} Photons}  \\
\colhead{ (Z/Z$_\odot$)} &
\colhead{log(number s$^{-1}$)} &
\colhead{log(number s$^{-1}$)} &
\colhead{log(number s$^{-1}$)} 
}

\startdata
0.05 & 53.342 & 52.739  & 49.516\\
0.2 & 53.261 & 52.576  & 49.450\\
0.4 & 53.220 & 52.465  & 49.171 \\
1.0  & 53.131  & 52.275 & 48.469 \\
2.0 & 53.018 & 52.034 & 47.937
\enddata
\tablecomments{The number of ionizing photons from a constant 1~\sfr Starburst99 model in three energy regimes, and five metallicities. The second column gives the number of photons per second capable of ionizing H (energy greater than 13.6~eV), the third column gives the number of photons per second  capable of ionizing He once (energy greater than 24.6~eV), and the forth column gives the number of photons per second  capable of ionizing He a second time (energy greater than 54.4~eV). Since stellar metallicity more heavily effects the higher energy photons, there are less ionizing photons for larger stellar metallicities.}
\label{tab:starburst99}
\end{deluxetable*}

\begin{figure*}
\includegraphics[width = \textwidth]{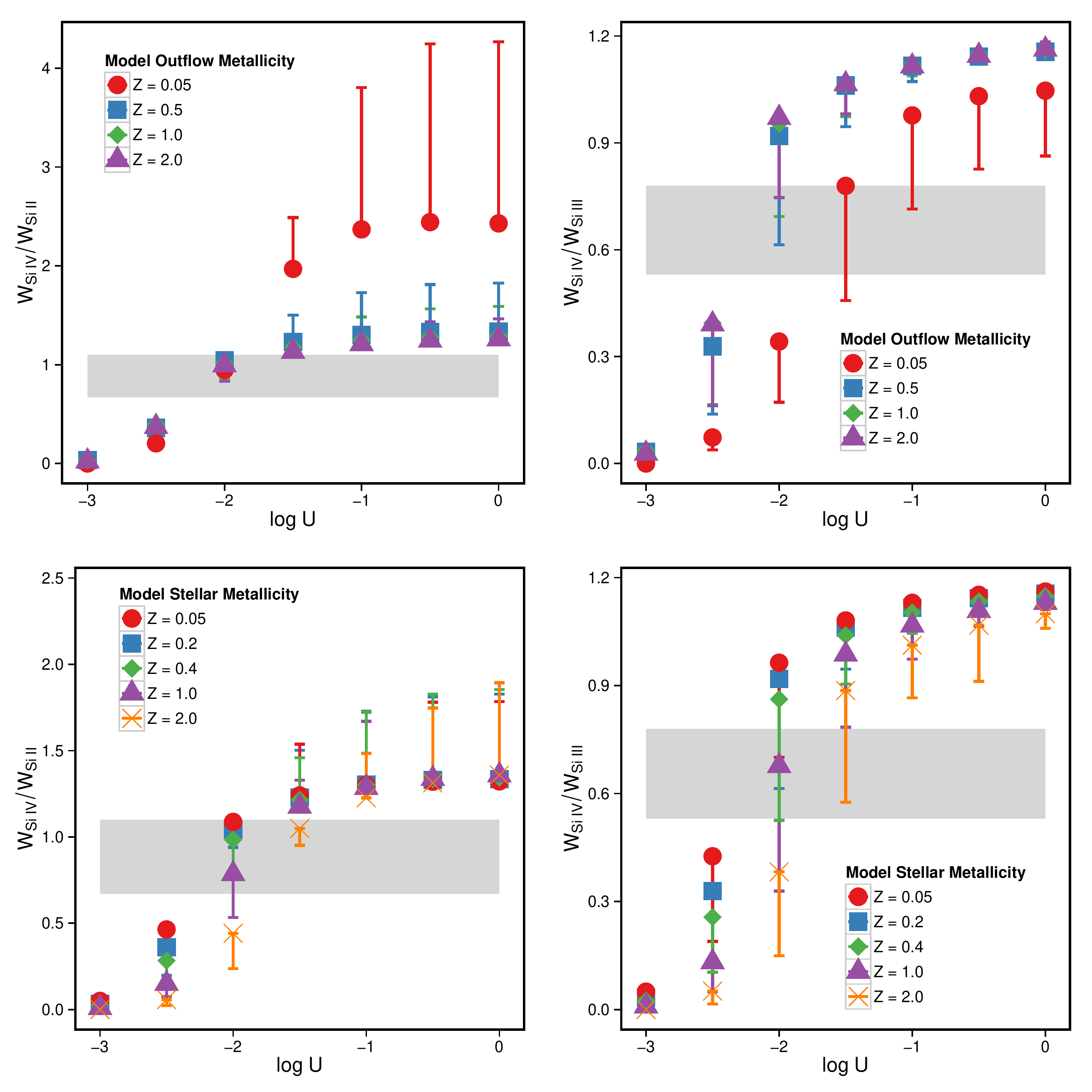}
\caption{Cloudy simulated curves of the equivalent width ratios with varying ionization parameters (U; see equation \autoref{eq:U}) for \siivp/\siii (left) and \siivp/\siiii (right). The gray shaded areas are the range of observed values from \autoref{fig:ewratios}. In the top panels we scale the outflow metallicity by 0.05, 0.5, 1.0, 1.5, and 2.0~Z$_\odot$, while keeping the Starburst99 stellar continuum metallicity fixed at 0.2~Z$_\odot$.  In the bottom panels we vary the stellar continuum metallicity, while keeping the outflow metallicity constant at 0.5~Z$_\odot$. The solid points are equivalent ratios measured from mock absorption lines created using seven blended absorption features, where each component has one-seventh of the total Cloudy column density, a b-parameter of 20~\kmsp, and a spacing of 20~\kmsp. The error bars are the equivalent width ratios for a single component profile with a line width of 150~\kmsp. Values of U, outflow metallicity, and stellar metallicity are constrained when each of the two ratios are within the gray region. The log(U) value is constrained between -2.25 and -1.5, while the outflow metallicity must be larger than 0.5~Z$_\odot$.}
\label{fig:cloudy}
\end{figure*}

This ionization structure gives vital clues for how outflows are heated and cooled, the density structure of the outflow, and the fraction of total mass in each transition (the ionization fraction). However, the gas can be ionized in many different ways: shocks, photons, cosmic rays, conduction between hot and cool gas, turbulent mixing layers, or many other ways (see \citet{agn} and \cite{wakker12} for a review of many of these). The physical conditions of the outflow cannot be determined unless the ionization mechanism is determined. In the following subsections we create equivalent width ratios from shock and photo-ionization models, and use these models to explore the physical conditions within galactic outflows. 

\subsubsection{Creating Model Equivalent Width Ratios}
\label{create}

The column density ratios define the ionization structure of the galactic outflow. Unfortunately, since the lines are often saturated, we must use the equivalent width (W) ratios as a proxy of the column density ratios. For this reason, we do not compare the model column densities to observed column densities, rather we use synthetic equivalent widths created using models of each ionization mechanism. The synthetic spectra suffer from saturation effects, similar to the observations: as the model column density increases, the equivalent widths enter the square root regime, and increase slowly with increasing column density.

Below, we create line profiles using column densities from models of shocks and photo-ionization. These profiles are created using a constant C$_f$ of 0.8, the column densities from the models, and two different assumptions for how the column density is distributed in velocity space.  First we assume that all of the model column density is within a single component with a line width of 150~\kmsp, and a velocity resolution of 70~\kms (these values are near the median of the sample).  Second, we assume that the absorption is composed of seven individual components, each with one-seventh of the total model column density. Each of the seven components have a b-parameter of 20~\kms and a separation of 20~\kmsp. We use a b-parameter of 20~\kms because it is the width measured in high-resolution Eschelle observations of galactic outflows \citep{scwhartz}. From tests with synthetic spectra, we find that the W ratio does not appreciably change after adding five components, and we conservatively use seven components to create the models. In \autoref{fig:ewratios} and \autoref{tab:ewmodels}, we highlight the multiple component models because absorption from outflows is likely produced by many optically-thick absorbers along the line-of-sight.

\subsubsection{Shock Models}
\label{shock}

When a fluid travels faster than the sound speed through a medium, a shock front develops. The shock compresses and heats the gas, which then efficiently radiatively cools. We use the \citet{allen} shock models, which assume that the shock is a steady flow with a defined shock velocity, preshock density, and magnetic field. We create a grid of equivalent width ratios using a range of these values, and preform a $\chi^2$ test to determine how well the models describe the observations. In the second and third columns of \autoref{tab:ewmodels} we list the observed and predicted equivalent width ratios, assuming the seven component absorption model, a shock velocity of 500~\kmsp, a preshock density of 100~cm$^{-3}$, and a magnetic field of 1~$\mu$G. The predicted shock ratios are also shown in \autoref{fig:ewratios} as dashed lines. The low ionization lines (\oi and \siiip) agree with the shock models, but the high ionization lines (\siiv and \siiiip) are under predicted. Additionally, the models predict a detectable amount of \cip, while \ci is not detected (see \autoref{tab:detection}). The shock ionization models poorly match the observed W ratios.

\subsubsection{Photo-ionization Models}
\label{photo}

Meanwhile, high-mass stars emit photons capable of ionizing the outflow.  When the recombination rate is equal to the ionization rate a photo-ionization equilibrium is established. We test whether the W ratios are consistent with photo-ionization equilibrium using Cloudy, version 13.03 \citep{ferland}. We use the output column densities from Cloudy to produce synthetic line profiles and W ratios in the two ways described in \autoref{create}. The Cloudy models assume that the different gas phases are co-spatial, an important result from the kinematics and the velocity distributions above (\autoref{ewratios}). 

We use a spherically expanding geometry, with a constant hydrogen density. The expanding spherical geometry is not meant to reproduce the outflow profiles, rather it allows for redshifted radiation from the backside of the outflow to ionize gas along the line of sight. The spherical geometry is effectively plane-parallel when the distance between the source and the sphere is much larger than the thickness of the sphere, which is the case for the models presented here \citep{ferland, erb12}.

We use Cloudy\rq{}s pre-loaded H~{\sc II} abundances, which are near the observed Milky Way ISM gas phase abundances  \citep{baldwin1991, savage91, osterbrock1992, rubin1993}. We scale these abundance values to test the effect of the gas phase metallicity on the equivalent width ratios. The abundance models include dust grains, with the Orion Nebular grain distribution \citep{baldwin1991}, which allows for gas to be depleted onto grains (important for Si), and for the dust to absorb, and destroy, UV photons. We input a Starburst99 \citep{claus99, claus2010} stellar continuum spectrum which uses a continuous star formation rate of 1~M$_\odot$~yr$^{-1}$, an age of 6~Myr (the median from the stellar continuum fitting in Paper {\sc I}), a covering fraction of 0.8 (found in \autoref{ewscaling}), and five different stellar metallicities (0.05, 0.2, 0.4, 1.0, and 2.0~Z$_\odot$). We stop the Cloudy simulations when the temperature drops below 4000~K, which is Cloudy\rq{}s default stopping criteria.

We create a grid of Cloudy models by varying Hydrogen density (log(n$_o$)) from -1 to 4, ionization parameter (log(U)) from -3 to 0, the outflow metallicity (Z$_o$) from 0.05 to 2, and the stellar metallicity (Z$_s$) according to the five Starburst99 metallicities. The ionization parameter is defined as
\begin{equation}
U = \frac{n_\gamma}{n_o} = \frac{Q}{4 \pi R^2 c n_o}
\label{eq:U}
\end{equation}
Where n$_\gamma$ is the number density of ionizing photons, Q is the number of H ionizing photons (see \autoref{tab:starburst99}), and $R$ is the distance between the source and the inner edge of the cloud. The output of each Cloudy model is a set of ionic column densities which we turn into equivalent widths following the two methods outlined in \autoref{create}. In \autoref{fig:ewratios} and \autoref{tab:ewmodels} we show photo-ionization models that reproduce the median W ratios of the sample. The photo-ionization models describe the observed equivalent width ratios well for a variety of values of U, Z$_o$, and Z$_s$. 

The fact that the ionization structure is described by photo-ionization is surprising. Shocks describe the optical emission lines \citep{sharp, ho}, but the emission lines arise in very different environments than the absorption lines. Emission traces the densest environments (scaling as n$_o^2$), while absorption traces lower density environments. Consequently, \citet{wood} find that H-$\alpha$ emission traces cluster scale outflows, with lower velocities, while UV absorption lines trace galaxy scale outflows, with larger velocities. 

\subsubsection{Constraining the Conditions Within Outflows}
\label{constraints}

To constrain the physical properties of the outflows we need to consider the full range of photo-ionization models that are consistent with the data.  In \autoref{fig:cloudy} we illustrate the impact of U, Z$_s$, and Z$_o$ on the model \siivp/\siiii and \siivp/\siii W ratios for both the one and seven absorption component models.   We find that at  $log(U) < -1.5$, there is a strong dependence of the model W ratios on  U \citep{tielens},  while at larger values of U the W ratios plateau as both lines saturate.  The range of observed W ratios are indicated by gray bands in \autoref{fig:cloudy}. A model that lies within the gray bands for both the \siivp/\siiii and \siivp/\siii plots fits the data. We do not show \siiip/\oi as this line ratio is insensitive to ionization parameter.

The top two panels of \autoref{fig:cloudy} show the impact of changing the Z$_o$ while keeping the stellar metallicity fixed at Z$_s$ = 0.2~Z$_\odot$. Over the range of U values consistent with the data, Z$_o$ has a relatively modest impact on the W ratios. The effect of the outflow metallicity is most apparent for very low metallicities, where the data rule out models with Z$_o$ < 0.5~Z$_\odot$, because a single U is inconsistent with both the \siivp/\siiii and \siivp/\siii W ratios.

The choice of how the column density is distributed in velocity space does have a small impact on the derived parameters.  For example in the top two panels of \autoref{fig:cloudy},  the seven component model (solid points) that best matches the data has log(U) = -2.25, while the best fitting single component model (the error bars) has log(U) = -2.0.  The W ratios for the 7 component model with Z$_o$  = 0.5 and log(U) = -2.25 are given in \autoref{tab:ewmodels}.
 
The lower panels of \autoref{fig:cloudy} show the effect of changing the stellar continuum metallicity. Changes in the stellar metallicity broaden the possible log(U) values, with larger stellar metallicities requiring larger U values to be within the gray bands. This is largely because line blanketing in high metallicity stars reduces the stellar effective temperature, and the number of ionizing photons \citep{claus99, claus2012}. 

Combining both the one and seven component profiles, Cloudy models with log(U) between-2.25 and -1.5 match the observations for the range of stellar and outflow metallicities tested. These ionization parameters are marginally larger than the log(U)$\sim-2.5$ typically observed in local galaxies \citep{charlot}, but are consistent with the higher values seen in H~{\sc II} regions, and H~{\sc II} galaxies\citep{campbell, snijders}.  Therefore, we conclude that photo-ionization by the stellar continuum is the most likely ionization mechanism, and photo-ionization determines the ionization structure of the outflows. 

\autoref{fig:cloudy} shows that the stellar metallicity can shift the ionization parameter by up to 0.5~dex.  The metallicity of the stellar continuum predominately effects the highest energy photons: as the stellar metallicity increases, the number of photons capable of doubly ionizing He drops by a factor of 38 (see \autoref{tab:starburst99}). A regression shows that the number of ionizing photons in the three zones go as:
\begin{equation}
\begin{aligned}
Q_\text{H~{\sc II}} &=\frac{1.29 \times 10^{53}~\text{photons s$^{-1}$}}{Z_S^{0.2}} \left(\frac{\text{SFR}}{1~\text{\sfrp}}\right)  \\
Q_\text{He~{\sc II}} &=\frac{1.73 \times 10^{52} ~\text{photons s$^{-1}$}}{Z_S^{0.43}}\left(\frac{\text{SFR}}{1~\text{\sfrp}}\right)  \\ 
Q_\text{He~{\sc III}} &=\frac{3.05 \times 10^{48} ~\text{photons s$^{-1}$}}{Z_S^{1.0}}\left(\frac{\text{SFR}}{1~\text{\sfrp}}\right)  \\
\end{aligned}
\end{equation}
If we assume that most of the \siiv and \siiii are created in the He~{\sc II} and H~{\sc II} regions, respectively, then at a constant SFR the ratio of \siivp/\siiii scales as Z$_S^{-0.23}$, and the \siivp/\siiii ratio drops by a factor of 0.59 from 0.2 to 2.0~Z$_\odot$. This accounts for the smooth decrease in \siivp/\siiii from 0.78 at low \mstar to 0.53 at higher \mstar (see \autoref{fig:ewratios}). This result demonstrates that the observed W ratios depend on the properties of the stellar continuum (the shape and normalization of the source), and that the ionizing source shapes the physical conditions within galactic outflows.

\subsubsection{Ionization Fractions for Mass Outflow Rate Calculations}
\label{dominant}
We use our Cloudy models to calculate the fraction of each element in a given ionization stage.   The W ratios constrain log(U) between -2.25 and -1.5 (see \autoref{fig:cloudy}). To bracket the ionization fractions we use two extremes: (1) a log(U)~=~-2.25, a stellar metallicity of 0.2~Z$_\odot$, and an outflow metallicity of 0.5~Z$_\odot$; (2) log(U)~=~-1.5, a stellar continuum metallicity of 2~Z$_\odot$, and an outflow metallicity of 1.5~Z$_\odot$. Cloudy gives the range of outflow ionization fractions ($\chi$ = N(X$_i$)/$\Sigma$ N(X$_i$), where N(X$_i$) is the column density of the ith ionization state of element X) for the four transitions as:
\begin{equation}
\begin{aligned}
&\chi_\text{O {\sc I}} = 0.04 \rightarrow 0.004 \\
&\chi_\text{Si {\sc II}} = 0.20 \rightarrow 0.12 \\
&\chi_\text{Si {\sc III}} = 0.68 \rightarrow 0.80 \\
&\chi_\text{Si {\sc IV}} = 0.12 \rightarrow 0.08 \\
\label{eq:ionfrac}
\end{aligned}
\end{equation}
Where the left number is for the sub-solar stellar metallicity model, and the right number is for the super-solar stellar metallicity model. Other ionization fractions for commonly used outflow tracers are: 
\begin{equation}
\begin{aligned}
&\chi_\text{H~{\sc II}} = 0.96 \rightarrow 0.995 \\
&\chi_\text{Na {\sc I}} = 8 \times 10^{-5} \rightarrow 2.9 \times 10^{-4} \\
&\chi_\text{Fe {\sc II}} = 0.06 \rightarrow 0.02 \\
&\chi_\text{Mg {\sc II}} = 0.13 \rightarrow 0.05 \\
\label{eq:ionfracother}
\end{aligned}
\end{equation}
Note that nearly 100\% of the total H is ionized. In \autoref{coolgas}, we caution that care should be taken when extrapolating these ionization fractions to very dusty outflows, where the dust opacities and distributions are dramatically different. The ionization fractions range by a factor of 1.15 to 10, depending on the transition, and a proper analysis for each galaxy is required to derive accurate ionization fractions. The large variation in the \oi ionization fraction explains most of the scatter seen in the \oi W plot (see \autoref{fig:ewm}).

Approximately 74$\pm$6\% of the total Si is in the \siiii transition. However the \siiii optical depth is difficult to constrain because it is a singlet and the column density is degenerate with the velocity width and covering fraction. Meanwhile, \siiv is a doublet and has an ionization fraction that moderately varies with galaxy properties. Therefore, in the absence of a full model of the ionization structure, the \siiv lines should be used with a constant ionization fraction of $0.10~\pm~0.02$.

When calculating the total amount of Hydrogen in an outflow, previous studies typically assume either no ionization fraction or a constant ionization fraction to convert the column density of a single ion to a total Hydrogen column density \citep{rupkee2005, weiner}. However, \autoref{eq:ionfrac} shows that the ionization fraction can vary by a factor of 10 for common outflow tracers. The absorption line that is used has a large impact on the uncertainty of the mass outflow rate (\autoref{eq:ionfrac}), and we recommend using the \siiv doublet to reduce this uncertainty.

\subsubsection{Where is the cooler gas?}
\label{coolgas}

The photo-ionization modeling does have a noticeable shortcoming: it cannot produce adequate amounts of cool gas. In the optical, Na~{\sc I} often traces outflows from dusty star forming galaxies, but the photo-ionization modeling predicts only trace amounts of Na~{\sc I} (\autoref{eq:ionfracother}).  How is Na~{\sc I} observed outflowing from some galaxies? 

We have ignored a crucial component: dust. Large dust column densities are required to shield Na~{\sc I} from high-energy photons \citep{heckman2000, murray07, chen10}.  While the Cloudy modeling includes dust, large dust columns change the photo-ionization equilibrium conditions.  \citet{chen10} find that Na~{\sc I} outflows do not arise from galaxies with E(B-V) less than 0.3, nearly 40\% higher than the median of our sample (0.22). There are only nine galaxies in this sample with E(B-V) larger than 0.3. The low dust content is a sample selection bias: UV bright galaxies must have low dust extinctions to be observed. We urge caution when extrapolating the ionization fractions to very dusty outflows.  

The ionization fractions of the cooler ions also vary more with stellar metallicity than the higher ions. These results agree with \citet{murray07}, who find that the large variation in Na~{\sc I} W and column densities are largely due to differences in the ionization fraction of Na~{\sc I}. Therefore, we suggest full photo-ionization modeling whenever calculating mass outflow rates.

\section{CONCLUSION}
\label{conclusion}
Here, we study galactic outflows of 37 nearby star forming galaxies. Using UV absorption lines observed with the Cosmic Origins Spectrograph on the Hubble Space Telescope, we characterize the different  ionization states of galactic outflows. We study how the equivalent widths (W), line widths ($\Delta$), covering fractions (C$_f$), optical depths ($\tau_0$), and outflow velocities scale with the stellar mass (\mstarp) and star formation rate (SFR) of their host galaxies. We conclude that
\begin{enumerate}

\item \oip, \siiip, \siiiip, and \siiv absorption lines are frequently detected as outflows, with a fairly constant detection fraction near 80\%. We never detect \ci in the outflows, and only detect \nv in 5 high SFR galaxies (\autoref{detect}).

\item W scales shallowly, but significantly with \mstar and SFR (W~$\propto$\mstarp$^{0.12}$), especially for the \siiii and \siiv transitions (\autoref{ewscaling}). The \oi and \siii relations have considerably more scatter than the higher ionization lines.

\item The \siiv W scaling is driven by an increasing $\Delta$ with increasing \mstarp. C$_f$ and $\tau$ remain roughly fixed with \mstarp, with values of 0.82 and 2.5, respectively (\autoref{fig:doublet}).

\item The equivalent width ratios between \siiip, \siiiip, and \siiv have low dispersion, and remain flat with \mstar (\autoref{fig:ewratios}). In contrast the relations that involve \oi have a factor of two scatter at constant \mstarp.

\item The velocities of all the transitions scale similarly with host galaxy properties as SFR$^{0.12}$, and \mstarp$^{0.13}$ (\autoref{fig:massscaling} and \autoref{fig:sfrscaling}). These relations are much shallower than typically assumed in galaxy simulations.

\item The outflow velocity ratios of the different transitions are flat with \mstar and SFR, but the stronger transitions have larger outflow velocities (\autoref{fig:vratios}). This implies that there is high velocity gas that has a covering fraction (or column density) below our detection limit.

\item Photo-ionization modeling reproduces the observed equivalent width ratios, while shock models do not match the observations. The ionization parameters (U) are constrained between log(U) of -2.25 and -1.5 (see \autoref{fig:cloudy}).To match the observations, the outflow metallicities (Z$_o$) must be greater than or equal to 0.5~Z$_\odot$.

\item Care must be taken when calculating mass outflow rates because galaxy-to-galaxy differences in ionization fractions can add a factor 1.16-10 uncertainty to the mass outflow rates, depending on which transitions are used. Where possible, detailed photo-ionization modeling should be undertaken to derive ionization fractions.

\item \siiii is the dominant ionization state in the outflows, but there is only a single, heavily saturated transition in the COS bandpass. When detailed photo-ionization modeling is  impossible, we recommend using  the \siiv doublet to estimate the mass outflow rate. The \siiv doublet allows for robust column density measurements, while the low variability in the ionization fraction means a nearly constant \siiv ionization fraction can be used with a 20\% error.

\end{enumerate}
In future work, we will analyze the highest quality data within our sample, and measure the ionic column densities to tightly constrain the ionization parameter of the outflows. Using these ionization models, we will derive the ionization fractions of the metal ions, and estimate the total gas mass within the outflows. These new total outflow masses will improve upon previous estimates by removing the substantial source of uncertainty from the ionization fractions and metallicities.

\begin{deluxetable*}{lcccc}
\tabletypesize{\scriptsize}
\tablewidth{0pt}
\tablecaption{Galaxy Sample}
\tablehead{
\colhead{(1)} &
\colhead{(2)} & 
\colhead{(3)} & 
\colhead{(4)}  & 
\colhead{(5)} \\ 
\colhead{Galaxy Name} &
\colhead{log(M$_\ast$/M$_\odot$)} & 
\colhead{SFR} & 
\colhead{E(B-V)$_\text{int}$} & 
\colhead{Morphology} \\ 
\colhead{} &
\colhead{} & 
\colhead{(\sfrp)} & 
\colhead{(mags)}  & 
\colhead{} \\ 
}
\startdata
IRAS~08339+6517  &   10.5  &  13.59 &   0.27  &M \\
NGC~3256  &   10.9  &  50.83 &   0.48  &M  \\
NGC~6090  &   10.7  &  25.15 &   0.32  &M  \\
NGC~7552  &   10.6  &  13.37 &   0.63  &M  \\
Haro~11  &   10.1  &  26.45 &   0.13  &M  \\
J0055-0021  &   10.6  &  71.15 &   0.17  &C  \\
J0150+1260  &   10.5  &  60.40 &   0.19  &M \\
J0926+4427  &    9.9  &  22.07 &   0.09  &M  \\
J0938+5428  &   10.0  &  20.28 &   0.13  &M  \\
J2103-0728  &   10.9  & 136.82 &   0.24  &S  \\
1Zw~18  &    7.2  &   0.02 &   0.09  &I  \\
M~83  &   10.6  &   3.65 &   0.28  &S  \\
NGC~3690  &   10.9  &  90.64 &   0.19  &M  \\
NGC~4214  &    9.0  &   0.13 &   0.27  &I  \\
NGC~4449  &    9.2  &   0.24 &   0.31  &I  \\
NGC~4670  &    9.1  &   0.34 &   0.26  &S  \\
NGC~5253 1  &    8.5  &   0.48 &   0.27  &I  \\
NGC~52532  &    8.5  &   0.48 &   0.20  &I  \\
SBS~1415+437  &    6.9  &   0.02 &   0.10  &I  \\
KISSR~1578  &    9.5  &   3.72 &   0.13  &S  \\
KISSR~218  &    9.9  &   0.99 &   0.40  &S  \\
KISSR~242  &    9.5  &   5.12 &   0.16  &S  \\
KISSR~108  &    8.8  &   0.23 &   0.19  &I  \\
KISSR~178  &   10.0  &   2.55 &   0.29  &S  \\
KISSR~182  &    9.0  &   0.28 &   0.26  &I  \\
NGC~7714  &   10.3  &   9.17 &   0.31  &M  \\
J1144+4012  &   10.4  &  11.82 &   0.32  &C  \\
J1429+1653  &   10.3  &  28.63 &   0.16  &C  \\
J1416+1223  &   10.6  &  46.60 &   0.22  &C  \\
J1415+0540  &   10.1  &  12.21 &   0.24  &C  \\
J1525+0757  &   10.3  &  12.97 &   0.22  &M  \\
J1429+0643  &   10.1  &  61.07 &   0.13  &C  \\
J1112+5503  &   10.7  &  58.47 &   0.28  &C  \\
J1025+3622  &    9.8  &  13.03 &   0.12  &C \\
MRK~1486  &    9.3  &   3.60 &   0.22  &S  \\
J0824+2806  &   10.6  &  20.73 &   0.30  &M  \\
J0907+5327  &    9.0  &   0.86 &   0.16  &I  \\
J1250+0734  &   10.9  &  16.71 &   0.33  &S  \\
J1307+5427  &    9.5  &   5.69 &   0.26  &I  \\
J1315+6207  &   10.6  &  41.08 &   0.45  &M  \\
J1403+0628  &   10.9  &  21.27 &   0.26  &S  \\
GP1244+0216  &    9.4  &  26.20 &   0.17  &C  \\
GP1054+5238  &   9.5  &  22.40 &   0.05  &C \\
\enddata
\tablecomments{List of host galaxy properties for the sample. The second column gives the logarithm of the stellar mass, the third column gives the star formation rate, the fourth column gives the extinction measured from the stellar continuum, and the fifth column gives the morphology of the galaxies. There are four morphological options: Merger/Interacting (M), Spirals (S), Irregular (I), and Compact (C).}
\label{tab:samplegal}
\end{deluxetable*}

\begin{deluxetable*}{lcccccc}
\tabletypesize{\scriptsize}
\tablewidth{0pt}
\tablecaption{Central Velocities (v$_\text{cen}$)}
\tablehead{
\colhead{Galaxy Name} &
\colhead{v$_\text{O~I}$} & 
\colhead{v$_\text{Si~II}$} & 
\colhead{v$_\text{Si~III}$} & 
\colhead{v$_\text{Si~IV}$} & 
\colhead{v$_\text{C~IV}$} & 
\colhead{v$_\text{N~V}$}  \\ 
\colhead{} &
\colhead{(\kmsp)} & 
\colhead{(\kmsp)} & 
\colhead{(\kmsp)} & 
\colhead{(\kmsp)} & 
\colhead{(\kmsp)} & 
\colhead{(\kmsp)} 
}
\startdata
IRAS~08339+6517  &        -514$\pm$          18&        -417$\pm$          12 &        -414$\pm$          11&        -282$\pm$          11&         -           &         -           \\
NGC~3256  &        -517$\pm$          23&        -504$\pm$          16 &         -           &        -418$\pm$          26&         -           &        -538$\pm$          36\\
NGC~6090  &        -146$\pm$           6&        -166$\pm$           6 &        -189$\pm$           5&        -132$\pm$           6&         -           &         -           \\
NGC~7552  &         -           &        -396$\pm$          11 &         -           &        -344$\pm$          13&         -           &        -550$\pm$          23\\
Haro~11  &         -           &        -134$\pm$           9 &         -           &        -151$\pm$           6&        -187$\pm$           9&         -           \\
J0055-0021  &        -121$\pm$          24&        -167$\pm$          24 &        -205$\pm$          26&         -           &         -           &         -           \\
J0150+1260  &         -           &        -157$\pm$          22 &         -           &        -182$\pm$          22&        -188$\pm$          35&         -           \\
J0926+4427  &         -           &        -304$\pm$          20 &        -346$\pm$          17&        -266$\pm$          21&         -           &         -           \\
J0938+5428  &         -21$\pm$          14&         -37$\pm$          14 &         -33$\pm$          15&         -52$\pm$          17&         -87$\pm$          35&         -           \\
J2103-0728  &        -195$\pm$          89&        -173$\pm$          29 &        -291$\pm$          27&        -218$\pm$          43&        -139$\pm$          86&        -291$\pm$          44\\
1Zw~18  &         -           &         -20$\pm$           7 &         -85$\pm$           8&         -           &         -           &         -           \\
M~83  &         -           &         -71$\pm$           1 &        -108$\pm$           4&         -           &         -           &         -           \\
NGC~3690  &         -           &         -            &         -           &         -85$\pm$          11&         -           &         -           \\
NGC~4214  &         -           &         -75$\pm$          14 &         -96$\pm$          21&         -65$\pm$          12&         -           &         -           \\
NGC~4449  &        -101$\pm$           4&         -87$\pm$           9 &         -           &         -           &         -           &         -           \\
NGC~4670  &         -           &         -            &         -           &         -35$\pm$           6&         -           &         -           \\
NGC~5253   &         -           &         -24$\pm$           4 &         -22$\pm$          18&          -9$\pm$           5&         -           &         -           \\
NGC~5253 &         -           &         -32$\pm$           9 &         -44$\pm$          10&         -           &         -           &         -           \\
SBS~1415+437  &         -           &         -58$\pm$           3 &         -           &         -           &         -           &         -           \\
KISSR~1578  &         -           &        -103$\pm$           7 &        -115$\pm$           8&        -128$\pm$           9&         -           &         -           \\
KISSR~218  &        -133$\pm$          28&        -140$\pm$          45 &        -143$\pm$          33&        -197$\pm$          36&         -           &         -           \\
KISSR~242  &         -79$\pm$          10&        -123$\pm$          13 &        -174$\pm$          11&        -130$\pm$          14&         -           &         -           \\
KISSR~108  &         -71$\pm$          18&         -90$\pm$          29 &        -121$\pm$          21&         -           &         -           &         -           \\
KISSR~178  &         -70$\pm$          34&         -62$\pm$          30 &         -           &         -           &         -           &         -           \\
KISSR~182  &         -35$\pm$          14&         -            &         -           &         -           &         -           &         -           \\
NGC~7714  &         -62$\pm$           5&         -57$\pm$           6 &         -           &         -63$\pm$           6&         -51$\pm$           8&         -34$\pm$          11\\
J1144+4012  &        -154$\pm$          18&        -187$\pm$          26 &         -           &        -193$\pm$          26&        -236$\pm$          53&         -           \\
J1429+1653  &         -           &        -126$\pm$          20 &         -           &        -100$\pm$          22&         -           &         -           \\
J1416+1223  &         -           &        -179$\pm$          21 &        -185$\pm$          22&        -232$\pm$          17&        -268$\pm$          39&        -359$\pm$          21\\
J1415+0540  &         -           &         -            &         -           &         -73$\pm$          20&         -           &         -           \\
J1525+0757  &         -           &        -367$\pm$          24 &        -365$\pm$          30&        -290$\pm$          25&        -379$\pm$          29&         -           \\
J1429+0643  &         -           &        -181$\pm$          16 &        -206$\pm$          30&        -185$\pm$          24&         -           &         -           \\
J1112+5503  &         -           &        -342$\pm$          24 &        -364$\pm$          30&        -381$\pm$          22&        -412$\pm$          44&         -           \\
J1025+3622  &         -           &        -178$\pm$          23 &        -171$\pm$          25&        -163$\pm$          15&        -177$\pm$          42&         -           \\
MRK~1486  &        -152$\pm$           8&        -165$\pm$           9 &        -162$\pm$           9&        -139$\pm$           8&         -           &         -           \\
J0824+2806  &         -           &         -            &         -           &         -83$\pm$          11&         -           &         -           \\
J0907+5327  &         -           &         -36$\pm$          15 &         -83$\pm$          14&         -65$\pm$          24&         -           &         -           \\
J1250+0734  &        -180$\pm$          15&        -174$\pm$          16 &        -191$\pm$          21&        -133$\pm$          19&         -           &         -           \\
J1307+5427  &         -           &         -            &         -           &         -41$\pm$          33&         -           &         -           \\
J1315+6207  &         -92$\pm$          18&         -98$\pm$          17 &        -110$\pm$          20&         -76$\pm$          17&         -           &         -           \\
J1403+0628  &        -188$\pm$          28&        -187$\pm$          28 &         -           &         -           &         -           &         -           \\
GP1244+0216  &         -48$\pm$          44&         -54$\pm$          52 &         -62$\pm$          33&         -           &         -           &         -           \\
GP1054+5238  &         -           &        -213$\pm$          39 &        -249$\pm$          34&         -           &         -           &         -           \\

\enddata
\tablecomments{Table of the measured central velocity (\vcenp) for the six studied transitions. We leave a hyphen (-) for transitions that are not measured. The \vcen is not always measured due to geocoronal lines, strong Milky Way absorption, low signal, insufficient wavelength coverage, or chip gaps.}
\label{tab:samplevcen}
\end{deluxetable*}

\begin{deluxetable*}{ccccccc}
\tabletypesize{\scriptsize}
\tablewidth{0pt}
\tablecaption{Velocities at 90\% of the Continuum (v$_\text{90}$)}
\tablehead{
\colhead{Galaxy Name} &
\colhead{v$_\text{O~I}$} & 
\colhead{v$_\text{Si~II}$} & 
\colhead{v$_\text{Si~III}$} & 
\colhead{v$_\text{Si~IV}$} & 
\colhead{v$_\text{C~IV}$} & 
\colhead{v$_\text{N~V}$}  \\ 
\colhead{} &
\colhead{(\kmsp)} & 
\colhead{(\kmsp)} & 
\colhead{(\kmsp)} & 
\colhead{(\kmsp)} & 
\colhead{(\kmsp)} & 
\colhead{(\kmsp)} 
}
\startdata
 IRAS~08339+6517  &        -761$\pm$         125&        -904$\pm$         106 &       -1165$\pm$         102&        -796$\pm$          77&         -           &         -           \\
NGC~3256  &       -1176$\pm$         134&       -1194$\pm$         120 &         -           &        -990$\pm$          89&         -           &       -1002$\pm$          95\\
NGC~6090  &        -424$\pm$          96&        -655$\pm$          76 &        -681$\pm$          11&        -599$\pm$          56&         -           &         -           \\
NGC~7552  &         -           &       -1043$\pm$          47 &         -           &        -886$\pm$          69&         -           &        -909$\pm$          94\\
Haro~11  &         -           &        -332$\pm$          22 &         -           &        -473$\pm$          19&        -533$\pm$          42&         -           \\
J0055-0021  &        -513$\pm$          73&        -670$\pm$          74 &       -1064$\pm$         199&         -           &         -           &         -           \\
J0150+1260  &         -           &        -437$\pm$          26 &         -           &        -447$\pm$          48&        -434$\pm$          91&         -           \\
J0926+4427  &         -           &        -439$\pm$          26 &        -705$\pm$          65&        -607$\pm$          77&         -           &         -           \\
J0938+5428  &        -356$\pm$          45&        -524$\pm$          43 &        -538$\pm$          51&        -408$\pm$          60&        -505$\pm$         149&         -           \\
J2103-0728  &        -873$\pm$         225&        -772$\pm$         117 &        -983$\pm$          73&        -835$\pm$         156&        -689$\pm$         165&        -913$\pm$         137\\
1Zw~18  &         -           &        -151$\pm$           9 &        -300$\pm$           5&         -           &         -           &         -           \\
M~83  &         -           &        -295$\pm$          11 &        -377$\pm$          10&         -           &         -           &         -           \\
NGC~3690  &         -           &         -            &         -           &        -410$\pm$          62&         -           &         -           \\
NGC~4214  &         -           &        -394$\pm$          44 &        -404$\pm$          31&        -344$\pm$          54&         -           &         -           \\
NGC~4449  &        -353$\pm$          11&        -326$\pm$          15 &         -           &         -           &         -           &         -           \\
NGC~4670  &         -           &         -            &         -           &        -393$\pm$          45&         -           &         -           \\
NGC~5253 1  &         -           &        -497$\pm$           5 &        -525$\pm$          89&        -279$\pm$          18&         -           &         -           \\
NGC~5253 2  &         -           &        -539$\pm$         141 &        -574$\pm$         122&         -           &         -           &         -           \\
SBS~1415+437  &         -           &        -200$\pm$          14 &         -           &         -           &         -           &         -           \\
KISSR~1578  &         -           &        -357$\pm$          18 &        -475$\pm$          76&        -395$\pm$          40&         -           &         -           \\
KISSR~218  &        -503$\pm$          72&        -626$\pm$          92 &        -687$\pm$         110&        -620$\pm$         125&         -           &         -           \\
KISSR~242  &        -263$\pm$          12&        -430$\pm$          45 &        -638$\pm$         112&        -380$\pm$          52&         -           &         -           \\
KISSR~108  &        -317$\pm$          34&        -471$\pm$          66 &        -471$\pm$          56&         -           &         -           &         -           \\
KISSR~178  &        -438$\pm$          82&        -420$\pm$          63 &         -           &         -           &         -           &         -           \\
KISSR~182  &        -253$\pm$          24&         -            &         -           &         -           &         -           &         -           \\
NGC~7714  &        -360$\pm$           9&        -412$\pm$          19 &         -           &        -456$\pm$          53&        -437$\pm$          69&        -250$\pm$          17\\
J1144+4012  &        -693$\pm$          63&        -688$\pm$          57 &         -           &        -588$\pm$         108&        -518$\pm$         103&         -           \\
J1429+1653  &         -           &        -419$\pm$          31 &         -           &        -401$\pm$          75&         -           &         -           \\
J1416+1223  &         -           &        -876$\pm$          77 &        -965$\pm$          79&        -815$\pm$         121&        -740$\pm$         140&        -573$\pm$          23\\
J1415+0540  &         -           &         -            &         -           &        -488$\pm$          75&         -           &         -           \\
J1525+0757  &         -           &        -690$\pm$          20 &        -955$\pm$          95&        -675$\pm$          42&        -688$\pm$          85&         -           \\
J1429+0643  &         -           &        -545$\pm$          31 &        -812$\pm$         101&        -609$\pm$         151&         -           &         -           \\
J1112+5503  &         -           &        -956$\pm$          73 &       -1139$\pm$         200&        -971$\pm$         100&       -1037$\pm$         231&         -           \\
J1025+3622  &         -           &        -512$\pm$          37 &        -829$\pm$         142&        -468$\pm$          58&        -512$\pm$         124&         -           \\
MRK~1486  &        -407$\pm$          18&        -497$\pm$          45 &        -605$\pm$          89&        -545$\pm$          36&         -           &         -           \\
J0824+2806  &         -           &         -            &         -           &        -324$\pm$          31&         -           &         -           \\
J0907+5327  &         -           &        -265$\pm$          31 &        -335$\pm$          30&        -241$\pm$          71&         -           &         -           \\
J1250+0734  &        -423$\pm$          14&        -445$\pm$          15 &        -607$\pm$          84&        -410$\pm$          62&         -           &         -           \\
J1307+5427  &         -           &         -            &         -           &        -515$\pm$         111&         -           &         -           \\
J1315+6207  &        -636$\pm$         187&        -866$\pm$         162 &        -903$\pm$         136&        -513$\pm$          78&         -           &         -           \\
J1403+0628  &        -483$\pm$          43&        -481$\pm$          40 &         -           &         -           &         -           &         -           \\
GP1244+0216  &        -275$\pm$          50&        -615$\pm$         112 &        -648$\pm$         136&         -           &         -           &         -           \\
GP1054+5238  &         -           &        -534$\pm$          45 &        -823$\pm$         120&         -           &         -           &         -          \\
\enddata
\tablecomments{Table of the measured equivalent widths (\vn) for the six studied transitions. We leave a hyphen (-) for transitions that are not measured. \vn is not always measured due to geocoronal lines, strong Milky Way absorption, low signal, insufficient wavelength coverage, or chip gaps.}
\label{tab:samplev90}
\end{deluxetable*}

\begin{deluxetable*}{lcccccc}
\tabletypesize{\scriptsize}
\tablewidth{0pt}
\tablecaption{Equivalent Widths}
\tablehead{
\colhead{Galaxy Name} &
\colhead{W$_\text{O~I}$} & 
\colhead{W$_\text{Si~II}$} & 
\colhead{W$_\text{Si~III}$} & 
\colhead{W$_\text{Si~IV}$} & 
\colhead{W$_\text{C~IV}$} & 
\colhead{W$_\text{N~V}$}  \\ 
\colhead{} &
\colhead{(\AA)} & 
\colhead{(\AA)} & 
\colhead{(\AA)} & 
\colhead{(\AA)} & 
\colhead{(\AA)} & 
\colhead{(\AA)} 
}
\startdata
IRAS~08339+6517  &   0.64$\pm$   0.11&   1.53$\pm$   0.11 &   3.18$\pm$   0.12&   1.69$\pm$   0.11&  -   &  -   \\
NGC~3256  &   2.42$\pm$   0.19&   2.62$\pm$   0.16 &  -   &   2.23$\pm$   0.19&  -   &   0.98$\pm$   0.18\\
NGC~6090  &   1.47$\pm$   0.10&   1.86$\pm$   0.10 &   2.47$\pm$   0.10&   1.77$\pm$   0.10&  -   &  -   \\
NGC~7552  &  -   &   3.18$\pm$   0.12 &  -   &   2.85$\pm$   0.13&  -   &   1.50$\pm$   0.14\\
Haro~11  &  -   &   0.70$\pm$   0.10 &  -   &   1.52$\pm$   0.11&   1.93$\pm$   0.12&  -   \\
J0055-0021  &   1.70$\pm$   0.24&   2.59$\pm$   0.27 &   4.03$\pm$   0.60&  -   &  -   &  -   \\
J0150+1260  &  -   &   1.42$\pm$   0.21 &  -   &   1.47$\pm$   0.20&   1.38$\pm$   0.36&  -   \\
J0926+4427  &  -   &   0.44$\pm$   0.13 &   1.28$\pm$   0.18&   1.43$\pm$   0.20&  -   &  -   \\
J0938+5428  &   0.98$\pm$   0.12&   1.84$\pm$   0.14 &   2.42$\pm$   0.20&   1.59$\pm$   0.14&   1.76$\pm$   0.27&  -   \\
J2103-0728  &   0.74$\pm$   0.17&   1.66$\pm$   0.19 &   3.40$\pm$   0.25&   1.85$\pm$   0.22&   2.33$\pm$   0.47&   1.73$\pm$   0.21\\
1Zw~18  &  -   &   0.36$\pm$   0.10 &   0.43$\pm$   0.10&  -   &  -   &  -   \\
M~83  &  -   &   0.84$\pm$   0.10 &   1.16$\pm$   0.10&  -   &  -   &  -   \\
NGC~3690  &  -   &  -    &  -   &   1.24$\pm$   0.11&  -   &  -   \\
NGC~4214  &  -   &   0.93$\pm$   0.12 &   1.08$\pm$   0.14&   0.85$\pm$   0.12&  -   &  -   \\
NGC~4449  &   1.30$\pm$   0.10&   1.12$\pm$   0.11 &  -   &  -   &  -   &  -   \\
NGC~4670  &  -   &  -    &  -   &   1.25$\pm$   0.10&  -   &  -   \\
NGC~5253 1  &  -   &   0.80$\pm$   0.11 &   1.52$\pm$   0.15&   1.10$\pm$   0.10&  -   &  -   \\
NGC~5253 2  &  -   &   0.95$\pm$   0.11 &   1.29$\pm$   0.11&  -   &  -   &  -   \\
SBS~1415+437  &  -   &   0.50$\pm$   0.10 &  -   &  -   &  -   &  -   \\
KISSR~1578  &  -   &   1-   0.11 &   1.63$\pm$   0.12&   1.15$\pm$   0.13&  -   &  -   \\
KISSR~218  &   1.56$\pm$   0.20&   1.83$\pm$   0.37 &   2.11$\pm$   0.41&   1.53$\pm$   0.29&  -   &  -   \\
KISSR~242  &   0.68$\pm$   0.11&   1.18$\pm$   0.14 &   1.75$\pm$   0.15&   1.03$\pm$   0.15&  -   &  -   \\
KISSR~108  &   1.26$\pm$   0.19&   1.86$\pm$   0.28 &   1.41$\pm$   0.21&  -   &  -   &  -   \\
KISSR~178  &   1.88$\pm$   0.37&   1.85$\pm$   0.37 &  -   &  -   &  -   &  -   \\
KISSR~182  &   0.88$\pm$   0.16&  -    &  -   &  -   &  -   &  -   \\
NGC~7714  &   0.98$\pm$   0.10&   1.88$\pm$   0.10 &  -   &   1.85$\pm$   0.10&   2.47$\pm$   0.12&   0.45$\pm$   0.10\\
J1144+4012  &   2.23$\pm$   0.20&   2.43$\pm$   0.31 &  -   &   1.29$\pm$   0.19&   1.39$\pm$   0.45&  -   \\
J1429+1653  &  -   &   0.85$\pm$   0.15 &  -   &   1.36$\pm$   0.21&  -   &  -   \\
J1416+1223  &  -   &   2.97$\pm$   0.17 &   3.88$\pm$   0.28&   2.55$\pm$   0.15&   3.10$\pm$   0.38&   0.72$\pm$   0.17\\
J1415+0540  &  -   &  -    &  -   &   1.36$\pm$   0.18&  -   &  -   \\
J1525+0757  &  -   &   1.81$\pm$   0.25 &   2.94$\pm$   0.85&   1.67$\pm$   0.16&   1.93$\pm$   0.37&  -   \\
J1429+0643  &  -   &   1.35$\pm$   0.14 &   2.48$\pm$   0.31&   1.85$\pm$   0.20&  -   &  -   \\
J1112+5503  &  -   &   3.02$\pm$   0.23 &   3.94$\pm$   0.49&   2.43$\pm$   0.17&   3.46$\pm$   0.39&  -   \\
J1025+3622  &  -   &   1.38$\pm$   0.19 &   2.65$\pm$   0.29&   1.41$\pm$   0.15&   1.81$\pm$   0.40&  -   \\
MRK~1486  &   0.89$\pm$   0.11&   1.49$\pm$   0.13 &   1.99$\pm$   0.13&   1.09$\pm$   0.12&  -   &  -   \\
J0824+2806  &  -   &  -    &  -   &   1.36$\pm$   0.13&  -   &  -   \\
J0907+5327  &  -   &   0.90$\pm$   0.14 &   1.06$\pm$   0.15&   0.83$\pm$   0.18&  -   &  -   \\
J1250+0734  &   1.36$\pm$   0.15&   1.64$\pm$   0.20 &   2.15$\pm$   0.28&   1.34$\pm$   0.20&  -   &  -   \\
J1307+5427  &  -   &  -    &  -   &   1.20$\pm$   0.18&  -   &  -   \\
J1315+6207  &   1.98$\pm$   0.15&   2.36$\pm$   0.35 &   2.85$\pm$   0.24&   1.60$\pm$   0.15&  -   &  -   \\
J1403+0628  &   1.63$\pm$   0.31&   1.59$\pm$   0.30 &  -   &  -   &  -   &  -   \\
GP1244+0216  &   1.18$\pm$   0.28&   1.37$\pm$   0.26 &   1.61$\pm$   0.24&  -   &  -   &  -   \\
GP1054+5238  &  -   &   0.74$\pm$   0.20 &   2.29$\pm$   0.31&  -   &  -   &  -   \\

\enddata
\tablecomments{Table of the measured equivalent widths (W) for the six studied transitions: \oip~1302~\AA, \siiip~1260~\AA, \siiiip~1206~\AA, \siivp~1393, \civp~1548~\AA, and \nvp~1238~\AA. We leave a hyphen (-) for transitions that are not measured. The W is not always measured due to geocoronal lines, strong Milky Way absorption, low signal, insufficient wavelength coverage, or chip gaps.}
\label{tab:sampleew}
\end{deluxetable*}

\begin{deluxetable}{lrrr}
\tabletypesize{\scriptsize}
\tablewidth{0pt}
\tablecaption{Properties of Si~{\sc IV}}
\tablehead{
\colhead{(1)} &
\colhead{(2)} & 
\colhead{(3)} & 
\colhead{(4)}  \\ 
\colhead{Galaxy Name} &
\colhead{$\Delta$} & 
\colhead{C$_f$} & 
\colhead{$\tau$}  \\ 
\colhead{} &
\colhead{(\kmsp)} & 
\colhead{} & 
\colhead{} 
}
\startdata
IRAS~08339+6517  &         299$\pm$           7  &   0.54&   1.72\\
NGC~3256         &         287$\pm$          38  &   0.83&   0.97\\
NGC~6090         &         174$\pm$           7  &   0.89&   3.32\\
NGC~7552         &         266$\pm$          11  &   0.80&   2.82\\
Haro~11          &         132$\pm$           2  &   0.86&   3.04\\
J0150+1260      &         134$\pm$          12  &   0.87&   2.73\\
J0926+4427      &         173$\pm$          13  &   0.82&   2.39\\
J0938+5428      &         170$\pm$          18  &   0.76&   2.68\\
J2103-0728      &         232$\pm$          21  &   0.83&   1.97\\
NGC~3690         &         160$\pm$           6  &   0.57&   2.31\\
NGC~4214         &         130$\pm$          26  &   0.72&   2.40\\
NGC~4670         &         128$\pm$           2  &   0.86&   2.24\\
NGC~5253       &         117$\pm$           3  &   0.80&   2.65\\
KISSR~1578       &         106$\pm$           8  &   0.89&   2.32\\
KISSR~218        &         183$\pm$          37  &   0.77&   2.71\\
KISSR~242        &         107$\pm$          17  &   0.84&   2.34\\
NGC~7714         &         168$\pm$           8  &   0.93&   3.34\\
J1144+4012      &         177$\pm$          35  &   0.72&   2.49\\
J1429+1653      &         113$\pm$          15  &   0.88&   2.40\\
J1416+1223      &         262$\pm$          10  &   0.81&   2.71\\
J1415+0540      &         141$\pm$          16  &   0.80&   2.55\\
J1525+0757      &         194$\pm$          10  &   0.84&   2.60\\
J1429+0643      &         178$\pm$          13  &   0.79&   2.59\\
J1112+5503      &         265$\pm$          10  &   0.81&   2.24\\
J1025+3622      &         132$\pm$           8  &   0.87&   2.78\\
MRK~1486         &         136$\pm$          10  &   0.82&   2.35\\
J0824+2806      &         121$\pm$          10  &   0.87&   2.34\\
J0907+5327      &          88$\pm$          27  &   0.81&   2.80\\
J1250+0734      &         152$\pm$          17  &   0.84&   2.55\\
J1307+5427      &         129$\pm$          23  &   0.79&   2.64\\
J1315+6207      &         173$\pm$          51  &   0.85&   2.95\\
\enddata
\tablecomments{Measured values of the components that make up the \siivp~1402~\AA\ line profile (see \autoref{ewscaling}). Columns two, three and four give the line widths ($\Delta$), covering fraction (C$_f$), and optical depths ($\tau$) of the \siiv transitions.}
\label{tab:samplesi4}
\end{deluxetable}

\begin{deluxetable}{lcccc}
\tabletypesize{\scriptsize}
\tablewidth{0pt}
\tablecaption{Equivalent Width Ratios}
\tablehead{
\colhead{Galaxy Name} &
\colhead{Si~{\sc II}/O~{\sc I}} & 
\colhead{Si~{\sc IV}/O~{\sc I}} & 
\colhead{Si~{\sc IV}/Si~{\sc II}} &
\colhead{Si~{\sc IV}/Si~{\sc III}} 
}
\startdata
IRAS~08339+6517  &   2.39$\pm$   0.43&   2.64$\pm$   0.47&   1.10$\pm$   0.10&   0.53$\pm$   0.04\\
NGC~3256  &   1.08$\pm$   0.11&   0.92$\pm$   0.11&   0.85$\pm$   0.09&  -   \\
NGC~6090  &   1.26$\pm$   0.11&   1.20$\pm$   0.11&   0.95$\pm$   0.08&   0.71$\pm$   0.05\\
NGC~7552  &  -   &  -   &   0.90$\pm$   0.05&  -   \\
J0055-0021  &   1.53$\pm$   0.27&  -   &  -   &  -   \\
J0150+1260  &  -   &  -   &   1.03$\pm$   0.21&  -   \\
J0926+4427  &  -   &  -   &  -   &   1.12$\pm$   0.22\\
J0938+5428  &   1.89$\pm$   0.27&   1.63$\pm$   0.24&   0.86$\pm$   0.10&   0.66$\pm$   0.08\\
J2103-0728  &   2.26$\pm$   0.58&   2.51$\pm$   0.65&   1.11$\pm$   0.18&   0.54$\pm$   0.07\\
NGC~4214  &  -   &  -   &   0.91$\pm$   0.17&   0.78$\pm$   0.15\\
NGC~4449  &   0.86$\pm$   0.11&  -   &  -   &  -   \\
NGC~5253 1  &  -   &  -   &   1.37$\pm$   0.23&   0.72$\pm$   0.10\\
KISSR~1578  &  -   &   3.65$\pm$   1.28&   1.15$\pm$   0.18&   0.70$\pm$   0.09\\
KISSR~218  &   1.17$\pm$   0.28&   0.98$\pm$   0.22&   0.84$\pm$   0.23&   0.73$\pm$   0.20\\
KISSR~242  &   1.74$\pm$   0.35&   1.52$\pm$   0.33&   0.87$\pm$   0.17&   0.59$\pm$   0.10\\
KISSR~108  &   1.47$\pm$   0.31&   0.74$\pm$   0.20&   0.50$\pm$   0.14&   0.66$\pm$   0.18\\
KISSR~178  &   0.98$\pm$   0.28&  -   &  -   &  -   \\
KISSR~182  &  -   &   0.96$\pm$   0.27&  -   &  -   \\
NGC~7714  &   1.92$\pm$   0.23&   1.89$\pm$   0.22&   0.98$\pm$   0.08&  -   \\
J1144+4012  &   1.09$\pm$   0.17&   0.58$\pm$   0.10&   0.53$\pm$   0.10&  -   \\
J1429+1653  &  -   &  -   &   1.60$\pm$   0.38&  -   \\
J1416+1223  &  -   &  -   &   0.86$\pm$   0.07&   0.66$\pm$   0.06\\
J1525+0757  &  -   &  -   &   0.92$\pm$   0.15&   0.57$\pm$   0.17\\
J1429+0643  &  -   &  -   &   1.37$\pm$   0.20&   0.75$\pm$   0.12\\
J1112+5503  &  -   &  -   &   0.80$\pm$   0.08&   0.62$\pm$   0.09\\
J1025+3622  &  -   &  -   &   1.02$\pm$   0.18&   0.53$\pm$   0.08\\
MRK~1486  &   1.67$\pm$   0.25&   1.23$\pm$   0.20&   0.73$\pm$   0.10&   0.55$\pm$   0.07\\
J0907+5327  &  -   &  -   &   0.92$\pm$   0.25&   0.78$\pm$   0.21\\
J1250+0734  &   1.20$\pm$   0.20&   0.98$\pm$   0.18&   0.82$\pm$   0.16&   0.62$\pm$   0.12\\
J1315+6207  &   1.19$\pm$   0.20&   0.81$\pm$   0.10&   0.68$\pm$   0.12&   0.56$\pm$   0.07\\
J1403+0628  &   0.97$\pm$   0.26&  -   &  -   &  -   \\
GP1244+0216  &   1.16$\pm$   0.35&  -   &  -   &  -   \\

\enddata
\label{tab:sampleewrat}
\tablecomments{Table of the equivalent width ratios for the four transitions used to measure the ionization structure of the outflows. We leave a hyphen (-) for transitions that are not measured. The equivalent width ratio is not always measured due to geocoronal lines, strong Milky Way absorption, low signal, insufficient wavelength coverage, or chip gaps.}
\end{deluxetable}

\section*{Acknowledgments}
We thank the anonymous referee for helpful comments that strengthened the manuscript. 

We thank Bart Wakker for the extensive help in extraction and reduction of the COS data, as well as helpful discussions about the data.

Support for program 13239 was provided by NASA through a grant from the Space Telescope Science Institute, which is operated by the Association of Universities for Research in Astronomy, Inc., under NASA contract NAS 5-26555. Some of the data presented in this paper were obtained from the Mikulski Archive for Space Telescopes (MAST). STScI is operated by the Association of Universities for Research in Astronomy, Inc., under NASA contract NAS 5-26555.  

This research has made use of the NASA/IPAC Extragalactic Database (NED), which is operated by the Jet Propulsion Laboratory, California Institute of Technology, under contract with the National Aeronautics and Space Administration. This research has made use of the NASA/ IPAC Infrared Science Archive, which is operated by the Jet Propulsion Laboratory, California Institute of Technology, under contract with the National Aeronautics and Space Administration. This publication makes use of data products from the Wide-field Infrared Survey Explorer, which is a joint project of the University of California, Los Angeles, and the Jet Propulsion Laboratory/California Institute of Technology, funded by the National Aeronautics and Space Administration.

Funding for the Sloan Digital Sky Survey (SDSS) has been provided by the Alfred P. Sloan Foundation, the Participating Institutions, the National Aeronautics and Space Administration, the National Science Foundation, the U.S. Department of Energy, the Japanese Monbukagakusho, and the Max Planck Society. The SDSS Web site is http:\\www.sdss.org/. The SDSS is managed by the Astrophysical Research Consortium (ARC) for the Participating Institutions. The Participating Institutions are The University of Chicago, Fermilab, the Institute for Advanced Study, the Japan Participation Group, The Johns Hopkins University, Los Alamos National Laboratory, the Max-Planck-Institute for Astronomy (MPIA), the Max-Planck-Institute for Astrophysics (MPA), New Mexico State University, University of Pittsburgh, Princeton University, the United States Naval Observatory, and the University of Washington.

AW acknowledges support from the ERC via an Advanced Grant under grant agreement no. 321323NEOGAL.

\bibliographystyle{apj}
\bibliography{chisholm_photo}

\begin{thebibliography}{}
\expandafter\ifx\csname natexlab\endcsname\relax\def\natexlab#1{#1}\fi

\bibitem[{{Alexandroff} {et~al.}(2015){Alexandroff}, {Heckman}, {Borthakur},
  {Overzier}, \& {Leitherer}}]{alexandroff}
{Alexandroff}, R., {Heckman}, T., {Borthakur}, S., {Overzier}, R., \&
  {Leitherer}, C. 2015, ArXiv e-prints, arXiv:1504.02446

\bibitem[{{Allen} {et~al.}(2008){Allen}, {Groves}, {Dopita}, {Sutherland}, \&
  {Kewley}}]{allen}
{Allen}, M.~G., {Groves}, B.~A., {Dopita}, M.~A., {Sutherland}, R.~S., \&
  {Kewley}, L.~J. 2008, \apjs, 178, 20

\bibitem[{{Andrews} \& {Martini}(2013)}]{andrews13}
{Andrews}, B.~H., \& {Martini}, P. 2013, \apj, 765, 140

\bibitem[{{Baldwin} {et~al.}(1991){Baldwin}, {Ferland}, {Martin}, {Corbin},
  {Cota}, {Peterson}, \& {Slettebak}}]{baldwin1991}
{Baldwin}, J.~A., {Ferland}, G.~J., {Martin}, P.~G., {et~al.} 1991, \apj, 374,
  580

\bibitem[{{Ben Zhu} {et~al.}(2015){Ben Zhu}, {Comparat}, {Kneib}, {Delubac},
  {Raichoor}, {Dawson}, {Newman}, {Y{\`e}che}, {Zhou}, \&
  {Schneider}}]{guantun}
{Ben Zhu}, G., {Comparat}, J., {Kneib}, J.-P., {et~al.} 2015, ArXiv e-prints,
  arXiv:1507.07979

\bibitem[{{Blanton} \& {Roweis}(2007)}]{kcorrect}
{Blanton}, M.~R., \& {Roweis}, S. 2007, \aj, 133, 734

\bibitem[{{Brinchmann} {et~al.}(2004){Brinchmann}, {Charlot}, {Heckman},
  {Kauffmann}, {Tremonti}, \& {White}}]{brinchmann2004}
{Brinchmann}, J., {Charlot}, S., {Heckman}, T.~M., {et~al.} 2004, ArXiv
  Astrophysics e-prints, astro-ph/0406220

\bibitem[{{Buat} {et~al.}(2011){Buat}, {Giovannoli}, {Takeuchi}, {Heinis},
  {Yuan}, {Burgarella}, {Noll}, \& {Iglesias-P{\'a}ramo}}]{buat11}
{Buat}, V., {Giovannoli}, E., {Takeuchi}, T.~T., {et~al.} 2011, \aap, 529, A22

\bibitem[{{Calzetti} {et~al.}(2000){Calzetti}, {Armus}, {Bohlin}, {Kinney},
  {Koornneef}, \& {Storchi-Bergmann}}]{calzetti}
{Calzetti}, D., {Armus}, L., {Bohlin}, R.~C., {et~al.} 2000, \apj, 533, 682

\bibitem[{{Campbell}(1988)}]{campbell}
{Campbell}, A. 1988, \apj, 335, 644

\bibitem[{{Cardamone} {et~al.}(2009){Cardamone}, {Schawinski}, {Sarzi},
  {Bamford}, {Bennert}, {Urry}, {Lintott}, {Keel}, {Parejko}, {Nichol},
  {Thomas}, {Andreescu}, {Murray}, {Raddick}, {Slosar}, {Szalay}, \&
  {Vandenberg}}]{cardamone}
{Cardamone}, C., {Schawinski}, K., {Sarzi}, M., {et~al.} 2009, \mnras, 399,
  1191

\bibitem[{{Castor} {et~al.}(1975){Castor}, {McCray}, \& {Weaver}}]{castor75}
{Castor}, J., {McCray}, R., \& {Weaver}, R. 1975, \apjl, 200, L107

\bibitem[{{Charlot} \& {Longhetti}(2001)}]{charlot}
{Charlot}, S., \& {Longhetti}, M. 2001, \mnras, 323, 887

\bibitem[{{Chen} {et~al.}(2010){Chen}, {Tremonti}, {Heckman}, {Kauffmann},
  {Weiner}, {Brinchmann}, \& {Wang}}]{chen10}
{Chen}, Y.-M., {Tremonti}, C.~A., {Heckman}, T.~M., {et~al.} 2010, \aj, 140,
  445

\bibitem[{{Chisholm} {et~al.}(2015){Chisholm}, {Tremonti}, {Leitherer}, {Chen},
  {Wofford}, \& {Lundgren}}]{chisholm}
{Chisholm}, J., {Tremonti}, C.~A., {Leitherer}, C., {et~al.} 2015, \apj, 811,
  149

\bibitem[{{Cutri} {et~al.}(2012){Cutri}, {Wright}, {Conrow}, {Bauer},
  {Benford}, {Brandenburg}, {Dailey}, {Eisenhardt}, {Evans}, {Fajardo-Acosta},
  {Fowler}, {Gelino}, {Grillmair}, {Harbut}, {Hoffman}, {Jarrett},
  {Kirkpatrick}, {Leisawitz}, {Liu}, {Mainzer}, {Marsh}, {Masci}, {McCallon},
  {Padgett}, {Ressler}, {Royer}, {Skrutskie}, {Stanford}, {Wyatt}, {Tholen},
  {Tsai}, {Wachter}, {Wheelock}, {Yan}, {Alles}, {Beck}, {Grav}, {Masiero},
  {McCollum}, {McGehee}, {Papin}, \& {Wittman}}]{wisesupp}
{Cutri}, R.~M., {Wright}, E.~L., {Conrow}, T., {et~al.} 2012, {Explanatory
  Supplement to the WISE All-Sky Data Release Products}, Tech. rep.

\bibitem[{{Dav{\'e}} {et~al.}(2012){Dav{\'e}}, {Finlator}, \&
  {Oppenheimer}}]{dave12}
{Dav{\'e}}, R., {Finlator}, K., \& {Oppenheimer}, B.~D. 2012, \mnras, 421, 98

\bibitem[{{de Mello} {et~al.}(2000){de Mello}, {Leitherer}, \&
  {Heckman}}]{demello}
{de Mello}, D.~F., {Leitherer}, C., \& {Heckman}, T.~M. 2000, \apj, 530, 251

\bibitem[{{Draine}(2011)}]{draine}
{Draine}, B.~T. 2011, {Physics of the Interstellar and Intergalactic Medium}

\bibitem[{{Erb}(2015)}]{erb15}
{Erb}, D.~K. 2015, \nat, 523, 169

\bibitem[{{Erb} {et~al.}(2012){Erb}, {Quider}, {Henry}, \& {Martin}}]{erb12}
{Erb}, D.~K., {Quider}, A.~M., {Henry}, A.~L., \& {Martin}, C.~L. 2012, \apj,
  759, 26

\bibitem[{{Feigelson} \& {Jogesh Babu}(2012)}]{feigelson}
{Feigelson}, E.~D., \& {Jogesh Babu}, G. 2012, {Modern Statistical Methods for
  Astronomy}

\bibitem[{{Ferland} {et~al.}(2013){Ferland}, {Porter}, {van Hoof}, {Williams},
  {Abel}, {Lykins}, {Shaw}, {Henney}, \& {Stancil}}]{ferland}
{Ferland}, G.~J., {Porter}, R.~L., {van Hoof}, P.~A.~M., {et~al.} 2013, rmxaa,
  49, 137

\bibitem[{{Ferrara} \& {Tolstoy}(2000)}]{ferrara2000}
{Ferrara}, A., \& {Tolstoy}, E. 2000, \mnras, 313, 291

\bibitem[{{Finlator} \& {Dav{\'e}}(2008)}]{finlator08}
{Finlator}, K., \& {Dav{\'e}}, R. 2008, \mnras, 385, 2181

\bibitem[{{Fox} {et~al.}(2013){Fox}, {Richter}, {Wakker}, {Lehner}, {Howk},
  {Ben Bekhti}, {Bland-Hawthorn}, \& {Lucas}}]{fox2013}
{Fox}, A.~J., {Richter}, P., {Wakker}, B.~P., {et~al.} 2013, \apj, 772, 110

\bibitem[{{Fox} {et~al.}(2014){Fox}, {Wakker}, {Barger}, {Hernandez},
  {Richter}, {Lehner}, {Bland-Hawthorn}, {Charlton}, {Westmeier}, {Thom},
  {Tumlinson}, {Misawa}, {Howk}, {Haffner}, {Ely}, {Rodriguez-Hidalgo}, \&
  {Kumari}}]{fox14}
{Fox}, A.~J., {Wakker}, B.~P., {Barger}, K.~A., {et~al.} 2014, ArXiv e-prints,
  arXiv:1404.5514

\bibitem[{{France} {et~al.}(2009){France}, {Beasley}, {Keeney}, {Danforth},
  {Froning}, {Green}, \& {Shull}}]{france09}
{France}, K., {Beasley}, M., {Keeney}, B.~A., {et~al.} 2009, \apjl, 707, L27

\bibitem[{{France} {et~al.}(2010){France}, {Nell}, {Green}, \&
  {Leitherer}}]{france2010}
{France}, K., {Nell}, N., {Green}, J.~C., \& {Leitherer}, C. 2010, \apjl, 722,
  L80

\bibitem[{{Gil de Paz} {et~al.}(2007){Gil de Paz}, {Boissier}, {Madore},
  {Seibert}, {Joe}, {Boselli}, {Wyder}, {Thilker}, {Bianchi}, {Rey}, {Rich},
  {Barlow}, {Conrow}, {Forster}, {Friedman}, {Martin}, {Morrissey}, {Neff},
  {Schiminovich}, {Small}, {Donas}, {Heckman}, {Lee}, {Milliard}, {Szalay}, \&
  {Yi}}]{galexatlas}
{Gil de Paz}, A., {Boissier}, S., {Madore}, B.~F., {et~al.} 2007, \apjs, 173,
  185

\bibitem[{{Green} {et~al.}(2012){Green}, {Froning}, {Osterman}, {Ebbets},
  {Heap}, {Leitherer}, {Linsky}, {Savage}, {Sembach}, {Shull}, {Siegmund},
  {Snow}, {Spencer}, {Stern}, {Stocke}, {Welsh}, {B{\'e}land}, {Burgh},
  {Danforth}, {France}, {Keeney}, {McPhate}, {Penton}, {Andrews},
  {Brownsberger}, {Morse}, \& {Wilkinson}}]{cos}
{Green}, J.~C., {Froning}, C.~S., {Osterman}, S., {et~al.} 2012, \apj, 744, 60

\bibitem[{{Grimes} {et~al.}(2009){Grimes}, {Heckman}, {Aloisi}, {Calzetti},
  {Leitherer}, {Martin}, {Meurer}, {Sembach}, \& {Strickland}}]{grimes09}
{Grimes}, J.~P., {Heckman}, T., {Aloisi}, A., {et~al.} 2009, \apjs, 181, 272

\bibitem[{Hallin(1966)}]{nitrogen}
Hallin, R. 1966, Ark. Fys. (Stockholm), 31, 511

\bibitem[{{Hamann} {et~al.}(1997){Hamann}, {Barlow}, {Junkkarinen}, \&
  {Burbidge}}]{hamann}
{Hamann}, F., {Barlow}, T.~A., {Junkkarinen}, V., \& {Burbidge}, E.~M. 1997,
  \apj, 478, 80

\bibitem[{{Hayes} {et~al.}(2013){Hayes}, {{\"O}stlin}, {Schaerer}, {Verhamme},
  {Mas-Hesse}, {Adamo}, {Atek}, {Cannon}, {Duval}, {Guaita}, {Herenz}, {Kunth},
  {Laursen}, {Melinder}, {Orlitov{\'a}}, {Ot{\'{\i}}-Floranes}, \&
  {Sandberg}}]{hayes}
{Hayes}, M., {{\"O}stlin}, G., {Schaerer}, D., {et~al.} 2013, \apjl, 765, L27

\bibitem[{{Heckman} {et~al.}(2015){Heckman}, {Alexandroff}, {Borthakur},
  {Overzier}, \& {Leitherer}}]{heckman15}
{Heckman}, T.~M., {Alexandroff}, R.~M., {Borthakur}, S., {Overzier}, R., \&
  {Leitherer}, C. 2015, \apj, 809, 147

\bibitem[{{Heckman} {et~al.}(1990){Heckman}, {Armus}, \& {Miley}}]{heckman90}
{Heckman}, T.~M., {Armus}, L., \& {Miley}, G.~K. 1990, \apjs, 74, 833

\bibitem[{{Heckman} {et~al.}(2000){Heckman}, {Lehnert}, {Strickland}, \&
  {Armus}}]{heckman2000}
{Heckman}, T.~M., {Lehnert}, M.~D., {Strickland}, D.~K., \& {Armus}, L. 2000,
  \apjs, 129, 493

\bibitem[{{Heckman} {et~al.}(2011){Heckman}, {Borthakur}, {Overzier},
  {Kauffmann}, {Basu-Zych}, {Leitherer}, {Sembach}, {Martin}, {Rich},
  {Schiminovich}, \& {Seibert}}]{heckman2011}
{Heckman}, T.~M., {Borthakur}, S., {Overzier}, R., {et~al.} 2011, \apj, 730, 5

\bibitem[{{Henry} {et~al.}(2015){Henry}, {Scarlata}, {Martin}, \&
  {Erb}}]{henry}
{Henry}, A., {Scarlata}, C., {Martin}, C.~L., \& {Erb}, D. 2015, \apj, 809, 19

\bibitem[{{Ho} {et~al.}(2014){Ho}, {Kewley}, {Dopita}, {Medling}, {Allen},
  {Bland-Hawthorn}, {Bloom}, {Bryant}, {Croom}, {Fogarty}, {Goodwin}, {Green},
  {Konstantopoulos}, {Lawrence}, {L{\'o}pez-S{\'a}nchez}, {Owers}, {Richards},
  \& {Sharp}}]{ho}
{Ho}, I.-T., {Kewley}, L.~J., {Dopita}, M.~A., {et~al.} 2014, \mnras, 444, 3894

\bibitem[{{Hopkins} {et~al.}(2014){Hopkins}, {Kere{\v s}}, {O{\~n}orbe},
  {Faucher-Gigu{\`e}re}, {Quataert}, {Murray}, \& {Bullock}}]{hopkins14}
{Hopkins}, P.~F., {Kere{\v s}}, D., {O{\~n}orbe}, J., {et~al.} 2014, \mnras,
  445, 581

\bibitem[{{Hopkins} {et~al.}(2012){Hopkins}, {Quataert}, \&
  {Murray}}]{hopkins12a}
{Hopkins}, P.~F., {Quataert}, E., \& {Murray}, N. 2012, \mnras, 421, 3488

\bibitem[{{Izotov} {et~al.}(2011){Izotov}, {Guseva}, \& {Thuan}}]{izotov}
{Izotov}, Y.~I., {Guseva}, N.~G., \& {Thuan}, T.~X. 2011, \apj, 728, 161

\bibitem[{{James} {et~al.}(2014){James}, {Aloisi}, {Heckman}, {Sohn}, \&
  {Wolfe}}]{james}
{James}, B.~L., {Aloisi}, A., {Heckman}, T.~M., {Sohn}, S.~T., \& {Wolfe},
  M.~A. 2014, ArXiv e-prints, arXiv:1408.4420

\bibitem[{{Jarosik} {et~al.}(2011){Jarosik}, {Bennett}, {Dunkley}, {Gold},
  {Greason}, {Halpern}, {Hill}, {Hinshaw}, {Kogut}, {Komatsu}, {Larson},
  {Limon}, {Meyer}, {Nolta}, {Odegard}, {Page}, {Smith}, {Spergel}, {Tucker},
  {Weiland}, {Wollack}, \& {Wright}}]{wmap}
{Jarosik}, N., {Bennett}, C.~L., {Dunkley}, J., {et~al.} 2011, \apjs, 192, 14

\bibitem[{{Jarrett} {et~al.}(2011){Jarrett}, {Cohen}, {Masci}, {Wright},
  {Stern}, {Benford}, {Blain}, {Carey}, {Cutri}, {Eisenhardt}, {Lonsdale},
  {Mainzer}, {Marsh}, {Padgett}, {Petty}, {Ressler}, {Skrutskie}, {Stanford},
  {Surace}, {Tsai}, {Wheelock}, \& {Yan}}]{jarrett2011}
{Jarrett}, T.~H., {Cohen}, M., {Masci}, F., {et~al.} 2011, \apj, 735, 112

\bibitem[{{Jarrett} {et~al.}(2013){Jarrett}, {Masci}, {Tsai}, {Petty},
  {Cluver}, {Assef}, {Benford}, {Blain}, {Bridge}, {Donoso}, {Eisenhardt},
  {Koribalski}, {Lake}, {Neill}, {Seibert}, {Sheth}, {Stanford}, \&
  {Wright}}]{jarrett2013}
{Jarrett}, T.~H., {Masci}, F., {Tsai}, C.~W., {et~al.} 2013, \aj, 145, 6

\bibitem[{{Jaskot} \& {Oey}(2014)}]{jaskot}
{Jaskot}, A.~E., \& {Oey}, M.~S. 2014, \apjl, 791, L19

\bibitem[{{Jenkins}(2009)}]{jenkins}
{Jenkins}, E.~B. 2009, \apj, 700, 1299

\bibitem[{{Kauffmann} {et~al.}(2003){Kauffmann}, {Heckman}, {White}, {Charlot},
  {Tremonti}, {Brinchmann}, {Bruzual}, {Peng}, {Seibert}, {Bernardi},
  {Blanton}, {Brinkmann}, {Castander}, {Cs{\'a}bai}, {Fukugita}, {Ivezic},
  {Munn}, {Nichol}, {Padmanabhan}, {Thakar}, {Weinberg}, \&
  {York}}]{kauffmann2003}
{Kauffmann}, G., {Heckman}, T.~M., {White}, S.~D.~M., {et~al.} 2003, \mnras,
  341, 33

\bibitem[{Kaufman(1982)}]{s}
Kaufman, V. 1982, Phys. Scr., 26, 439

\bibitem[{{Kennicutt}(1998)}]{kennicutt}
{Kennicutt}, Jr., R.~C. 1998, \apj, 498, 541

\bibitem[{{Kornei} {et~al.}(2012){Kornei}, {Shapley}, {Martin}, {Coil}, {Lotz},
  {Schiminovich}, {Bundy}, \& {Noeske}}]{kornei12}
{Kornei}, K.~A., {Shapley}, A.~E., {Martin}, C.~L., {et~al.} 2012, \apj, 758,
  135

\bibitem[{Kramida {et~al.}(2014)Kramida, {Yu.~Ralchenko}, Reader, \& {and NIST
  ASD Team}}]{nist}
Kramida, A., {Yu.~Ralchenko}, Reader, J., \& {and NIST ASD Team}. 2014, {NIST
  Atomic Spectra Database (ver. 5.2), [Online]. Available:
  {\tt{http://physics.nist.gov/asd}} [2014, October 23]. National Institute of
  Standards and Technology, Gaithersburg, MD.}

\bibitem[{{Larson}(1974)}]{larson74}
{Larson}, R.~B. 1974, \mnras, 169, 229

\bibitem[{{Leitherer} {et~al.}(2013){Leitherer}, {Chandar}, {Tremonti},
  {Wofford}, \& {Schaerer}}]{claus2012}
{Leitherer}, C., {Chandar}, R., {Tremonti}, C.~A., {Wofford}, A., \&
  {Schaerer}, D. 2013, \apj, 772, 120

\bibitem[{{Leitherer} {et~al.}(2010){Leitherer}, {Ortiz Ot{\'a}lvaro},
  {Bresolin}, {Kudritzki}, {Lo Faro}, {Pauldrach}, {Pettini}, \&
  {Rix}}]{claus2010}
{Leitherer}, C., {Ortiz Ot{\'a}lvaro}, P.~A., {Bresolin}, F., {et~al.} 2010,
  \apjs, 189, 309

\bibitem[{{Leitherer} {et~al.}(1999){Leitherer}, {Schaerer}, {Goldader},
  {Gonz{\'a}lez Delgado}, {Robert}, {Kune}, {de Mello}, {Devost}, \&
  {Heckman}}]{claus99}
{Leitherer}, C., {Schaerer}, D., {Goldader}, J.~D., {et~al.} 1999, \apjs, 123,
  3

\bibitem[{{Markwardt}(2009)}]{mpfit}
{Markwardt}, C.~B. 2009, in Astronomical Society of the Pacific Conference
  Series, Vol. 411, Astronomical Data Analysis Software and Systems XVIII, ed.
  D.~A. {Bohlender}, D.~{Durand}, \& P.~{Dowler}, 251

\bibitem[{{Martin}(2005)}]{martin2005}
{Martin}, C.~L. 2005, \apj, 621, 227

\bibitem[{{Martin} \& {Bouch{\'e}}(2009)}]{martin09}
{Martin}, C.~L., \& {Bouch{\'e}}, N. 2009, \apj, 703, 1394

\bibitem[{{Martin} {et~al.}(2012){Martin}, {Shapley}, {Coil}, {Kornei},
  {Bundy}, {Weiner}, {Noeske}, \& {Schiminovich}}]{martin12}
{Martin}, C.~L., {Shapley}, A.~E., {Coil}, A.~L., {et~al.} 2012, \apj, 760, 127

\bibitem[{{Martin} {et~al.}(2005){Martin}, {Fanson}, {Schiminovich},
  {Morrissey}, {Friedman}, {Barlow}, {Conrow}, {Grange}, {Jelinsky},
  {Milliard}, {Siegmund}, {Bianchi}, {Byun}, {Donas}, {Forster}, {Heckman},
  {Lee}, {Madore}, {Malina}, {Neff}, {Rich}, {Small}, {Surber}, {Szalay},
  {Welsh}, \& {Wyder}}]{galex}
{Martin}, D.~C., {Fanson}, J., {Schiminovich}, D., {et~al.} 2005, \apjl, 619,
  L1

\bibitem[{{McKee} \& {Ostriker}(1977)}]{mckee77}
{McKee}, C.~F., \& {Ostriker}, J.~P. 1977, \apj, 218, 148

\bibitem[{{Meynet} {et~al.}(1994){Meynet}, {Maeder}, {Schaller}, {Schaerer}, \&
  {Charbonnel}}]{geneva94}
{Meynet}, G., {Maeder}, A., {Schaller}, G., {Schaerer}, D., \& {Charbonnel}, C.
  1994, \aaps, 103, 97

\bibitem[{Moore(1970)}]{carbon}
Moore, C.~E. 1970, in Nat. Stand. Ref. Data Ser., NSRDS-NBS 3 (Sect. 3) (U.S.:
  Nat. Bur. Stand.)

\bibitem[{Moore(1976)}]{oxygen}
---. 1976, in Nat. Stand. Ref. Data Ser., NSRDS-NBS 3 (Sect. 7) (U.S.: Nat.
  Bur. Stand.)

\bibitem[{{Moster} {et~al.}(2010){Moster}, {Somerville}, {Maulbetsch}, {van den
  Bosch}, {Macci{\`o}}, {Naab}, \& {Oser}}]{moster10}
{Moster}, B.~P., {Somerville}, R.~S., {Maulbetsch}, C., {et~al.} 2010, \apj,
  710, 903

\bibitem[{{Murray} {et~al.}(2007){Murray}, {Martin}, {Quataert}, \&
  {Thompson}}]{murray07}
{Murray}, N., {Martin}, C.~L., {Quataert}, E., \& {Thompson}, T.~A. 2007, \apj,
  660, 211

\bibitem[{{Murray} {et~al.}(2005){Murray}, {Quataert}, \&
  {Thompson}}]{murray05}
{Murray}, N., {Quataert}, E., \& {Thompson}, T.~A. 2005, \apj, 618, 569

\bibitem[{Nave \& Johansson(2013)}]{iron}
Nave, G., \& Johansson, S. 2013, Astrophys. J., Suppl. Ser., 204, 1

\bibitem[{{Oppenheimer} \& {Dav{\'e}}(2006)}]{oppenheimer06}
{Oppenheimer}, B.~D., \& {Dav{\'e}}, R. 2006, \mnras, 373, 1265

\bibitem[{{Osterbrock} \& {Ferland}(2006)}]{agn}
{Osterbrock}, D.~E., \& {Ferland}, G.~J. 2006, {Astrophysics of gaseous nebulae
  and active galactic nuclei}

\bibitem[{{Osterbrock} {et~al.}(1992){Osterbrock}, {Tran}, \&
  {Veilleux}}]{osterbrock1992}
{Osterbrock}, D.~E., {Tran}, H.~D., \& {Veilleux}, S. 1992, \apj, 389, 305

\bibitem[{{{\"O}stlin} {et~al.}(2014){{\"O}stlin}, {Hayes}, {Duval},
  {Sandberg}, {Rivera-Thorsen}, {Marquart}, {Orlitova}, {Adamo}, {Melinder},
  {Guaita}, {Atek}, {Cannon}, {Gruyters}, {Herenz}, {Kunth}, {Laursen},
  {Mas-Hesse}, {Micheva}, {Pardy}, {Roth}, {Schaerer}, \& {Verhamme}}]{ostlin}
{{\"O}stlin}, G., {Hayes}, M., {Duval}, F., {et~al.} 2014, ArXiv e-prints,
  arXiv:1409.8347

\bibitem[{{Overzier} {et~al.}(2009){Overzier}, {Heckman}, {Tremonti}, {Armus},
  {Basu-Zych}, {Gon{\c c}alves}, {Rich}, {Martin}, {Ptak}, {Schiminovich},
  {Ford}, {Madore}, \& {Seibert}}]{overzier09}
{Overzier}, R.~A., {Heckman}, T.~M., {Tremonti}, C., {et~al.} 2009, \apj, 706,
  203

\bibitem[{{Prochaska} {et~al.}(2011){Prochaska}, {Kasen}, \&
  {Rubin}}]{prochaska2011}
{Prochaska}, J.~X., {Kasen}, D., \& {Rubin}, K. 2011, \apj, 734, 24

\bibitem[{{Querejeta} {et~al.}(2015){Querejeta}, {Meidt}, {Schinnerer},
  {Cisternas}, {Mu{\~n}oz-Mateos}, {Sheth}, {Knapen}, {van de Ven}, {Norris},
  {Peletier}, {Laurikainen}, {Salo}, {Holwerda}, {Athanassoula}, {Bosma},
  {Groves}, {Ho}, {Gadotti}, {Zaritsky}, {Regan}, {Hinz}, {Gil de Paz},
  {Menendez-Delmestre}, {Seibert}, {Mizusawa}, {Kim}, {Erroz-Ferrer}, {Laine},
  \& {Comer{\'o}n}}]{querejeta}
{Querejeta}, M., {Meidt}, S.~E., {Schinnerer}, E., {et~al.} 2015, \apjs, 219, 5

\bibitem[{{Richter} {et~al.}(2013){Richter}, {Fox}, {Wakker}, {Lehner}, {Howk},
  {Bland-Hawthorn}, {Ben Bekhti}, \& {Fechner}}]{richter2013}
{Richter}, P., {Fox}, A.~J., {Wakker}, B.~P., {et~al.} 2013, \apj, 772, 111

\bibitem[{{Rivera-Thorsen} {et~al.}(2015){Rivera-Thorsen}, {Hayes},
  {{\"O}stlin}, {Duval}, {Orlitov{\'a}}, {Verhamme}, {Mas-Hesse}, {Schaerer},
  {Cannon}, {Ot{\'{\i}}-Floranes}, {Sandberg}, {Guaita}, {Adamo}, {Atek},
  {Herenz}, {Kunth}, {Laursen}, \& {Melinder}}]{Rivera}
{Rivera-Thorsen}, T.~E., {Hayes}, M., {{\"O}stlin}, G., {et~al.} 2015, \apj,
  805, 14

\bibitem[{{Rubin} {et~al.}(2014){Rubin}, {Prochaska}, {Koo}, {Phillips},
  {Martin}, \& {Winstrom}}]{rubin13}
{Rubin}, K.~H.~R., {Prochaska}, J.~X., {Koo}, D.~C., {et~al.} 2014, \apj, 794,
  156

\bibitem[{{Rubin} {et~al.}(1993){Rubin}, {Dufour}, \& {Walter}}]{rubin1993}
{Rubin}, R.~H., {Dufour}, R.~J., \& {Walter}, D.~K. 1993, \apj, 413, 242

\bibitem[{{Rupke} {et~al.}(2005){Rupke}, {Veilleux}, \& {Sanders}}]{rupkee2005}
{Rupke}, D.~S., {Veilleux}, S., \& {Sanders}, D.~B. 2005, \apjs, 160, 87

\bibitem[{Sansonetti {et~al.}(2004)Sansonetti, Kerber, Reader, \&
  Rosa}]{hydrogen}
Sansonetti, C.~J., Kerber, F., Reader, J., \& Rosa, M.~R. 2004, Astrophys. J.,
  Suppl. Ser., 153, 555

\bibitem[{{Savage} \& {Sembach}(1991)}]{savage91}
{Savage}, B.~D., \& {Sembach}, K.~R. 1991, \apj, 379, 245

\bibitem[{{Schwartz} \& {Martin}(2004)}]{scwhartz}
{Schwartz}, C.~M., \& {Martin}, C.~L. 2004, \apj, 610, 201

\bibitem[{{Sembach} \& {Savage}(1992)}]{Sembach1992}
{Sembach}, K.~R., \& {Savage}, B.~D. 1992, \apjs, 83, 147

\bibitem[{{Shapiro} \& {Field}(1976)}]{shapiro}
{Shapiro}, P.~R., \& {Field}, G.~B. 1976, \apj, 205, 762

\bibitem[{{Sharp} \& {Bland-Hawthorn}(2010)}]{sharp}
{Sharp}, R.~G., \& {Bland-Hawthorn}, J. 2010, \apj, 711, 818

\bibitem[{Shenstone(1961)}]{si2}
Shenstone, A.~G. 1961, Proc. R. Soc. London, Ser. A, 261, 153

\bibitem[{{Snijders} {et~al.}(2007){Snijders}, {Kewley}, \& {van der
  Werf}}]{snijders}
{Snijders}, L., {Kewley}, L.~J., \& {van der Werf}, P.~P. 2007, \apj, 669, 269

\bibitem[{{Spitzer}(1978)}]{spitzer}
{Spitzer}, L. 1978, {Physical processes in the interstellar medium}

\bibitem[{{Springel} \& {Hernquist}(2003)}]{springel03}
{Springel}, V., \& {Hernquist}, L. 2003, \mnras, 339, 312

\bibitem[{{Steidel} {et~al.}(2010){Steidel}, {Erb}, {Shapley}, {Pettini},
  {Reddy}, {Bogosavljevi{\'c}}, {Rudie}, \& {Rakic}}]{steidel10}
{Steidel}, C.~C., {Erb}, D.~K., {Shapley}, A.~E., {et~al.} 2010, \apj, 717, 289

\bibitem[{{Tielens}(2005)}]{tielens}
{Tielens}, A.~G.~G.~M. 2005, {The Physics and Chemistry of the Interstellar
  Medium}

\bibitem[{Toresson(1960)}]{si4}
Toresson, Y.~G. 1960, Ark. Fys. (Stockholm), 17, 179

\bibitem[{Toresson(1961)}]{si3}
---. 1961, Ark. Fys. (Stockholm), 18, 389

\bibitem[{{Tremonti} {et~al.}(2004){Tremonti}, {Heckman}, {Kauffmann},
  {Brinchmann}, {Charlot}, {White}, {Seibert}, {Peng}, {Schlegel}, {Uomoto},
  {Fukugita}, \& {Brinkmann}}]{tremonti04}
{Tremonti}, C.~A., {Heckman}, T.~M., {Kauffmann}, G., {et~al.} 2004, \apj, 613,
  898

\bibitem[{{Veilleux} {et~al.}(2005){Veilleux}, {Cecil}, \&
  {Bland-Hawthorn}}]{veilleux}
{Veilleux}, S., {Cecil}, G., \& {Bland-Hawthorn}, J. 2005, \araa, 43, 769

\bibitem[{{Wakker} {et~al.}(2015){Wakker}, {Hernandez}, {French}, {Kim},
  {Oppenheimer}, \& {Savage}}]{wakker}
{Wakker}, B.~P., {Hernandez}, A.~K., {French}, D., {et~al.} 2015, ArXiv
  e-prints, arXiv:1504.02539

\bibitem[{{Wakker} {et~al.}(2012){Wakker}, {Savage}, {Fox}, {Benjamin}, \&
  {Shapiro}}]{wakker12}
{Wakker}, B.~P., {Savage}, B.~D., {Fox}, A.~J., {Benjamin}, R.~A., \&
  {Shapiro}, P.~R. 2012, \apj, 749, 157

\bibitem[{{Weaver} {et~al.}(1977){Weaver}, {McCray}, {Castor}, {Shapiro}, \&
  {Moore}}]{weaver77}
{Weaver}, R., {McCray}, R., {Castor}, J., {Shapiro}, P., \& {Moore}, R. 1977,
  \apj, 218, 377

\bibitem[{{Weiner} {et~al.}(2009){Weiner}, {Coil}, {Prochaska}, {Newman},
  {Cooper}, {Bundy}, {Conselice}, {Dutton}, {Faber}, {Koo}, {Lotz}, {Rieke}, \&
  {Rubin}}]{weiner}
{Weiner}, B.~J., {Coil}, A.~L., {Prochaska}, J.~X., {et~al.} 2009, \apj, 692,
  187

\bibitem[{{Wofford} {et~al.}(2013){Wofford}, {Leitherer}, \&
  {Salzer}}]{wofford2013}
{Wofford}, A., {Leitherer}, C., \& {Salzer}, J. 2013, \apj, 765, 118

\bibitem[{{Wood} {et~al.}(2015){Wood}, {Tremonti}, {Calzetti}, {Leitherer},
  {Chisholm}, \& {Gallagher}}]{wood}
{Wood}, C.~M., {Tremonti}, C.~A., {Calzetti}, D., {et~al.} 2015, \mnras, 452,
  2712

\bibitem[{{Wright} {et~al.}(2010){Wright}, {Eisenhardt}, {Mainzer}, {Ressler},
  {Cutri}, {Jarrett}, {Kirkpatrick}, {Padgett}, {McMillan}, {Skrutskie},
  {Stanford}, {Cohen}, {Walker}, {Mather}, {Leisawitz}, {Gautier}, {McLean},
  {Benford}, {Lonsdale}, {Blain}, {Mendez}, {Irace}, {Duval}, {Liu}, {Royer},
  {Heinrichsen}, {Howard}, {Shannon}, {Kendall}, {Walsh}, {Larsen}, {Cardon},
  {Schick}, {Schwalm}, {Abid}, {Fabinsky}, {Naes}, \& {Tsai}}]{wise}
{Wright}, E.~L., {Eisenhardt}, P.~R.~M., {Mainzer}, A.~K., {et~al.} 2010, \aj,
  140, 1868

\end{thebibliography}

\end{document}